\documentclass[11pt]{article}
\pdfoutput=1

\usepackage{jheppub}
\usepackage[utf8]{inputenc}
\usepackage{booktabs}
\usepackage{graphicx}
\usepackage{bbold}
\usepackage{subcaption}
\usepackage{mathtools}

\usepackage{xkeyval}

\usepackage{tikz,fp}
\usetikzlibrary{external,fixedpointarithmetic,decorations.pathmorphing,positioning,calc,fadings,decorations.markings,decorations.pathreplacing,patterns,arrows}

\tikzset{
  label distance=-3pt,
  >=stealth,
  inner sep=1.5pt,
  boson/.style={
    decoration={snake, segment length=2mm, amplitude=0.5mm},
    decorate,
  },
  graviton/.style={
  	line width=0.1pt,
    decoration={snake, segment length=0.7mm, amplitude=0.2mm},
    decorate,
  },
  quark/.style={
    postaction={decorate},
    decoration={markings,mark=at position .5 with {\draw[->] (-0.01,0) -- (0.07,0);}}
  },
  aquark/.style={
    postaction={decorate},
    decoration={markings,mark=at position .5 with {\draw[->] (0.01,0) -- (-0.07,0);}}
  },
  gluonOld/.style={
    line width=0.7pt,
    decorate,
    decoration={coil,aspect=-0.5,amplitude=-2pt,segment length=2.2pt}
  },
  gluon/.style={
    line width=0.5pt,
    decorate,
    decoration={coil,aspect=-0.75,amplitude=-1.5pt,segment length=2pt}
  },
  agluon/.style={
    line width=0.5pt,
    decorate,
    decoration={coil,aspect=0.75,amplitude=1.5pt,segment length=2pt}
  },
  scalar/.style={
    dashed
  },
  dscalar/.style={
    dashed,
    postaction={decorate},
    decoration={markings,mark=at position 0.65 with {\draw[->] (-0.01,0) -- (0.07,0);}}
  },
  adscalar/.style={
    dashed,
    postaction={decorate},
    decoration={markings,mark=at position 0.65 with {\draw[->] (-0.01,0) -- (0.07,0);}}
  },
  line/.style={
  },
  doubleline/.style={
    double
  },
  middleArrow/.style={
    draw=white,
    decoration={markings,mark=at position 0.65 with{\arrow[scale=1,black]{>}}},
    decorate
  },
  dot/.style={
  	postaction={decorate},
    decoration={markings,mark=between positions 0.5 and 0.5 step 1 with {\draw[fill] (0,0) circle [radius=1.5pt];}}  
  },
  dots/.style={
  	postaction={decorate},
    decoration={markings,mark=between positions 0.33 and 0.66 step 0.33 with {\draw[fill] (0,0) circle [radius=1.5pt];}}  
  },
  dotss/.style={
  	postaction={decorate},
    decoration={markings,mark=between positions 0.25 and 0.75 step 0.25 with {\draw[fill] (0,0) circle [radius=1.5pt];}}  
  },
  blob/.style={
  	pattern=north west lines,
    draw=black
  },
  blobBlack/.style={
    draw=black,
    fill=black
  },
  blobWhite/.style={
    draw=black,
    fill=white
  }
}

\makeatletter
\def\gSetStdExtTypes#1{\presetkeys{gTypes}{eA=#1,eB=#1,eC=#1,eD=#1,eE=#1}{}}
\def\gSetStdIntTypes#1{\presetkeys{gTypes}{iA=#1,iB=#1,iC=#1,iD=#1,iE=#1,iF=#1,iG=#1,iH=#1,iI=#1,iJ=#1}{}}
\def\gSetStdTypes#1{\gSetStdExtTypes{#1}\gSetStdIntTypes{#1}}

\define@cmdkeys{general}{scale}
\define@cmdkeys{general}{rotate}
\define@cmdkeys{general}{font}
\define@cmdkeys{general}{yshift}
\presetkeys{general}{scale=1, rotate=0, font=\scriptsize, yshift=0pt}{}

\define@cmdkeys{gTypes}{eA,eB,eC,eD,eE,iA,iB,iC,iD,iE,iF,iG,iH,iI,iJ}
\gSetStdTypes{line}
\define@key{pre}{all}{\gSetStdTypes{#1}}
\define@key{pre}{allE}{\gSetStdExtTypes{#1}}
\define@key{pre}{allI}{\gSetStdIntTypes{#1}}

\define@cmdkeys{gLabels}{eLA,eLB,eLC,eLD,eLE,iLA,iLB,iLC,iLD,iLE,iLF,iLG,iLH,iLI,iLJ}
\presetkeys{gLabels}{eLA={},eLB={},eLC={},eLD={},eLE={},iLA={},iLB={},iLC={},iLD={},iLE={},iLF={},iLG={},iLH={},iLI={},iLJ={}}{}

\define@boolkeys{dots}{lDots,rDots}
\presetkeys{dots}{lDots=false, rDots=false}{}

\define@cmdkeys{blobs}{blobA,blobB,blobC}
\presetkeys{blobs}{blobA=blob,blobB=blob,blobC=blob}{}

\def\gTreeTri{\@ifnextchar[\@gTreeTri{\@gTreeTri[]}}
\def\@gTreeTri[#1]#2{%
  \setkeys*{pre}{#1}%
  \setrmkeys{general,gTypes,gLabels}%
  \begin{tikzpicture}
    [line width=1pt,
    baseline={([yshift=-0.5ex+\cmdKV@general@yshift]current bounding box.center)},
    scale=\cmdKV@general@scale,
    rotate=\cmdKV@general@rotate,
    font=\cmdKV@general@font]
    \draw[\cmdKV@gTypes@eA] (0.7,0.3) -- (0.4,0.3) node[label=(-180+\cmdKV@general@rotate):\cmdKV@gLabels@eLA] {};
    \draw[\cmdKV@gTypes@eB] (0.7,0.3) -- (0.9,0.6) node[label=(\cmdKV@general@rotate):\cmdKV@gLabels@eLB] {};
    \draw[\cmdKV@gTypes@eC] (0.7,0.3) -- (0.9,0) node[label=(\cmdKV@general@rotate):\cmdKV@gLabels@eLC] {};
    #2
  \end{tikzpicture}
  \gSetStdTypes{line}
}

\def\gTreeS{\@ifnextchar[\@gTreeS{\@gTreeS[]}}
\def\@gTreeS[#1]#2{%
  \setkeys*{pre}{#1}%
  \setrmkeys{general,gTypes,gLabels}%
  \begin{tikzpicture}
    [line width=1pt,
    baseline={([yshift=-0.5ex+\cmdKV@general@yshift]current bounding box.center)},
    scale=\cmdKV@general@scale,
    rotate=\cmdKV@general@rotate,
    font=\cmdKV@general@font]
    \draw[\cmdKV@gTypes@eA] (0.2,0.3) -- (0,0) node[label=(-180+\cmdKV@general@rotate):\cmdKV@gLabels@eLA] {};
    \draw[\cmdKV@gTypes@eB] (0.2,0.3) -- (0,0.6) node[label=(180+\cmdKV@general@rotate):\cmdKV@gLabels@eLB] {};
    \draw[\cmdKV@gTypes@eC] (0.7,0.3) -- (0.9,0.6) node[label=(\cmdKV@general@rotate):\cmdKV@gLabels@eLC] {};
    \draw[\cmdKV@gTypes@eD] (0.7,0.3) -- (0.9,0) node[label=(\cmdKV@general@rotate):\cmdKV@gLabels@eLD] {};    
    \draw[label distance=0,\cmdKV@gTypes@iA] (0.2,0.3) -- node[label=(90+\cmdKV@general@rotate):\cmdKV@gLabels@iLA] {} (0.7,0.3);
    #2
  \end{tikzpicture}
  \gSetStdTypes{line}
}

\def\gTreeT{\@ifnextchar[\@gTreeT{\@gTreeT[]}}
\def\@gTreeT[#1]#2{%
  \setkeys*{pre}{#1}%
  \setrmkeys{general,gTypes,gLabels}%
  \begin{tikzpicture}
    [line width=1pt,
    baseline={([yshift=-0.5ex+\cmdKV@general@yshift]current bounding box.center)},
    scale=\cmdKV@general@scale,
    rotate=\cmdKV@general@rotate,
    font=\cmdKV@general@font]
    \draw[\cmdKV@gTypes@eA] (0.45,0.1) -- (0,0) node[label=(-180+\cmdKV@general@rotate):\cmdKV@gLabels@eLA] {};
    \draw[\cmdKV@gTypes@eB] (0.45,0.5) -- (0,0.6) node[label=(180+\cmdKV@general@rotate):\cmdKV@gLabels@eLB] {};
    \draw[\cmdKV@gTypes@eC] (0.45,0.5) -- (0.9,0.6) node[label=(\cmdKV@general@rotate):\cmdKV@gLabels@eLC] {};
    \draw[\cmdKV@gTypes@eD] (0.45,0.1) -- (0.9,0) node[label=(\cmdKV@general@rotate):\cmdKV@gLabels@eLD] {};
    \draw[label distance=0,\cmdKV@gTypes@iA] (0.45,0.5) -- node[label=(\cmdKV@general@rotate):\cmdKV@gLabels@iLA] {} (0.45,0.1);
    #2
  \end{tikzpicture}
  \gSetStdTypes{line}
}

\def\gTreeU{\@ifnextchar[\@gTreeU{\@gTreeU[]}}
\def\@gTreeU[#1]#2{%
  \setkeys*{pre}{#1}%
  \setrmkeys{general,gTypes,gLabels}%
  \begin{tikzpicture}
    [line width=1pt,
    baseline={([yshift=-0.5ex+\cmdKV@general@yshift]current bounding box.center)},
    scale=\cmdKV@general@scale,
    rotate=\cmdKV@general@rotate,
    font=\cmdKV@general@font]
    \draw[\cmdKV@gTypes@eA] (0.45,0.5) -- (0,0) node[label=(-180+\cmdKV@general@rotate):\cmdKV@gLabels@eLA] {};
    \foreach \x in {5,...,14} \fill[opacity=0.12,white] (0.27,0.3) circle (\x/100);
    \draw[\cmdKV@gTypes@eB] (0.45,0.1) -- (0,0.6) node[label=(180+\cmdKV@general@rotate):\cmdKV@gLabels@eLB] {};
    \draw[\cmdKV@gTypes@eC] (0.45,0.5) -- (0.9,0.6) node[label=(\cmdKV@general@rotate):\cmdKV@gLabels@eLC] {};
    \draw[\cmdKV@gTypes@eD] (0.45,0.1) -- (0.9,0) node[label=(\cmdKV@general@rotate):\cmdKV@gLabels@eLD] {};
    \draw[label distance=0,\cmdKV@gTypes@iA] (0.45,0.5) -- node[label=(\cmdKV@general@rotate):\cmdKV@gLabels@iLA] {} (0.45,0.1);
    #2
  \end{tikzpicture}
  \gSetStdTypes{line}
}

\def\gTreeFourPt{\@ifnextchar[\@gTreeFourPt{\@gTreeFourPt[]}}
\def\@gTreeFourPt[#1]#2{%
  \setkeys*{pre}{#1}%
  \setrmkeys{general,gTypes,gLabels}%
  \begin{tikzpicture}
    [line width=1pt,
    baseline={([yshift=-0.5ex+\cmdKV@general@yshift]current bounding box.center)},
    scale=\cmdKV@general@scale,
    rotate=\cmdKV@general@rotate,
    font=\cmdKV@general@font]
    \draw[\cmdKV@gTypes@eA] (0,0) -- (-0.4,-0.4) node[label=(-180+\cmdKV@general@rotate):\cmdKV@gLabels@eLA] {};
    \draw[\cmdKV@gTypes@eB] (0,0) -- (-0.4,0.4) node[label=(180+\cmdKV@general@rotate):\cmdKV@gLabels@eLB] {};
    \draw[\cmdKV@gTypes@eC] (0,0) -- (0.4,0.4) node[label=(\cmdKV@general@rotate):\cmdKV@gLabels@eLC] {};
    \draw[\cmdKV@gTypes@eD] (0,0) -- (0.4,-0.4) node[label=(\cmdKV@general@rotate):\cmdKV@gLabels@eLD] {};
    #2
  \end{tikzpicture}
  \gSetStdTypes{line}
}

\def\gTadA{\@ifnextchar[\@gTadA{\@gTadA[]}}
\def\@gTadA[#1]#2{%
  \setkeys*{pre}{#1}%
  \setrmkeys{general,gTypes,gLabels}%
  \begin{tikzpicture}
    [line width=1pt,
    baseline={([yshift=-0.5ex+\cmdKV@general@yshift]current bounding box.center)},
    scale=\cmdKV@general@scale,
    rotate=\cmdKV@general@rotate,
    font=\cmdKV@general@font]
    \draw[\cmdKV@gTypes@eA] (-0.7,0.3) -- (-0.9,0) node[label=(180+\cmdKV@general@rotate):\cmdKV@gLabels@eLA] {};
    \draw[\cmdKV@gTypes@eB] (-0.7,0.3) -- (-0.9,0.6) node[label=(180+\cmdKV@general@rotate):\cmdKV@gLabels@eLB] {};
    \draw[\cmdKV@gTypes@eC] (0.15,0.45) -- (0.5,0.6) node[label=(\cmdKV@general@rotate):\cmdKV@gLabels@eLC] {};
    \draw[\cmdKV@gTypes@eD] (-0.3,0.3) -- (-0.2,0) node[label=(\cmdKV@general@rotate):\cmdKV@gLabels@eLD] {};
    \draw[label distance=0,\cmdKV@gTypes@iA] (-0.7,0.3) -- node[label=(90+\cmdKV@general@rotate):\cmdKV@gLabels@iLA] {} (-0.3,0.3);
    \draw[label distance=0,\cmdKV@gTypes@iB] (-0.3,0.3) -- node[label=(-80+\cmdKV@general@rotate):\cmdKV@gLabels@iLB] {} (0.15,0.45);
    \draw[label distance=0,\cmdKV@gTypes@iC] (0.15,0.45) -- node[label=(45+\cmdKV@general@rotate):\cmdKV@gLabels@iLC] {} (-0.18,0.69);
    \draw[label distance=-0.5,\cmdKV@gTypes@iD] (-0.18,0.69) .. controls ($(-0.18,0.69) + (-0.12,-0.12)$) and ($(-0.35,0.87) + (-0.12,-0.12)$) .. (-0.35,0.87) node[label=(180+\cmdKV@general@rotate):\cmdKV@gLabels@iLD] {} .. controls ($(-0.35,0.87) + (0.12,0.12)$) and ($(-0.18,0.69) + (0.12,0.12)$) .. (-0.18,0.69);
    #2
  \end{tikzpicture}
  \gSetStdTypes{line}
}

\def\gTadB{\@ifnextchar[\@gTadB{\@gTadB[]}}
\def\@gTadB[#1]#2{%
  \setkeys*{pre}{#1}%
  \setrmkeys{general,gTypes,gLabels}%
  \begin{tikzpicture}
    [line width=1pt,
    baseline={([yshift=-0.5ex+\cmdKV@general@yshift]current bounding box.center)},
    scale=\cmdKV@general@scale,
    rotate=\cmdKV@general@rotate,
    font=\cmdKV@general@font]
    \draw[\cmdKV@gTypes@eA] (0,0.3) -- (-0.2,0) node[label=(-180+\cmdKV@general@rotate):\cmdKV@gLabels@eLA] {};
    \draw[\cmdKV@gTypes@eB] (0,0.3) -- (-0.2,0.6) node[label=(180+\cmdKV@general@rotate):\cmdKV@gLabels@eLB] {};
    \draw[\cmdKV@gTypes@eC] (0.7,0.3) -- (0.9,0.6) node[label=(\cmdKV@general@rotate):\cmdKV@gLabels@eLC] {};
    \draw[\cmdKV@gTypes@eD] (0.7,0.3) -- (0.9,0) node[label=(\cmdKV@general@rotate):\cmdKV@gLabels@eLD] {};
    \draw[label distance=0,\cmdKV@gTypes@iA] (0,0.3) -- node[label=(-90+\cmdKV@general@rotate):\cmdKV@gLabels@iLA] {} (0.35,0.3);
    \draw[label distance=0,\cmdKV@gTypes@iB] (0.35,0.3) -- node[label=(\cmdKV@general@rotate):\cmdKV@gLabels@iLB] {} (0.35,0.6);
    \draw[label distance=-0.5,\cmdKV@gTypes@iC] (0.35,0.6) .. controls (0.18,0.6) and (0.18,0.85) .. (0.35,0.85) .. controls (0.52,0.85) and (0.52,0.6) .. node[label=(\cmdKV@general@rotate):\cmdKV@gLabels@iLC] {} (0.35,0.6);
    \draw[label distance=0,\cmdKV@gTypes@iD] (0.35,0.3) -- node[label=(-90+\cmdKV@general@rotate):\cmdKV@gLabels@iLD] {} (0.7,0.3);
    #2
  \end{tikzpicture}
  \gSetStdTypes{line}
}
\def\gBubA{\@ifnextchar[\@gBubA{\@gBubA[]}}
\def\@gBubA[#1]#2{%
  \setkeys*{pre}{#1}%
  \setrmkeys{general,gTypes,gLabels}%
  \begin{tikzpicture}
    [line width=1pt,
    baseline={([yshift=-0.5ex+\cmdKV@general@yshift]current bounding box.center)},
    scale=\cmdKV@general@scale,
    rotate=\cmdKV@general@rotate,
    font=\cmdKV@general@font]
    \draw[\cmdKV@gTypes@eA] (0.5,0.7) -- (0.2,0.7) node[label=(180+\cmdKV@general@rotate):\cmdKV@gLabels@eLA] {};
    \draw[\cmdKV@gTypes@eB] (0.9,0.7) -- (1.2,0.7) node[label=(\cmdKV@general@rotate):\cmdKV@gLabels@eLB] {};
    \draw[label distance=0,\cmdKV@gTypes@iA] (0.5,0.7) .. controls (0.5,0.95) and (0.9,0.95) .. node[label=(90+\cmdKV@general@rotate):\cmdKV@gLabels@iLA] {} (0.9,0.7);
    \draw[label distance=0,\cmdKV@gTypes@iB] (0.9,0.7) .. controls (0.9,0.45) and (0.5,0.45) .. node[label=(-90+\cmdKV@general@rotate):\cmdKV@gLabels@iLB] {} (0.5,0.7);
    #2
  \end{tikzpicture}
  \gSetStdTypes{line}
}
\def\gBubB{\@ifnextchar[\@gBubB{\@gBubB[]}}
\def\@gBubB[#1]#2{%
  \setkeys*{pre}{#1}%
  \setrmkeys{general,gTypes,gLabels}%
  \begin{tikzpicture}
    [line width=1pt,
    baseline={([yshift=-0.5ex+\cmdKV@general@yshift]current bounding box.center)},
    scale=\cmdKV@general@scale,
    rotate=\cmdKV@general@rotate,
    font=\cmdKV@general@font]
    \draw[\cmdKV@gTypes@eA] (0.2,0.7) -- (0,0.4) node[label=(-180+\cmdKV@general@rotate):\cmdKV@gLabels@eLA] {};
    \draw[\cmdKV@gTypes@eB] (0.2,0.7) -- (0,1) node[label=(180+\cmdKV@general@rotate):\cmdKV@gLabels@eLB] {};
    \draw[\cmdKV@gTypes@eC] (1.2,0.7) -- (1.4,1) node[label=(\cmdKV@general@rotate):\cmdKV@gLabels@eLC] {};
    \draw[\cmdKV@gTypes@eD] (1.2,0.7) -- (1.4,0.4) node[label=(\cmdKV@general@rotate):\cmdKV@gLabels@eLD] {};
    \draw[label distance=0,\cmdKV@gTypes@iA] (0.2,0.7) -- node[label=(90+\cmdKV@general@rotate):\cmdKV@gLabels@iLA] {} (0.5,0.7);
    \draw[label distance=0,\cmdKV@gTypes@iB] (0.5,0.7) .. controls (0.5,0.95) and (0.9,0.95) .. node[label=(90+\cmdKV@general@rotate):\cmdKV@gLabels@iLB] {} (0.9,0.7);
    \draw[label distance=0,\cmdKV@gTypes@iC] (0.9,0.7) .. controls (0.9,0.45) and (0.5,0.45) .. node[label=(-90+\cmdKV@general@rotate):\cmdKV@gLabels@iLC] {} (0.5,0.7);
    \draw[label distance=0,\cmdKV@gTypes@iD] (0.9,0.7) -- node[label=(90+\cmdKV@general@rotate):\cmdKV@gLabels@iLD] {} (1.2,0.7);
    #2
  \end{tikzpicture}
  \gSetStdTypes{line}
}

\def\gBubC{\@ifnextchar[\@gBubC{\@gBubC[]}}
\def\@gBubC[#1]#2{%
  \setkeys*{pre}{#1}%
  \setrmkeys{general,gTypes,gLabels}%
  \begin{tikzpicture}
    [line width=1pt,
    baseline={([yshift=-0.5ex+\cmdKV@general@yshift]current bounding box.center)},
    scale=\cmdKV@general@scale,
    rotate=\cmdKV@general@rotate,
    font=\cmdKV@general@font]
    \draw[\cmdKV@gTypes@eA] (-0.7,0.3) -- (-0.9,0) node[label=(180+\cmdKV@general@rotate):\cmdKV@gLabels@eLA] {};
    \draw[\cmdKV@gTypes@eB] (-0.7,0.3) -- (-0.9,0.6) node[label=(180+\cmdKV@general@rotate):\cmdKV@gLabels@eLB] {};
    \draw[\cmdKV@gTypes@eC] (0.3,0.5) -- (0.5,0.6) node[label=(\cmdKV@general@rotate):\cmdKV@gLabels@eLC] {};
    \draw[\cmdKV@gTypes@eD] (-0.3,0.3) -- (-0.2,0) node[label=(\cmdKV@general@rotate):\cmdKV@gLabels@eLD] {};
    \draw[label distance=0,\cmdKV@gTypes@iA] (-0.7,0.3) -- node[label=(90+\cmdKV@general@rotate):\cmdKV@gLabels@iLA] {} (-0.3,0.3);
    \draw[label distance=0,\cmdKV@gTypes@iB] (-0.3,0.3) -- node[label=(90+\cmdKV@general@rotate):\cmdKV@gLabels@iLB] {} (-0,0.4);
    \draw[label distance=0,\cmdKV@gTypes@iC] (0,0.4) .. controls (-0.03,0.6) and (0.22,0.65) .. node[label=(90+\cmdKV@general@rotate):\cmdKV@gLabels@iLC] {} (0.3,0.5);
    \draw[label distance=0,\cmdKV@gTypes@iD] (0.3,0.5) .. controls (0.38,0.3) and (0.08,0.21) .. node[label=(-90+\cmdKV@general@rotate):\cmdKV@gLabels@iLD] {} (0,0.4);
    #2
  \end{tikzpicture}
  \gSetStdTypes{line}
}
\def\gTriA{\@ifnextchar[\@gTriA{\@gTriA[]}}
\def\@gTriA[#1]#2{%
  \setkeys*{pre}{#1}%
  \setrmkeys{general,gTypes,gLabels}%
  \begin{tikzpicture}
    [line width=1pt,
    baseline={([yshift=-0.5ex+\cmdKV@general@yshift]current bounding box.center)},
    scale=\cmdKV@general@scale,
    rotate=\cmdKV@general@rotate,
    font=\cmdKV@general@font]
    \draw[\cmdKV@gTypes@eC] (0:0.3) -- (0:0.55) node[label=(\cmdKV@general@rotate):\cmdKV@gLabels@eLC] {};
    \draw[\cmdKV@gTypes@eA] (-120:0.3) -- (-120:0.55) node[label=(-180+\cmdKV@general@rotate):\cmdKV@gLabels@eLA] {};
    \draw[\cmdKV@gTypes@eB] (120:0.3) -- (120:0.55) node[label=(180+\cmdKV@general@rotate):\cmdKV@gLabels@eLB] {};    
    \draw[label distance=0,\cmdKV@gTypes@iA] (-120:0.3) -- node[label=(180+\cmdKV@general@rotate):\cmdKV@gLabels@iLA] {} (120:0.3);
    \draw[label distance=0,\cmdKV@gTypes@iB] (120:0.3) -- node[label=(60+\cmdKV@general@rotate):\cmdKV@gLabels@iLB] {} (0:0.3);
    \draw[label distance=0,\cmdKV@gTypes@iC] (0:0.3) -- node[label=(-60+\cmdKV@general@rotate):\cmdKV@gLabels@iLC] {} (-120:0.3);
    #2
  \end{tikzpicture}
  \gSetStdTypes{line}
}
\def\gTriB{\@ifnextchar[\@gTriB{\@gTriB[]}}
\def\@gTriB[#1]#2{%
  \setkeys*{pre}{#1}%
  \setrmkeys{general,gTypes,gLabels}%
  \begin{tikzpicture}
    [line width=1pt,
    baseline={([yshift=-0.5ex+\cmdKV@general@yshift]current bounding box.center)},
    scale=\cmdKV@general@scale,
    rotate=\cmdKV@general@rotate,
    font=\cmdKV@general@font]
    \draw[\cmdKV@gTypes@eA] (0.3,0.45) -- (0,0.3) node[label=(-180+\cmdKV@general@rotate):\cmdKV@gLabels@eLA] {};
    \draw[\cmdKV@gTypes@eB] (0.3,0.95) -- (0,1.1) node[label=(180+\cmdKV@general@rotate):\cmdKV@gLabels@eLB] {};
    \draw[\cmdKV@gTypes@eC] (1,0.7) -- (1.3,1.1) node[label=(\cmdKV@general@rotate):\cmdKV@gLabels@eLC] {};
    \draw[\cmdKV@gTypes@eD] (1,0.7) -- (1.3,0.3) node[label=(\cmdKV@general@rotate):\cmdKV@gLabels@eLD] {};
    \draw[label distance=0,\cmdKV@gTypes@iA] (0.3,0.45) -- node[label=(180+\cmdKV@general@rotate):\cmdKV@gLabels@iLA] {} (0.3,0.95);
    \draw[label distance=0,\cmdKV@gTypes@iB] (0.3,0.95) -- node[label=(70+\cmdKV@general@rotate):\cmdKV@gLabels@iLB] {} (0.7,0.7);
    \draw[label distance=0,\cmdKV@gTypes@iC] (0.7,0.7) -- node[label=(-70+\cmdKV@general@rotate):\cmdKV@gLabels@iLC] {} (0.3,0.45);
    \draw[label distance=0,\cmdKV@gTypes@iD] (1,0.7) -- node[label=(90+\cmdKV@general@rotate):\cmdKV@gLabels@iLD] {} (0.7,0.7);
    #2
  \end{tikzpicture}
  \gSetStdTypes{line}
}
\def\gBox{\@ifnextchar[\@gBox{\@gBox[]}}
\def\@gBox[#1]#2{%
  \setkeys*{pre}{#1}%
  \setrmkeys{general,gTypes,gLabels}%
  \begin{tikzpicture}
    [line width=1pt,
    baseline={([yshift=-0.5ex+\cmdKV@general@yshift]current bounding box.center)},
    scale=\cmdKV@general@scale,
    rotate=\cmdKV@general@rotate,
    font=\cmdKV@general@font]
    \draw[\cmdKV@gTypes@eA] (0.25,0.25) -- (0,0) node[label=(-180+\cmdKV@general@rotate):\cmdKV@gLabels@eLA] {};
    \draw[\cmdKV@gTypes@eB] (0.25,0.75) -- (0,1) node[label=(180+\cmdKV@general@rotate):\cmdKV@gLabels@eLB] {};
    \draw[\cmdKV@gTypes@eC] (0.75,0.75) -- (1,1) node[label=(\cmdKV@general@rotate):\cmdKV@gLabels@eLC] {};
    \draw[\cmdKV@gTypes@eD] (0.75,0.25) -- (1,0) node[label=(\cmdKV@general@rotate):\cmdKV@gLabels@eLD] {};
    \draw[label distance=0,\cmdKV@gTypes@iA] (0.25,0.25) -- node[label=(180+\cmdKV@general@rotate):\cmdKV@gLabels@iLA] {} (0.25,0.75);
    \draw[label distance=0,\cmdKV@gTypes@iB] (0.25,0.75) -- node[label=(90+\cmdKV@general@rotate):\cmdKV@gLabels@iLB] {} (0.75,0.75);
    \draw[label distance=0,\cmdKV@gTypes@iC] (0.75,0.75) -- node[label=(\cmdKV@general@rotate):\cmdKV@gLabels@iLC] {} (0.75,0.25);
    \draw[label distance=0,\cmdKV@gTypes@iD] (0.75,0.25) -- node[label=(-90+\cmdKV@general@rotate):\cmdKV@gLabels@iLD] {} (0.25,0.25);
    #2
  \end{tikzpicture}
  \gSetStdTypes{line}
}
\def\gPenta{\@ifnextchar[\@gPenta{\@gPenta[]}}
\def\@gPenta[#1]#2{%
  \setkeys*{pre}{#1}%
  \setrmkeys{general,gTypes,gLabels}%
  \begin{tikzpicture}
    [line width=1pt,
    baseline={([yshift=-0.5ex+\cmdKV@general@yshift]current bounding box.center)},
    scale=\cmdKV@general@scale,
    rotate=\cmdKV@general@rotate,
    font=\cmdKV@general@font]
    \draw[\cmdKV@gTypes@eA] (-144:0.35) -- (-144:0.75) node[label=(-144+\cmdKV@general@rotate):\cmdKV@gLabels@eLA] {};
    \draw[\cmdKV@gTypes@eB] (144:0.35) -- (144:0.75) node[label=(144+\cmdKV@general@rotate):\cmdKV@gLabels@eLB] {};
    \draw[\cmdKV@gTypes@eC] (72:0.35) -- (72:0.75) node[label=(72+\cmdKV@general@rotate):\cmdKV@gLabels@eLC] {};
    \draw[\cmdKV@gTypes@eD] (0:0.35) -- (0:0.75) node[label=(\cmdKV@general@rotate):\cmdKV@gLabels@eLD] {};
    \draw[\cmdKV@gTypes@eE] (-72:0.35) -- (-72:0.75) node[label=(-72+\cmdKV@general@rotate):\cmdKV@gLabels@eLE] {};
    \draw[label distance=0,\cmdKV@gTypes@iA] (-144:0.35) -- node[label=(180+\cmdKV@general@rotate):\cmdKV@gLabels@iLA] {} (144:0.35);
    \draw[label distance=0,\cmdKV@gTypes@iB] (144:0.35) -- node[label=(108+\cmdKV@general@rotate):\cmdKV@gLabels@iLB] {} (72:0.35);
    \draw[label distance=0,\cmdKV@gTypes@iC] (72:0.35) -- node[label=(36+\cmdKV@general@rotate):\cmdKV@gLabels@iLC] {} (0:0.35);
    \draw[label distance=0,\cmdKV@gTypes@iD] (0:0.35) -- node[label=(-36+\cmdKV@general@rotate):\cmdKV@gLabels@iLD] {} (-72:0.35);
    \draw[label distance=0,\cmdKV@gTypes@iE] (-72:0.35) -- node[label=(-108+\cmdKV@general@rotate):\cmdKV@gLabels@iLE] {} (-144:0.35);
    #2
  \end{tikzpicture}
  \gSetStdTypes{line}
}

\def\gBoxBox{\@ifnextchar[\@gBoxBox{\@gBoxBox[]}}
\def\@gBoxBox[#1]#2{%
  \setkeys*{pre}{#1}%
  \setrmkeys{general,gTypes,gLabels}%
  \begin{tikzpicture}
    [line width=1pt,
    baseline={([yshift=-0.5ex+\cmdKV@general@yshift]current bounding box.center)},
    scale=\cmdKV@general@scale,
    rotate=\cmdKV@general@rotate,
    font=\cmdKV@general@font]
    \draw[\cmdKV@gTypes@eA] (0.3,0.3) -- (0,0) node[label=(-180+\cmdKV@general@rotate):\cmdKV@gLabels@eLA] {};
    \draw[\cmdKV@gTypes@eB] (0.3,0.9) -- (0,1.2) node[label=(180+\cmdKV@general@rotate):\cmdKV@gLabels@eLB] {};
    \draw[\cmdKV@gTypes@eC] (1.5,0.9) -- (1.8,1.2) node[label=(\cmdKV@general@rotate):\cmdKV@gLabels@eLC] {};
    \draw[\cmdKV@gTypes@eD] (1.5,0.3) -- (1.8,0) node[label=(\cmdKV@general@rotate):\cmdKV@gLabels@eLD] {};
    \draw[label distance=0,\cmdKV@gTypes@iA] (0.3,0.3) -- node[label=(180+\cmdKV@general@rotate):\cmdKV@gLabels@iLA] {} (0.3,0.9);
    \draw[label distance=0,\cmdKV@gTypes@iB] (0.3,0.9) -- node[label=(90+\cmdKV@general@rotate):\cmdKV@gLabels@iLB] {} (0.9,0.9);
    \draw[label distance=0,\cmdKV@gTypes@iC] (0.9,0.9) -- node[label=(90+\cmdKV@general@rotate):\cmdKV@gLabels@iLC] {} (1.5,0.9);
    \draw[label distance=0,\cmdKV@gTypes@iD] (1.5,0.9) -- node[label=(\cmdKV@general@rotate):\cmdKV@gLabels@iLD] {} (1.5,0.3);
    \draw[label distance=0,\cmdKV@gTypes@iE] (1.5,0.3) -- node[label=(-90+\cmdKV@general@rotate):\cmdKV@gLabels@iLE] {} (0.9,0.3);
    \draw[label distance=0,\cmdKV@gTypes@iF] (0.9,0.3) -- node[label=(-90+\cmdKV@general@rotate):\cmdKV@gLabels@iLF] {} (0.3,0.3);
    \draw[label distance=0,\cmdKV@gTypes@iG] (0.9,0.9) -- node[label=(\cmdKV@general@rotate):\cmdKV@gLabels@iLG] {} (0.9,0.3);
    #2
  \end{tikzpicture}
  \gSetStdTypes{line}
}
\def\gBoxBoxNP{\@ifnextchar[\@gBoxBoxNP{\@gBoxBoxNP[]}}
\def\@gBoxBoxNP[#1]#2{%
  \setkeys*{pre}{#1}%
  \setrmkeys{general,gTypes,gLabels}%
  \begin{tikzpicture}
    [line width=1pt,
    baseline={([yshift=-0.5ex+\cmdKV@general@yshift]current bounding box.center)},
    scale=\cmdKV@general@scale,
    rotate=\cmdKV@general@rotate,
    font=\cmdKV@general@font]
    \draw[\cmdKV@gTypes@eA] (-1.9,0.3) -- (-2.2,0) node[label=(180+\cmdKV@general@rotate):\cmdKV@gLabels@eLA] {};
    \draw[\cmdKV@gTypes@eB] (-1.9,1.1) -- (-2.2,1.4) node[label=(180+\cmdKV@general@rotate):\cmdKV@gLabels@eLB] {};
    \draw[\cmdKV@gTypes@eC] (-0.3,0.7) -- (0,0.7) node[label=(\cmdKV@general@rotate):\cmdKV@gLabels@eLC] {};
    \draw[\cmdKV@gTypes@eD] (-1.1,0.7) -- (-1.4,0.7) node[label=(180+\cmdKV@general@rotate):\cmdKV@gLabels@eLD] {};
    \draw[label distance=0,\cmdKV@gTypes@iA] (-1.9,0.3) -- node[label=(180+\cmdKV@general@rotate):\cmdKV@gLabels@iLA] {} (-1.9,1.1);
    \draw[label distance=0,\cmdKV@gTypes@iB] (-1.9,1.1) -- node[label=(90+\cmdKV@general@rotate):\cmdKV@gLabels@iLB] {} (-0.7,1.1);
    \draw[label distance=0,\cmdKV@gTypes@iC] (-0.7,1.1) -- node[label=(45+\cmdKV@general@rotate):\cmdKV@gLabels@iLC] {} (-0.3,0.7);
    \draw[label distance=0,\cmdKV@gTypes@iD] (-0.3,0.7) -- node[label=(-45+\cmdKV@general@rotate):\cmdKV@gLabels@iLD] {} (-0.7,0.3);
    \draw[label distance=0,\cmdKV@gTypes@iE] (-0.7,0.3) -- node[label=(-90+\cmdKV@general@rotate):\cmdKV@gLabels@iLE] {} (-1.9,0.3);
    \draw[label distance=0,\cmdKV@gTypes@iF] (-0.7,1.1) -- node[label=(180+\cmdKV@general@rotate):\cmdKV@gLabels@iLF] {} (-1.1,0.7);
    \draw[label distance=0,\cmdKV@gTypes@iG] (-1.1,0.7) -- node[label=(-180+\cmdKV@general@rotate):\cmdKV@gLabels@iLG] {} (-0.7,0.3);
    #2
  \end{tikzpicture}
  \gSetStdTypes{line}
}

\def\gBoxBoxNPB{\@ifnextchar[\@gBoxBoxNPB{\@gBoxBoxNPB[]}}
\def\@gBoxBoxNPB[#1]#2{%
  \setkeys*{pre}{#1}%
  \setrmkeys{general,gTypes,gLabels}%
  \begin{tikzpicture}
    [line width=1pt,
    baseline={([yshift=-0.5ex+\cmdKV@general@yshift]current bounding box.center)},
    scale=\cmdKV@general@scale,
    rotate=\cmdKV@general@rotate,
    font=\cmdKV@general@font]
    \draw[\cmdKV@gTypes@eA] (0.3,0.3) -- (0,0) node[label=(-180+\cmdKV@general@rotate):\cmdKV@gLabels@eLA] {};
    \draw[\cmdKV@gTypes@eB] (0.3,0.9) -- (0,1.2) node[label=(180+\cmdKV@general@rotate):\cmdKV@gLabels@eLB] {};
    \draw[\cmdKV@gTypes@eC] (1.5,0.9) -- (1.8,1.2) node[label=(\cmdKV@general@rotate):\cmdKV@gLabels@eLC] {};
    \draw[\cmdKV@gTypes@eD] (1.5,0.3) -- (1.8,0) node[label=(\cmdKV@general@rotate):\cmdKV@gLabels@eLD] {};
    \draw[label distance=0,\cmdKV@gTypes@iA] (0.3,0.3) -- node[label=(180+\cmdKV@general@rotate):\cmdKV@gLabels@iLA] {} (0.3,0.9);
    \draw[label distance=0,\cmdKV@gTypes@iB] (0.3,0.9) -- node[label=(90+\cmdKV@general@rotate):\cmdKV@gLabels@iLB] {} (0.9,0.9);
    \draw[label distance=0,\cmdKV@gTypes@iC] (0.9,0.9) -- node[label=(90+\cmdKV@general@rotate):\cmdKV@gLabels@iLC] {} (1.5,0.3);
    \foreach \x in {5,...,14} \fill[opacity=0.12,white] (1.2,0.6) circle (\x/100);
    \draw[label distance=0,\cmdKV@gTypes@iD] (1.5,0.3) -- node[label=(\cmdKV@general@rotate):\cmdKV@gLabels@iLD] {} (1.5,0.9);
    \draw[label distance=0,\cmdKV@gTypes@iE] (1.5,0.9) -- node[label=(-90+\cmdKV@general@rotate):\cmdKV@gLabels@iLE] {} (0.9,0.3);
    \draw[label distance=0,\cmdKV@gTypes@iF] (0.9,0.3) -- node[label=(-90+\cmdKV@general@rotate):\cmdKV@gLabels@iLF] {} (0.3,0.3);
    \draw[label distance=-1,\cmdKV@gTypes@iG] (0.9,0.9) -- node[label=(180+\cmdKV@general@rotate):\cmdKV@gLabels@iLG] {} (0.9,0.3);
    #2
  \end{tikzpicture}
  \gSetStdTypes{line}
}

\def\gBoxBoxNPC{\@ifnextchar[\@gBoxBoxNPC{\@gBoxBoxNPC[]}}
\def\@gBoxBoxNPC[#1]#2{%
  \setkeys*{pre}{#1}%
  \setrmkeys{general,gTypes,gLabels}%
  \begin{tikzpicture}
    [line width=1pt,
    baseline={([yshift=-0.5ex+\cmdKV@general@yshift]current bounding box.center)},
    scale=\cmdKV@general@scale,
    rotate=\cmdKV@general@rotate,
    font=\cmdKV@general@font]
    \draw[\cmdKV@gTypes@eA] (0.3,0.3) -- (0,0) node[label=(-180+\cmdKV@general@rotate):\cmdKV@gLabels@eLA] {};
    \draw[\cmdKV@gTypes@eB] (0.3,0.9) -- (0,1.2) node[label=(180+\cmdKV@general@rotate):\cmdKV@gLabels@eLB] {};
    \draw[\cmdKV@gTypes@eC] (1.5,0.9) -- (1.8,1.2) node[label=(\cmdKV@general@rotate):\cmdKV@gLabels@eLC] {};
    \draw[\cmdKV@gTypes@eD] (1.5,0.3) -- (1.8,0) node[label=(\cmdKV@general@rotate):\cmdKV@gLabels@eLD] {};
    \draw[label distance=0,\cmdKV@gTypes@iA] (0.3,0.3) -- node[label=(180+\cmdKV@general@rotate):\cmdKV@gLabels@iLA] {} (0.3,0.9);
    \draw[label distance=0,\cmdKV@gTypes@iB] (0.3,0.9) -- node[label=(90+\cmdKV@general@rotate):\cmdKV@gLabels@iLB] {} (0.9,0.9);
    \draw[label distance=0,\cmdKV@gTypes@iC] (0.9,0.9) -- node[label=(90+\cmdKV@general@rotate):\cmdKV@gLabels@iLC] {} (1.5,0.9);
    \draw[label distance=0,\cmdKV@gTypes@iD] (1.5,0.9) -- node[label=(\cmdKV@general@rotate):\cmdKV@gLabels@iLD] {} (0.9,0.3);
    \foreach \x in {5,...,14} \fill[opacity=0.12,white] (1.2,0.6) circle (\x/100);
    \draw[label distance=0,\cmdKV@gTypes@iE] (1.5,0.3) -- node[label=(-90+\cmdKV@general@rotate):\cmdKV@gLabels@iLE] {} (0.9,0.3);
    \draw[label distance=0,\cmdKV@gTypes@iF] (0.9,0.3) -- node[label=(-90+\cmdKV@general@rotate):\cmdKV@gLabels@iLF] {} (0.3,0.3);
    \draw[label distance=-1,\cmdKV@gTypes@iG] (0.9,0.9) -- node[label=(180+\cmdKV@general@rotate):\cmdKV@gLabels@iLG] {} (1.5,0.3);
    #2
  \end{tikzpicture}
  \gSetStdTypes{line}
}

\def\gBoxTriNPA{\@ifnextchar[\@gBoxTriNPA{\@gBoxTriNPA[]}}
\def\@gBoxTriNPA[#1]#2{%
  \setkeys*{pre}{#1}%
  \setrmkeys{general,gTypes,gLabels}%
  \begin{tikzpicture}
    [line width=1pt,
    baseline={([yshift=-0.5ex+\cmdKV@general@yshift]current bounding box.center)},
    scale=\cmdKV@general@scale,
    rotate=\cmdKV@general@rotate,
    font=\cmdKV@general@font]
    \draw[\cmdKV@gTypes@eA] (0.3,0.6) -- (0,0) node[label=(-180+\cmdKV@general@rotate):\cmdKV@gLabels@eLA] {};
    \draw[\cmdKV@gTypes@eB] (0.3,0.6) -- (0,1.2) node[label=(180+\cmdKV@general@rotate):\cmdKV@gLabels@eLB] {};
    \draw[\cmdKV@gTypes@eC] (1.5,0.9) -- (1.8,1.2) node[label=(\cmdKV@general@rotate):\cmdKV@gLabels@eLC] {};
    \draw[\cmdKV@gTypes@eD] (1.5,0.3) -- (1.8,0) node[label=(\cmdKV@general@rotate):\cmdKV@gLabels@eLD] {};
    \draw[label distance=0,\cmdKV@gTypes@iA] (0.3,0.6) -- node[label=(90+\cmdKV@general@rotate):\cmdKV@gLabels@iLA] {} (0.9,0.9);
    \draw[label distance=0,\cmdKV@gTypes@iB] (0.9,0.9) -- node[label=(90+\cmdKV@general@rotate):\cmdKV@gLabels@iLB] {} (1.5,0.9);
    \draw[label distance=-1,\cmdKV@gTypes@iC] (0.9,0.9) -- node[label=(180+\cmdKV@general@rotate):\cmdKV@gLabels@iLC] {} (1.5,0.3);
    \foreach \x in {5,...,14} \fill[opacity=0.12,white] (1.2,0.6) circle (\x/100);
    \draw[label distance=0,\cmdKV@gTypes@iD] (1.5,0.3) -- node[label=(-90+\cmdKV@general@rotate):\cmdKV@gLabels@iLD] {} (0.9,0.3);
    \draw[label distance=0,\cmdKV@gTypes@iE] (0.9,0.3) -- node[label=(-90+\cmdKV@general@rotate):\cmdKV@gLabels@iLE] {} (0.3,0.6);
    \draw[label distance=-1,\cmdKV@gTypes@iF] (1.5,0.9) -- node[label=(180+\cmdKV@general@rotate):\cmdKV@gLabels@iLF] {} (0.9,0.3);
    #2
  \end{tikzpicture}
  \gSetStdTypes{line}
}

\def\gTriPenta{\@ifnextchar[\@gTriPenta{\@gTriPenta[]}}
\def\@gTriPenta[#1]#2{%
  \setkeys*{pre}{#1}%
  \setrmkeys{general,gTypes,gLabels}%
  \begin{tikzpicture}
    [line width=1pt,
    baseline={([yshift=-0.5ex+\cmdKV@general@yshift]current bounding box.center)},
    scale=\cmdKV@general@scale,
    rotate=\cmdKV@general@rotate,
    font=\cmdKV@general@font]
    \draw[\cmdKV@gTypes@eA] (-108:0.5) -- (-135:1) node[label=(145+\cmdKV@general@rotate):\cmdKV@gLabels@eLA] {};
    \draw[\cmdKV@gTypes@eB] (180:0.5) -- (180:1) node[label=(180+\cmdKV@general@rotate):\cmdKV@gLabels@eLB] {};
    \draw[\cmdKV@gTypes@eC] (108:0.5) -- (135:1) node[label=(-145+\cmdKV@general@rotate):\cmdKV@gLabels@eLC] {};
    \draw[\cmdKV@gTypes@eD] (0:1.2) -- (0:1.5) node[label=(\cmdKV@general@rotate):\cmdKV@gLabels@eLD] {};
    \draw[label distance=0,\cmdKV@gTypes@iA] (-108:0.5) -- node[label=(-135+\cmdKV@general@rotate):\cmdKV@gLabels@iLA] {} (-180:0.5);
    \draw[label distance=0,\cmdKV@gTypes@iB] (180:0.5) -- node[label=(135+\cmdKV@general@rotate):\cmdKV@gLabels@iLB] {} (108:0.5);
    \draw[label distance=0,\cmdKV@gTypes@iC] (108:0.5) -- node[label=(72+\cmdKV@general@rotate):\cmdKV@gLabels@iLC] {} (36:0.5);
    \draw[label distance=0,\cmdKV@gTypes@iD] (36:0.5) -- node[label=(72+\cmdKV@general@rotate):\cmdKV@gLabels@iLD] {} (0:1.2);
    \draw[label distance=2,\cmdKV@gTypes@iE] (0:1.2) -- node[label=(-90+\cmdKV@general@rotate):\cmdKV@gLabels@iLE] {} (-36:0.5);
    \draw[label distance=2,\cmdKV@gTypes@iF] (-36:0.5) -- node[label=(-90+\cmdKV@general@rotate):\cmdKV@gLabels@iLF] {} (-108:0.5);
    \draw[label distance=0,\cmdKV@gTypes@iG] (36:0.5) -- node[label=(180+\cmdKV@general@rotate):\cmdKV@gLabels@iLG] {} (-36:0.5);
    #2
  \end{tikzpicture}
  \gSetStdTypes{line}
}
\def\gTriTri{\@ifnextchar[\@gTriTri{\@gTriTri[]}}
\def\@gTriTri[#1]#2{%
  \setkeys*{pre}{#1}%
  \setrmkeys{general,gTypes,gLabels}%
  \begin{tikzpicture}
    [line width=1pt,
    baseline={([yshift=-0.5ex+\cmdKV@general@yshift]current bounding box.center)},
    scale=\cmdKV@general@scale,
    rotate=\cmdKV@general@rotate,
    font=\cmdKV@general@font]
    \draw[\cmdKV@gTypes@eA] (0.3,0.3) -- (0,0) node[label=(-180+\cmdKV@general@rotate):\cmdKV@gLabels@eLA] {};
    \draw[\cmdKV@gTypes@eB] (0.3,1.1) -- (0,1.4) node[label=(180+\cmdKV@general@rotate):\cmdKV@gLabels@eLB] {};
    \draw[\cmdKV@gTypes@eC] (1.9,1.1) -- (2.2,1.4) node[label=(\cmdKV@general@rotate):\cmdKV@gLabels@eLC] {};
    \draw[\cmdKV@gTypes@eD] (1.9,0.3) -- (2.2,0) node[label=(\cmdKV@general@rotate):\cmdKV@gLabels@eLD] {};
    \draw[label distance=0,\cmdKV@gTypes@iA] (0.3,0.3) -- node[label=(180+\cmdKV@general@rotate):\cmdKV@gLabels@iLA] {} (0.3,1.1);
    \draw[label distance=0,\cmdKV@gTypes@iB] (0.3,1.1) -- node[label=(45+\cmdKV@general@rotate):\cmdKV@gLabels@iLB] {} (0.9,0.7);
    \draw[label distance=0,\cmdKV@gTypes@iC] (0.9,0.7) -- node[label=(-45+\cmdKV@general@rotate):\cmdKV@gLabels@iLC] {} (0.3,0.3);
    \draw[label distance=0,\cmdKV@gTypes@iD] (0.9,0.7) -- node[label=(90+\cmdKV@general@rotate):\cmdKV@gLabels@iLD] {} (1.3,0.7);
    \draw[label distance=0,\cmdKV@gTypes@iE] (1.3,0.7) -- node[label=(135+\cmdKV@general@rotate):\cmdKV@gLabels@iLE] {} (1.9,1.1);
    \draw[label distance=0,\cmdKV@gTypes@iF] (1.9,1.1) -- node[label=(\cmdKV@general@rotate):\cmdKV@gLabels@iLF] {} (1.9,0.3);
    \draw[label distance=0,\cmdKV@gTypes@iG] (1.9,0.3) -- node[label=(-135+\cmdKV@general@rotate):\cmdKV@gLabels@iLG] {} (1.3,0.7);
    #2
  \end{tikzpicture}
  \gSetStdTypes{line}
}

\def\gBoxBubA{\@ifnextchar[\@gBoxBubA{\@gBoxBubA[]}}
\def\@gBoxBubA[#1]#2{%
  \setkeys*{pre}{#1}%
  \setrmkeys{general,gTypes,gLabels}%
  \begin{tikzpicture}
    [line width=1pt,
    baseline={([yshift=-0.5ex+\cmdKV@general@yshift]current bounding box.center)},
    scale=\cmdKV@general@scale,
    rotate=\cmdKV@general@rotate,
    font=\cmdKV@general@font]
    \draw[\cmdKV@gTypes@eA] (0.1,0.1) -- (-0.1,-0.1) node[label=(-180+\cmdKV@general@rotate):\cmdKV@gLabels@eLA] {};
    \draw[\cmdKV@gTypes@eB] (0.1,0.9) -- (-0.1,1.1) node[label=(180+\cmdKV@general@rotate):\cmdKV@gLabels@eLB] {};
    \draw[\cmdKV@gTypes@eC] (0.9,0.9) -- (1.1,1.1) node[label=(\cmdKV@general@rotate):\cmdKV@gLabels@eLC] {};
    \draw[\cmdKV@gTypes@eD] (0.9,0.1) -- (1.1,-0.1) node[label=(\cmdKV@general@rotate):\cmdKV@gLabels@eLD] {};
    \draw[label distance=0,\cmdKV@gTypes@iA] (0.1,0.1) -- node[label=(180+\cmdKV@general@rotate):\cmdKV@gLabels@iLA] {} (0.1,0.9);
    \draw[label distance=0,\cmdKV@gTypes@iB] (0.1,0.9) -- node[label=(90+\cmdKV@general@rotate):\cmdKV@gLabels@iLB] {} (0.32,0.9);
    \draw[label distance=0,\cmdKV@gTypes@iC] (0.32,0.9) .. controls (0.32,1.15) and (0.68,1.15) .. node[label=(90+\cmdKV@general@rotate):\cmdKV@gLabels@iLC] {} (0.68,0.9);
    \draw[label distance=0,\cmdKV@gTypes@iD] (0.68,0.9) .. controls (0.68,0.65) and (0.32,0.65) .. node[label=(-90+\cmdKV@general@rotate):\cmdKV@gLabels@iLD] {} (0.32,0.9);
    \draw[label distance=0,\cmdKV@gTypes@iE] (0.68,0.9) -- node[label=(90+\cmdKV@general@rotate):\cmdKV@gLabels@iLE] {} (0.9,0.9);
    \draw[label distance=0,\cmdKV@gTypes@iF] (0.9,0.9) -- node[label=(\cmdKV@general@rotate):\cmdKV@gLabels@iLF] {} (0.9,0.1);
    \draw[label distance=0,\cmdKV@gTypes@iG] (0.9,0.1) -- node[label=(-90+\cmdKV@general@rotate):\cmdKV@gLabels@iLG] {} (0.1,0.1);
    #2
  \end{tikzpicture}
  \gSetStdTypes{line}
}

\def\gBoxBubB{\@ifnextchar[\@gBoxBubB{\@gBoxBubB[]}}
\def\@gBoxBubB[#1]#2{%
  \setkeys*{pre}{#1}%
  \setrmkeys{general,gTypes,gLabels}%
  \begin{tikzpicture}
    [line width=1pt,
    baseline={([yshift=-0.5ex+\cmdKV@general@yshift]current bounding box.center)},
    scale=\cmdKV@general@scale,
    rotate=\cmdKV@general@rotate,
    font=\cmdKV@general@font]
    \draw[label distance=-1,\cmdKV@gTypes@eA] (0,-0.4) -- (0,-0.7) node[label=(180+\cmdKV@general@rotate):\cmdKV@gLabels@eLA] {};
    \draw[\cmdKV@gTypes@eB] (-0.4,0) -- (-0.7,0) node[label=(180+\cmdKV@general@rotate):\cmdKV@gLabels@eLB] {};
    \draw[label distance=-1,\cmdKV@gTypes@eC] (0,0.4) -- (0,0.7) node[label=(180+\cmdKV@general@rotate):\cmdKV@gLabels@eLC] {};
    \draw[\cmdKV@gTypes@eD] (1.1,0) -- (1.4,0) node[label=(\cmdKV@general@rotate):\cmdKV@gLabels@eLD] {};
    \draw[label distance=0,\cmdKV@gTypes@iA] (0,-0.4) -- node[label=(-135+\cmdKV@general@rotate):\cmdKV@gLabels@iLA] {} (-0.4,0);
    \draw[label distance=0,\cmdKV@gTypes@iB] (-0.4,0) -- node[label=(135+\cmdKV@general@rotate):\cmdKV@gLabels@iLB] {} (0,0.4);
    \draw[label distance=0,\cmdKV@gTypes@iC] (0,0.4) -- node[label=(45+\cmdKV@general@rotate):\cmdKV@gLabels@iLC] {} (0.4,0);
    \draw[label distance=0,\cmdKV@gTypes@iD] (0.4,0) -- node[label=(-45+\cmdKV@general@rotate):\cmdKV@gLabels@iLD] {} (0,-0.4);
    \draw[label distance=0,\cmdKV@gTypes@iE] (0.4,0) --  node[label=(90+\cmdKV@general@rotate):\cmdKV@gLabels@iLE] {} (0.7,0);
    \draw[label distance=0,\cmdKV@gTypes@iF] (0.7,0) .. controls (0.7,0.26) and (1.1,0.26) .. node[label=(90+\cmdKV@general@rotate):\cmdKV@gLabels@iLF] {} (1.1,0);
    \draw[label distance=0,\cmdKV@gTypes@iG] (1.1,0) .. controls (1.1,-0.26) and (0.7,-0.26) ..  node[label=(-90+\cmdKV@general@rotate):\cmdKV@gLabels@iLG] {} (0.7,0);
    #2
  \end{tikzpicture}
  \gSetStdTypes{line}
}

\def\gPentaTad{\@ifnextchar[\@gPentaTad{\@gPentaTad[]}}
\def\@gPentaTad[#1]#2{%
  \setkeys*{pre}{#1}%
  \setrmkeys{general,gTypes,gLabels}%
  \begin{tikzpicture}
    [line width=1pt,
    baseline={([yshift=-0.5ex+\cmdKV@general@yshift]current bounding box.center)},
    scale=\cmdKV@general@scale,
    rotate=\cmdKV@general@rotate,
    font=\cmdKV@general@font]
    \draw[label distance=-0.5,\cmdKV@gTypes@eA] (108:0.35) -- (108:0.75) node[label=(180+\cmdKV@general@rotate):\cmdKV@gLabels@eLA] {};
    \draw[\cmdKV@gTypes@eB] (36:0.35) -- (36:0.75) node[label=(36+\cmdKV@general@rotate):\cmdKV@gLabels@eLB] {};
    \draw[\cmdKV@gTypes@eC] (-36:0.35) -- (-36:0.75) node[label=(-36+\cmdKV@general@rotate):\cmdKV@gLabels@eLC] {};
    \draw[label distance=-0.5,\cmdKV@gTypes@eD] (-108:0.35) -- (-108:0.75) node[label=(-180+\cmdKV@general@rotate):\cmdKV@gLabels@eLD] {};
    \draw[label distance=0,\cmdKV@gTypes@iA] (108:0.35) -- node[label=(72+\cmdKV@general@rotate):\cmdKV@gLabels@iLA] {} (36:0.35);
    \draw[label distance=0,\cmdKV@gTypes@iB] (36:0.35) -- node[label=(\cmdKV@general@rotate):\cmdKV@gLabels@iLB] {} (-36:0.35);
    \draw[label distance=0,\cmdKV@gTypes@iC] (-36:0.35) -- node[label=(-72+\cmdKV@general@rotate):\cmdKV@gLabels@iLC] {} (-108:0.35);
    \draw[label distance=0,\cmdKV@gTypes@iD] (-108:0.35) -- node[label=(-144+\cmdKV@general@rotate):\cmdKV@gLabels@iLD] {} (-180:0.35);
    \draw[label distance=0,\cmdKV@gTypes@iE] (180:0.35) -- node[label=(144+\cmdKV@general@rotate):\cmdKV@gLabels@iLE] {} (108:0.35);
    \draw[label distance=0,\cmdKV@gTypes@iF] (180:0.35) -- node[label=(90+\cmdKV@general@rotate):\cmdKV@gLabels@iLF] {} (180:0.75);
    \draw[label distance=-0.5,\cmdKV@gTypes@iG] (180:0.75) .. controls ($(180:0.75)+(0,-0.18)$) and ($(180:1)+(0,-0.18)$) .. (180:1) node[label=(180+\cmdKV@general@rotate):\cmdKV@gLabels@iLG] {} .. controls ($(180:1)+(0,0.18)$) and ($(180:0.75)+(0,0.18)$) .. (180:0.75);
    #2
  \end{tikzpicture}
  \gSetStdTypes{line}
}

\def\gBoxTadA{\@ifnextchar[\@gBoxTadA{\@gBoxTadA[]}}
\def\@gBoxTadA[#1]#2{%
  \setkeys*{pre}{#1}%
  \setrmkeys{general,gTypes,gLabels}%
  \begin{tikzpicture}
    [line width=1pt,
    baseline={([yshift=-0.5ex+\cmdKV@general@yshift]current bounding box.center)},
    scale=\cmdKV@general@scale,
    rotate=\cmdKV@general@rotate,
    font=\cmdKV@general@font]
    \draw[label distance=-0.5,\cmdKV@gTypes@eA] (0,0.4) -- (0,0.7) node[label=(\cmdKV@general@rotate):\cmdKV@gLabels@eLA] {};
    \draw[\cmdKV@gTypes@eB] (0.4,0) -- (0.7,0) node[label=(\cmdKV@general@rotate):\cmdKV@gLabels@eLB] {};
    \draw[label distance=-0.5,\cmdKV@gTypes@eC] (0,-0.4) -- (0,-0.7) node[label=(\cmdKV@general@rotate):\cmdKV@gLabels@eLC] {};
    \draw[\cmdKV@gTypes@eD] (-0.7,0) -- (-0.7,-0.3) node[label=(-90+\cmdKV@general@rotate):\cmdKV@gLabels@eLD] {};
    \draw[label distance=0,\cmdKV@gTypes@iA] (0,0.4) -- node[label=(45+\cmdKV@general@rotate):\cmdKV@gLabels@iLA] {} (0.4,0);
    \draw[label distance=0,\cmdKV@gTypes@iB] (0.4,0) -- node[label=(-45+\cmdKV@general@rotate):\cmdKV@gLabels@iLB] {} (0,-0.4);
    \draw[label distance=0,\cmdKV@gTypes@iC] (0,-0.4) -- node[label=(-135+\cmdKV@general@rotate):\cmdKV@gLabels@iLC] {} (-0.4,0);
    \draw[label distance=0,\cmdKV@gTypes@iD] (-0.4,0) -- node[label=(135+\cmdKV@general@rotate):\cmdKV@gLabels@iLD] {} (0,0.4);
    \draw[label distance=0,\cmdKV@gTypes@iE] (-0.4,0) --  node[label=(90+\cmdKV@general@rotate):\cmdKV@gLabels@iLE] {} (-0.7,0);
    \draw[label distance=0,\cmdKV@gTypes@iF] (-0.7,0) -- node[label=(90+\cmdKV@general@rotate):\cmdKV@gLabels@iLF] {} (-1.1,0);
    \draw[label distance=-0.5,\cmdKV@gTypes@iG] (-1.1,0) .. controls (-1.1,-0.18) and (-1.35,-0.18) .. (-1.35,0) node[label=(180+\cmdKV@general@rotate):\cmdKV@gLabels@iLG] {} .. controls (-1.35,0.18) and (-1.1,0.18) .. (-1.1,0);
    #2
  \end{tikzpicture}
  \gSetStdTypes{line}
}

\def\gSunsetA{\@ifnextchar[\@gSunsetA{\@gSunsetA[]}}
\def\@gSunsetA[#1]#2{%
  \setkeys*{pre}{#1}%
  \setrmkeys{general,gTypes,gLabels}%
  \begin{tikzpicture}
    [line width=1pt,
    baseline={([yshift=-0.5ex+\cmdKV@general@yshift]current bounding box.center)},
    scale=\cmdKV@general@scale,
    rotate=\cmdKV@general@rotate,
    font=\cmdKV@general@font]
    \draw[\cmdKV@gTypes@eA] (0.3,0.6) -- (0,0) node[label=(-180+\cmdKV@general@rotate):\cmdKV@gLabels@eLA] {};
    \draw[\cmdKV@gTypes@eB] (0.3,0.6) -- (0,1.2) node[label=(180+\cmdKV@general@rotate):\cmdKV@gLabels@eLB] {};
    \draw[\cmdKV@gTypes@eC] (1.5,0.6) -- (1.8,1.2) node[label=(\cmdKV@general@rotate):\cmdKV@gLabels@eLC] {};
    \draw[\cmdKV@gTypes@eD] (1.5,0.6) -- (1.8,0) node[label=(\cmdKV@general@rotate):\cmdKV@gLabels@eLD] {};
    \draw[label distance=0,\cmdKV@gTypes@iA] (0.3,0.6) to[bend left=90] node[label=(90+\cmdKV@general@rotate):\cmdKV@gLabels@iLA] {} (1.5,0.6);
    \draw[label distance=0,\cmdKV@gTypes@iB] (1.5,0.6) to[bend left=90] node[label=(-90+\cmdKV@general@rotate):\cmdKV@gLabels@iLB] {} (0.3,0.6);
    \draw[label distance=-2,\cmdKV@gTypes@iC] (0.3,0.6) -- node[label=(90+\cmdKV@general@rotate):\cmdKV@gLabels@iLC] {} (1.5,0.6);
    #2
  \end{tikzpicture}
  \gSetStdTypes{line}
}

\def\gTriBubA{\@ifnextchar[\@gTriBubA{\@gTriBubA[]}}
\def\@gTriBubA[#1]#2{%
  \setkeys*{pre}{#1}%
  \setrmkeys{general,gTypes,gLabels}%
  \begin{tikzpicture}
    [line width=1pt,
    baseline={([yshift=-0.5ex+\cmdKV@general@yshift]current bounding box.center)},
    scale=\cmdKV@general@scale,
    rotate=\cmdKV@general@rotate,
    font=\cmdKV@general@font]
    \draw[\cmdKV@gTypes@eA] (0.4,0.3) -- (0,0) node[label=(-180+\cmdKV@general@rotate):\cmdKV@gLabels@eLA] {};
    \draw[\cmdKV@gTypes@eB] (0.4,0.9) -- (0,1.2) node[label=(180+\cmdKV@general@rotate):\cmdKV@gLabels@eLB] {};
    \draw[\cmdKV@gTypes@eC] (1.4,0.6) -- (1.8,1.2) node[label=(\cmdKV@general@rotate):\cmdKV@gLabels@eLC] {};
    \draw[\cmdKV@gTypes@eD] (1.4,0.6) -- (1.8,0) node[label=(\cmdKV@general@rotate):\cmdKV@gLabels@eLD] {};
    \draw[label distance=0,\cmdKV@gTypes@iA] (0.4,0.3) to[bend left=40] node[label=(180+\cmdKV@general@rotate):\cmdKV@gLabels@iLA] {} (0.4,0.9);
    \draw[label distance=0,\cmdKV@gTypes@iB] (0.4,0.9) -- node[label=(90+\cmdKV@general@rotate):\cmdKV@gLabels@iLB] {} (1.4,0.6);
    \draw[label distance=0,\cmdKV@gTypes@iC] (1.4,0.6) -- node[label=(-90+\cmdKV@general@rotate):\cmdKV@gLabels@iLC] {} (0.4,0.3);
    \draw[label distance=-1,\cmdKV@gTypes@iD] (0.4,0.9) to[bend left=40] node[label=(\cmdKV@general@rotate):\cmdKV@gLabels@iLD] {} (0.4,0.3);
    #2
  \end{tikzpicture}
  \gSetStdTypes{line}
}

\def\gBubBubA{\@ifnextchar[\@gBubBubA{\@gBubBubA[]}}
\def\@gBubBubA[#1]#2{%
  \setkeys*{pre}{#1}%
  \setrmkeys{general,gTypes,gLabels}%
  \begin{tikzpicture}
    [line width=1pt,
    baseline={([yshift=-0.5ex+\cmdKV@general@yshift]current bounding box.center)},
    scale=\cmdKV@general@scale,
    rotate=\cmdKV@general@rotate,
    font=\cmdKV@general@font]
    \draw[\cmdKV@gTypes@eA] (0.3,0.6) -- (0,0) node[label=(-180+\cmdKV@general@rotate):\cmdKV@gLabels@eLA] {};
    \draw[\cmdKV@gTypes@eB] (0.3,0.6) -- (0,1.2) node[label=(180+\cmdKV@general@rotate):\cmdKV@gLabels@eLB] {};
    \draw[\cmdKV@gTypes@eC] (1.5,0.6) -- (1.8,1.2) node[label=(\cmdKV@general@rotate):\cmdKV@gLabels@eLC] {};
    \draw[\cmdKV@gTypes@eD] (1.5,0.6) -- (1.8,0) node[label=(\cmdKV@general@rotate):\cmdKV@gLabels@eLD] {};
    \draw[label distance=0,\cmdKV@gTypes@iA] (0.3,0.6) to[bend left=90] node[label=(90+\cmdKV@general@rotate):\cmdKV@gLabels@iLA] {} (0.9,0.6);
    \draw[label distance=0,\cmdKV@gTypes@iB] (0.9,0.6) to[bend left=90] node[label=(90+\cmdKV@general@rotate):\cmdKV@gLabels@iLB] {} (1.5,0.6);
    \draw[label distance=0,\cmdKV@gTypes@iC] (1.5,0.6) to[bend left=90] node[label=(-90+\cmdKV@general@rotate):\cmdKV@gLabels@iLC] {} (0.9,0.6);
    \draw[label distance=0,\cmdKV@gTypes@iD] (0.9,0.6) to[bend left=90] node[label=(-90+\cmdKV@general@rotate):\cmdKV@gLabels@iLD] {} (0.3,0.6);
    #2
  \end{tikzpicture}
  \gSetStdTypes{line}
}

\def\gBoxBubC{\@ifnextchar[\@gBoxBubC{\@gBoxBubC[]}}
\def\@gBoxBubC[#1]#2{%
  \setkeys*{pre}{#1}%
  \setrmkeys{general,gTypes,gLabels}%
  \begin{tikzpicture}
    [line width=1pt,
    baseline={([yshift=-0.5ex+\cmdKV@general@yshift]current bounding box.center)},
    scale=\cmdKV@general@scale,
    rotate=\cmdKV@general@rotate,
    font=\cmdKV@general@font]
    \draw[\cmdKV@gTypes@eA] (0.1,0.1) -- (-0.1,-0.1) node[label=(-180+\cmdKV@general@rotate):\cmdKV@gLabels@eLA] {};
    \draw[\cmdKV@gTypes@eB] (0.1,0.9) -- (-0.1,1.1) node[label=(180+\cmdKV@general@rotate):\cmdKV@gLabels@eLB] {};
    \draw[\cmdKV@gTypes@eC] (0.9,0.9) -- (1.1,1.1) node[label=(\cmdKV@general@rotate):\cmdKV@gLabels@eLC] {};
    \draw[\cmdKV@gTypes@eD] (0.9,0.1) -- (1.1,-0.1) node[label=(\cmdKV@general@rotate):\cmdKV@gLabels@eLD] {};
    \draw[label distance=0,\cmdKV@gTypes@iA] (0.1,0.1) -- node[label=(180+\cmdKV@general@rotate):\cmdKV@gLabels@iLA] {} (0.1,0.9);
    \draw[label distance=0,\cmdKV@gTypes@iB] (0.1,0.9) to[bend left] node[label=(90+\cmdKV@general@rotate):\cmdKV@gLabels@iLB] {} (0.9,0.9);
    \draw[label distance=0,\cmdKV@gTypes@iC] (0.9,0.9) to[bend left] node[label=(-90+\cmdKV@general@rotate):\cmdKV@gLabels@iLC] {} (0.1,0.9);
    \draw[label distance=0,\cmdKV@gTypes@iD] (0.9,0.9) -- node[label=(\cmdKV@general@rotate):\cmdKV@gLabels@iLD] {} (0.9,0.1);
    \draw[label distance=0,\cmdKV@gTypes@iE] (0.9,0.1) -- node[label=(-90+\cmdKV@general@rotate):\cmdKV@gLabels@iLE] {} (0.1,0.1);
    #2
  \end{tikzpicture}
  \gSetStdTypes{line}
}

\def\gBoxZ{\@ifnextchar[\@gBoxZ{\@gBoxZ[]}}
\def\@gBoxZ[#1]#2{%
  \setkeys*{pre}{#1}%
  \setrmkeys{general,gTypes,gLabels}%
  \begin{tikzpicture}
    [line width=1pt,
    baseline={([yshift=-0.5ex+\cmdKV@general@yshift]current bounding box.center)},
    scale=\cmdKV@general@scale,
    rotate=\cmdKV@general@rotate,
    font=\cmdKV@general@font]
    \draw[\cmdKV@gTypes@eA] (0.15,0.15) -- (-0.1,-0.1) node[label=(-180+\cmdKV@general@rotate):\cmdKV@gLabels@eLA] {};
    \draw[\cmdKV@gTypes@eB] (0.15,0.85) -- (-0.1,1.1) node[label=(180+\cmdKV@general@rotate):\cmdKV@gLabels@eLB] {};
    \draw[\cmdKV@gTypes@eC] (0.85,0.85) -- (1.1,1.1) node[label=(\cmdKV@general@rotate):\cmdKV@gLabels@eLC] {};
    \draw[\cmdKV@gTypes@eD] (0.85,0.15) -- (1.1,-0.1) node[label=(\cmdKV@general@rotate):\cmdKV@gLabels@eLD] {};
    \draw[label distance=0,\cmdKV@gTypes@iA] (0.15,0.15) -- node[label=(180+\cmdKV@general@rotate):\cmdKV@gLabels@iLA] {} (0.15,0.85);
    \draw[label distance=0,\cmdKV@gTypes@iB] (0.15,0.85) -- node[label=(90+\cmdKV@general@rotate):\cmdKV@gLabels@iLB] {} (0.85,0.85);
    \draw[label distance=0,\cmdKV@gTypes@iC] (0.85,0.85) -- node[label=(\cmdKV@general@rotate):\cmdKV@gLabels@iLC] {} (0.85,0.15);
    \draw[label distance=0,\cmdKV@gTypes@iD] (0.85,0.15) -- node[label=(-90+\cmdKV@general@rotate):\cmdKV@gLabels@iLD] {} (0.15,0.15);
    \draw[label distance=-4,\cmdKV@gTypes@iE] (0.15,0.15) -- node[label=(135+\cmdKV@general@rotate):\cmdKV@gLabels@iLE] {} (0.85,0.85);
    #2
  \end{tikzpicture}
  \gSetStdTypes{line}
}

\def\gBoxBoxBoxA{\@ifnextchar[\@gBoxBoxBoxA{\@gBoxBoxBoxA[]}}
\def\@gBoxBoxBoxA[#1]#2{%
  \setkeys*{pre}{#1}%
  \setrmkeys{general,gTypes,gLabels}%
  \begin{tikzpicture}
    [line width=1pt,
    baseline={([yshift=-0.5ex+\cmdKV@general@yshift]current bounding box.center)},
    scale=\cmdKV@general@scale,
    rotate=\cmdKV@general@rotate,
    font=\cmdKV@general@font]
    \draw[\cmdKV@gTypes@eA] (-0.3,0.3) -- (-0.6,0) node[label=(-180+\cmdKV@general@rotate):\cmdKV@gLabels@eLA] {};
    \draw[\cmdKV@gTypes@eB] (-0.3,0.9) -- (-0.6,1.2) node[label=(180+\cmdKV@general@rotate):\cmdKV@gLabels@eLB] {};
    \draw[\cmdKV@gTypes@eC] (1.5,0.9) -- (1.8,1.2) node[label=(\cmdKV@general@rotate):\cmdKV@gLabels@eLC] {};
    \draw[\cmdKV@gTypes@eD] (1.5,0.3) -- (1.8,0) node[label=(\cmdKV@general@rotate):\cmdKV@gLabels@eLD] {};
    \draw[label distance=0,\cmdKV@gTypes@iA] (-0.3,0.3) -- node[label=(180+\cmdKV@general@rotate):\cmdKV@gLabels@iLA] {} (-0.3,0.9);
    \draw[label distance=0,\cmdKV@gTypes@iB] (-0.3,0.9) -- node[label=(90+\cmdKV@general@rotate):\cmdKV@gLabels@iLB] {} (0.3,0.9);
    \draw[label distance=0,\cmdKV@gTypes@iC] (0.3,0.9) -- node[label=(90+\cmdKV@general@rotate):\cmdKV@gLabels@iLC] {} (0.9,0.9);
    \draw[label distance=0,\cmdKV@gTypes@iD] (0.9,0.9) -- node[label=(90+\cmdKV@general@rotate):\cmdKV@gLabels@iLD] {} (1.5,0.9);
    \draw[label distance=0,\cmdKV@gTypes@iE] (1.5,0.9) -- node[label=(\cmdKV@general@rotate):\cmdKV@gLabels@iLE] {} (1.5,0.3);
    \draw[label distance=0,\cmdKV@gTypes@iF] (1.5,0.3) -- node[label=(-90+\cmdKV@general@rotate):\cmdKV@gLabels@iLF] {} (0.9,0.3);
    \draw[label distance=0,\cmdKV@gTypes@iG] (0.9,0.3) -- node[label=(-90+\cmdKV@general@rotate):\cmdKV@gLabels@iLG] {} (0.3,0.3);
    \draw[label distance=0,\cmdKV@gTypes@iH] (0.3,0.3) -- node[label=(-90+\cmdKV@general@rotate):\cmdKV@gLabels@iLH] {} (-0.3,0.3);
    \draw[label distance=0,\cmdKV@gTypes@iI] (0.3,0.9) -- node[label=(\cmdKV@general@rotate):\cmdKV@gLabels@iLI] {} (0.3,0.3);
    \draw[label distance=0,\cmdKV@gTypes@iJ] (0.9,0.9) -- node[label=(\cmdKV@general@rotate):\cmdKV@gLabels@iLJ] {} (0.9,0.3);
    #2
  \end{tikzpicture}
  \gSetStdTypes{line}
}

\def\gBoxBoxBoxB{\@ifnextchar[\@gBoxBoxBoxB{\@gBoxBoxBoxB[]}}
\def\@gBoxBoxBoxB[#1]#2{%
  \setkeys*{pre}{#1}%
  \setrmkeys{general,gTypes,gLabels}%
  \begin{tikzpicture}
    [line width=1pt,
    baseline={([yshift=-0.5ex+\cmdKV@general@yshift]current bounding box.center)},
    scale=\cmdKV@general@scale,
    rotate=\cmdKV@general@rotate,
    font=\cmdKV@general@font]
    \draw[\cmdKV@gTypes@eA] (0.3,0.3) -- (0,0) node[label=(-180+\cmdKV@general@rotate):\cmdKV@gLabels@eLA] {};
    \draw[\cmdKV@gTypes@eB] (0.3,1.5) -- (0,1.8) node[label=(180+\cmdKV@general@rotate):\cmdKV@gLabels@eLB] {};
    \draw[\cmdKV@gTypes@eC] (1.5,1.5) -- (1.8,1.8) node[label=(\cmdKV@general@rotate):\cmdKV@gLabels@eLC] {};
    \draw[\cmdKV@gTypes@eD] (1.5,0.3) -- (1.8,0) node[label=(\cmdKV@general@rotate):\cmdKV@gLabels@eLD] {};
    \draw[label distance=0,\cmdKV@gTypes@iA] (0.3,0.3) -- node[label=(180+\cmdKV@general@rotate):\cmdKV@gLabels@iLA] {} (0.3,0.9);
    \draw[label distance=0,\cmdKV@gTypes@iB] (0.3,0.9) -- node[label=(180+\cmdKV@general@rotate):\cmdKV@gLabels@iLB] {} (0.3,1.5);
    \draw[label distance=0,\cmdKV@gTypes@iC] (0.3,1.5) -- node[label=(90+\cmdKV@general@rotate):\cmdKV@gLabels@iLC] {} (0.9,1.5);
    \draw[label distance=0,\cmdKV@gTypes@iD] (0.9,1.5) -- node[label=(90+\cmdKV@general@rotate):\cmdKV@gLabels@iLD] {} (1.5,1.5);
    \draw[label distance=0,\cmdKV@gTypes@iE] (1.5,1.5) -- node[label=(\cmdKV@general@rotate):\cmdKV@gLabels@iLE] {} (1.5,0.3);
    \draw[label distance=0,\cmdKV@gTypes@iF] (1.5,0.3) -- node[label=(-90+\cmdKV@general@rotate):\cmdKV@gLabels@iLF] {} (0.9,0.3);
    \draw[label distance=0,\cmdKV@gTypes@iG] (0.9,0.3) -- node[label=(-90+\cmdKV@general@rotate):\cmdKV@gLabels@iLG] {} (0.3,0.3);
    \draw[label distance=0,\cmdKV@gTypes@iH] (0.9,1.5) -- node[label=(\cmdKV@general@rotate):\cmdKV@gLabels@iLH] {} (0.9,0.9);
    \draw[label distance=0,\cmdKV@gTypes@iI] (0.9,0.9) -- node[label=(\cmdKV@general@rotate):\cmdKV@gLabels@iLI] {} (0.9,0.3);
    \draw[label distance=0,\cmdKV@gTypes@iJ] (0.9,0.9) -- node[label=(90+\cmdKV@general@rotate):\cmdKV@gLabels@iLJ] {} (0.3,0.9);
    #2
  \end{tikzpicture}
  \gSetStdTypes{line}
}

\def\gEFTTree{\@ifnextchar[\@gEFTTree{\@gEFTTree[]}}
\def\@gEFTTree[#1]#2{%
  \setkeys*{pre}{#1}%
  \setrmkeys{general,gTypes,gLabels}%
  \begin{tikzpicture}
    [line width=1pt,
    baseline={([yshift=-0.5ex+\cmdKV@general@yshift]current bounding box.center)},
    scale=\cmdKV@general@scale,
    rotate=\cmdKV@general@rotate,
    font=\cmdKV@general@font]
    \draw[\cmdKV@gTypes@iA] (0,0.5) node[label=(90+\cmdKV@general@rotate):\cmdKV@gLabels@eLA] {} -- node[label=(\cmdKV@general@rotate):\cmdKV@gLabels@iLA] {} (0,-0.5) node[label=(-90+\cmdKV@general@rotate):\cmdKV@gLabels@eLB] {};
    \draw[fill] (0,0.5) circle (0.02);
    \draw[fill] (0,-0.5) circle (0.02);
    #2
  \end{tikzpicture}
  \gSetStdTypes{line}
}

\def\gEFTV{\@ifnextchar[\@gEFTV{\@gEFTV[]}}
\def\@gEFTV[#1]#2{%
  \setkeys*{pre}{#1}%
  \setrmkeys{general,gTypes,gLabels}%
  \begin{tikzpicture}
    [line width=1pt,
    baseline={([yshift=-0.5ex+\cmdKV@general@yshift]current bounding box.center)},
    scale=\cmdKV@general@scale,
    rotate=\cmdKV@general@rotate,
    font=\cmdKV@general@font]
    \draw[\cmdKV@gTypes@iA] (0,0.5) node[label=(90+\cmdKV@general@rotate):\cmdKV@gLabels@eLA] {} -- node[label=(200+\cmdKV@general@rotate):\cmdKV@gLabels@iLA] {} (0.3,-0.5) node[label=(-90+\cmdKV@general@rotate):\cmdKV@gLabels@eLC] {};
    \draw[\cmdKV@gTypes@iB] (0.6,0.5) node[label=(90+\cmdKV@general@rotate):\cmdKV@gLabels@eLB] {} -- node[label=(-20+\cmdKV@general@rotate):\cmdKV@gLabels@iLB] {} (0.3,-0.5);
    \draw[fill] (0,0.5) circle (0.02);
    \draw[fill] (0.3,-0.5) circle (0.02);
    \draw[fill] (0.6,0.5) circle (0.02);
    #2
  \end{tikzpicture}
  \gSetStdTypes{line}
}

\def\gEFTY{\@ifnextchar[\@gEFTY{\@gEFTY[]}}
\def\@gEFTY[#1]#2{%
  \setkeys*{pre}{#1}%
  \setrmkeys{general,gTypes,gLabels}%
  \begin{tikzpicture}
    [line width=1pt,
    baseline={([yshift=-0.5ex+\cmdKV@general@yshift]current bounding box.center)},
    scale=\cmdKV@general@scale,
    rotate=\cmdKV@general@rotate,
    font=\cmdKV@general@font]
    \draw[\cmdKV@gTypes@iA] (0,0.5) node[label=(90+\cmdKV@general@rotate):\cmdKV@gLabels@eLA] {} -- node[label=(225+\cmdKV@general@rotate):\cmdKV@gLabels@iLA] {} (0.3,0);
    \draw[\cmdKV@gTypes@iB] (0.6,0.5) node[label=(90+\cmdKV@general@rotate):\cmdKV@gLabels@eLB] {} -- node[label=(-45+\cmdKV@general@rotate):\cmdKV@gLabels@iLB] {} (0.3,0);
    \draw[\cmdKV@gTypes@iC] (0.3,0) -- node[label=(\cmdKV@general@rotate):\cmdKV@gLabels@iLC] {} (0.3,-0.5) node[label=(-90+\cmdKV@general@rotate):\cmdKV@gLabels@eLC] {};
    \draw[fill] (0,0.5) circle (0.02);
    \draw[fill] (0.6,0.5) circle (0.02);
    \draw[fill] (0.3,-0.5) circle (0.02);
    #2
  \end{tikzpicture}
  \gSetStdTypes{line}
}

\def\gEFTPsi{\@ifnextchar[\@gEFTPsi{\@gEFTPsi[]}}
\def\@gEFTPsi[#1]#2{%
  \setkeys*{pre}{#1}%
  \setrmkeys{general,gTypes,gLabels}%
  \begin{tikzpicture}
    [line width=1pt,
    baseline={([yshift=-0.5ex+\cmdKV@general@yshift]current bounding box.center)},
    scale=\cmdKV@general@scale,
    rotate=\cmdKV@general@rotate,
    font=\cmdKV@general@font]
    \draw[\cmdKV@gTypes@iA] (0,0.5) node[label=(90+\cmdKV@general@rotate):\cmdKV@gLabels@eLA] {} -- node[label=(225+\cmdKV@general@rotate):\cmdKV@gLabels@iLA] {} (0.5,0);
    \draw[\cmdKV@gTypes@iB] (0.5,0.5) node[label=(90+\cmdKV@general@rotate):\cmdKV@gLabels@eLB] {} -- node[label=(\cmdKV@general@rotate):\cmdKV@gLabels@iLB] {} (0.5,0);
    \draw[\cmdKV@gTypes@iC] (1,0.5) node[label=(90+\cmdKV@general@rotate):\cmdKV@gLabels@eLC] {} -- node[label=(-45+\cmdKV@general@rotate):\cmdKV@gLabels@iLC] {} (0.5,0);
    \draw[\cmdKV@gTypes@iD] (0.5,0) -- node[label=(\cmdKV@general@rotate):\cmdKV@gLabels@iLD] {} (0.5,-0.5) node[label=(-90+\cmdKV@general@rotate):\cmdKV@gLabels@eLD] {};
    \draw[fill] (0,0.5) circle (0.02);
    \draw[fill] (0.5,0.5) circle (0.02);
    \draw[fill] (1,0.5) circle (0.02);
    \draw[fill] (0.5,-0.5) circle (0.02);
    #2
  \end{tikzpicture}
  \gSetStdTypes{line}
}

\def\gEFTVinY{\@ifnextchar[\@gEFTVinY{\@gEFTVinY[]}}
\def\@gEFTVinY[#1]#2{%
  \setkeys*{pre}{#1}%
  \setrmkeys{general,gTypes,gLabels}%
  \begin{tikzpicture}
    [line width=1pt,
    baseline={([yshift=-0.5ex+\cmdKV@general@yshift]current bounding box.center)},
    scale=\cmdKV@general@scale,
    rotate=\cmdKV@general@rotate,
    font=\cmdKV@general@font]
    \draw[\cmdKV@gTypes@iA] (0,0.5) node[label=(90+\cmdKV@general@rotate):\cmdKV@gLabels@eLA] {} -- node[label=(225+\cmdKV@general@rotate):\cmdKV@gLabels@iLA] {} (0.25,0.25);
    \draw[\cmdKV@gTypes@iE] (0.25,0.25) -- node[label=(225+\cmdKV@general@rotate):\cmdKV@gLabels@iLE] {} (0.5,0);
    \draw[\cmdKV@gTypes@iB] (0.5,0.5) node[label=(90+\cmdKV@general@rotate):\cmdKV@gLabels@eLB] {} -- node[label=(\cmdKV@general@rotate):\cmdKV@gLabels@iLB] {} (0.25,0.25);
    \draw[\cmdKV@gTypes@iC] (1,0.5) node[label=(90+\cmdKV@general@rotate):\cmdKV@gLabels@eLC] {} -- node[label=(-45+\cmdKV@general@rotate):\cmdKV@gLabels@iLC] {} (0.5,0);
    \draw[\cmdKV@gTypes@iD] (0.5,0) -- node[label=(\cmdKV@general@rotate):\cmdKV@gLabels@iLD] {} (0.5,-0.5) node[label=(-90+\cmdKV@general@rotate):\cmdKV@gLabels@eLD] {};
    \draw[fill] (0,0.5) circle (0.02);
    \draw[fill] (0.5,0.5) circle (0.02);
    \draw[fill] (1,0.5) circle (0.02);
    \draw[fill] (0.5,-0.5) circle (0.02);
    #2
  \end{tikzpicture}
  \gSetStdTypes{line}
}

\def\gEFTYinvY{\@ifnextchar[\@gEFTYinvY{\@gEFTYinvY[]}}
\def\@gEFTYinvY[#1]#2{%
  \setkeys*{pre}{#1}%
  \setrmkeys{general,gTypes,gLabels}%
  \begin{tikzpicture}
    [line width=1pt,
    baseline={([yshift=-0.5ex+\cmdKV@general@yshift]current bounding box.center)},
    scale=\cmdKV@general@scale,
    rotate=\cmdKV@general@rotate,
    font=\cmdKV@general@font]
    \draw[\cmdKV@gTypes@iA] (0,0.5) node[label=(90+\cmdKV@general@rotate):\cmdKV@gLabels@eLA] {} -- node[label=(225+\cmdKV@general@rotate):\cmdKV@gLabels@iLA] {} (0.3,0.166);
    \draw[\cmdKV@gTypes@iB] (0.6,0.5) node[label=(90+\cmdKV@general@rotate):\cmdKV@gLabels@eLB] {} -- node[label=(-45+\cmdKV@general@rotate):\cmdKV@gLabels@iLB] {} (0.3,0.166);
    \draw[\cmdKV@gTypes@iC] (0.3,-0.166) -- node[label=(135+\cmdKV@general@rotate):\cmdKV@gLabels@iLC] {} (0,-0.5) node[label=(-90+\cmdKV@general@rotate):\cmdKV@gLabels@eLC] {}; 
    \draw[\cmdKV@gTypes@iD] (0.3,-0.166) -- node[label=(45+\cmdKV@general@rotate):\cmdKV@gLabels@iLD] {} (0.6,-0.5) node[label=(-90+\cmdKV@general@rotate):\cmdKV@gLabels@eLD] {};
    \draw[\cmdKV@gTypes@iE] (0.3,0.166) -- node[label=(\cmdKV@general@rotate):\cmdKV@gLabels@iLE] {} (0.3,-0.166);
    \draw[fill] (0,0.5) circle (0.02);
    \draw[fill] (0.6,0.5) circle (0.02);
    \draw[fill] (0,-0.5) circle (0.02);
    \draw[fill] (0.6,-0.5) circle (0.02);
    #2
  \end{tikzpicture}
  \gSetStdTypes{line}
}

\def\gEFTX{\@ifnextchar[\@gEFTX{\@gEFTX[]}}
\def\@gEFTX[#1]#2{%
  \setkeys*{pre}{#1}%
  \setrmkeys{general,gTypes,gLabels}%
  \begin{tikzpicture}
    [line width=1pt,
    baseline={([yshift=-0.5ex+\cmdKV@general@yshift]current bounding box.center)},
    scale=\cmdKV@general@scale,
    rotate=\cmdKV@general@rotate,
    font=\cmdKV@general@font]
    \draw[\cmdKV@gTypes@iA] (0,0.5) node[label=(90+\cmdKV@general@rotate):\cmdKV@gLabels@eLA] {} -- node[label=(225+\cmdKV@general@rotate):\cmdKV@gLabels@iLA] {} (0.3,0);
    \draw[\cmdKV@gTypes@iB] (0.6,0.5) node[label=(90+\cmdKV@general@rotate):\cmdKV@gLabels@eLB] {} -- node[label=(-45+\cmdKV@general@rotate):\cmdKV@gLabels@iLB] {} (0.3,0);
    \draw[\cmdKV@gTypes@iC] (0.3,0) -- node[label=(135+\cmdKV@general@rotate):\cmdKV@gLabels@iLC] {} (0,-0.5) node[label=(-90+\cmdKV@general@rotate):\cmdKV@gLabels@eLC] {}; 
    \draw[\cmdKV@gTypes@iD] (0.3,0) -- node[label=(45+\cmdKV@general@rotate):\cmdKV@gLabels@iLD] {} (0.6,-0.5) node[label=(-90+\cmdKV@general@rotate):\cmdKV@gLabels@eLD] {};
    \draw[fill] (0,0.5) circle (0.02);
    \draw[fill] (0.6,0.5) circle (0.02);
    \draw[fill] (0,-0.5) circle (0.02);
    \draw[fill] (0.6,-0.5) circle (0.02);
    #2
  \end{tikzpicture}
  \gSetStdTypes{line}
}

\def\gEFTH{\@ifnextchar[\@gEFTH{\@gEFTH[]}}
\def\@gEFTH[#1]#2{%
  \setkeys*{pre}{#1}%
  \setrmkeys{general,gTypes,gLabels}%
  \begin{tikzpicture}
    [line width=1pt,
    baseline={([yshift=-0.5ex+\cmdKV@general@yshift]current bounding box.center)},
    scale=\cmdKV@general@scale,
    rotate=\cmdKV@general@rotate,
    font=\cmdKV@general@font]
    \draw[\cmdKV@gTypes@iA] (0,0.5) node[label=(90+\cmdKV@general@rotate):\cmdKV@gLabels@eLA] {} -- node[label=(180+\cmdKV@general@rotate):\cmdKV@gLabels@iLA] {} (0,0);
    \draw[\cmdKV@gTypes@iB] (0.6,0.5) node[label=(90+\cmdKV@general@rotate):\cmdKV@gLabels@eLB] {} -- node[label=(\cmdKV@general@rotate):\cmdKV@gLabels@iLB] {} (0.6,0);
    \draw[\cmdKV@gTypes@iC] (0.0,0) -- node[label=(180+\cmdKV@general@rotate):\cmdKV@gLabels@iLC] {} (0,-0.5) node[label=(-90+\cmdKV@general@rotate):\cmdKV@gLabels@eLC] {}; 
    \draw[\cmdKV@gTypes@iD] (0.6,0) -- node[label=(\cmdKV@general@rotate):\cmdKV@gLabels@iLD] {} (0.6,-0.5) node[label=(-90+\cmdKV@general@rotate):\cmdKV@gLabels@eLD] {};
    \draw[\cmdKV@gTypes@iE] (0,0) -- node[label=(90+\cmdKV@general@rotate):\cmdKV@gLabels@iLE] {} (0.6,0);
    \draw[fill] (0,0.5) circle (0.02);
    \draw[fill] (0.6,0.5) circle (0.02);
    \draw[fill] (0,-0.5) circle (0.02);
    \draw[fill] (0.6,-0.5) circle (0.02);
    #2
  \end{tikzpicture}
  \gSetStdTypes{line}
}

\def\cTree{\@ifnextchar[\@cTree{\@cTree[]}}
\def\@cTree[#1]#2{%
  \setkeys*{pre}{#1}%
  \setrmkeys{general,gTypes,gLabels,blobs}%
  \begin{tikzpicture}
    [line width=1pt,
    baseline={([yshift=-0.5ex+\cmdKV@general@yshift]current bounding box.center)},
    scale=\cmdKV@general@scale,
    rotate=\cmdKV@general@rotate,
    font=\cmdKV@general@font]
    \fill[\cmdKV@blobs@blobA] (0,0) ellipse (0.25 and 0.25);
    \draw[\cmdKV@gTypes@eA] ($ (0,0) + (-135:0.25) $) -- ($ (0,0) + (-135:0.6) $) node[label=(-180+\cmdKV@general@rotate):\cmdKV@gLabels@eLA] {};
    \draw[\cmdKV@gTypes@eB] ($ (0,0) + (135:0.25) $) -- ($ (0,0) + (135:0.6) $) node[label=(180+\cmdKV@general@rotate):\cmdKV@gLabels@eLB] {};
    \draw[\cmdKV@gTypes@eC] ($ (0,0) + (45:0.25) $) -- ($ (0,0) + (45:0.6) $) node[label=(\cmdKV@general@rotate):\cmdKV@gLabels@eLC] {};
    \draw[\cmdKV@gTypes@eD] ($ (0,0) + (-45:0.25) $) -- ($ (0,0) + (-45:0.6) $) node[label=(\cmdKV@general@rotate):\cmdKV@gLabels@eLD] {};
    \node[label=(90+\cmdKV@general@rotate):\cmdKV@gLabels@iLA] at (0,0.3) {};
    \node[label=(\cmdKV@general@rotate):\cmdKV@gLabels@iLB] at (0.3,0) {};
    #2
  \end{tikzpicture}
  \gSetStdTypes{line}
}

\def\cBoxA{\@ifnextchar[\@cBoxA{\@cBoxA[]}}
\def\@cBoxA[#1]#2{%
  \setkeys*{pre}{#1}%
  \setrmkeys{general,gTypes,gLabels,dots,blobs}%
  \begin{tikzpicture}
    [line width=1pt,
    baseline={([yshift=-0.5ex+\cmdKV@general@yshift]current bounding box.center)},
    scale=\cmdKV@general@scale,
    rotate=\cmdKV@general@rotate,
    font=\cmdKV@general@font]
    \fill[\cmdKV@blobs@blobA] (-0.4,0) ellipse (0.2 and 0.4);
    \fill[\cmdKV@blobs@blobB] (0.4,0) ellipse (0.2 and 0.4);
    \draw[\cmdKV@gTypes@eA] ($ (-0.4,0) + (-135:0.26) $) -- ($ (-0.4,0) + (-135:0.7) $) node[label=(-180+\cmdKV@general@rotate):\cmdKV@gLabels@eLA] {};
    \draw[\cmdKV@gTypes@eB] ($ (-0.4,0) + (135:0.26) $) -- ($ (-0.4,0) + (135:0.7) $) node[label=(180+\cmdKV@general@rotate):\cmdKV@gLabels@eLB] {};
    \draw[\cmdKV@gTypes@eC] ($ (0.4,0) + (45:0.26) $) -- ($ (0.4,0) + (45:0.7) $) node[label=(\cmdKV@general@rotate):\cmdKV@gLabels@eLC] {};
    \draw[\cmdKV@gTypes@eD] ($ (0.4,0) + (-45:0.26) $) -- ($ (0.4,0) + (-45:0.7) $) node[label=(\cmdKV@general@rotate):\cmdKV@gLabels@eLD] {};
    \draw[label distance=1,\cmdKV@gTypes@iA] ($ (-0.4,0) + (45:0.26) $) -- node[label=(90+\cmdKV@general@rotate):\cmdKV@gLabels@iLA] {}($ (0.4,0) + (135:0.26) $);
    \draw[label distance=1,\cmdKV@gTypes@iB] ($ (0.4,0) + (-135:0.26) $) -- node[label=(-90+\cmdKV@general@rotate):\cmdKV@gLabels@iLB] {}($ (-0.4,0) + (-45:0.26) $);
    \node[label=(180+\cmdKV@general@rotate):\cmdKV@gLabels@iLC] at (-0.6,0) {};
    \node[label=(\cmdKV@general@rotate):\cmdKV@gLabels@iLD] at (0.6,0) {};
    \ifKV@dots@lDots
      \draw[dotted] ($(-0.4,0) + (-150:0.7)$) to[bend left] ($(-0.4,0) + (150:0.7)$);%
    \fi
    \ifKV@dots@rDots
      \draw[dotted] ($(0.4,0) + (30:0.7)$) to[bend left] ($(0.4,0) + (-30:0.7)$);%
    \fi
    #2
  \end{tikzpicture}
  \gSetStdTypes{line}
}

\def\cBoxAA{\@ifnextchar[\@cBoxAA{\@cBoxAA[]}}
\def\@cBoxAA[#1]#2{%
  \setkeys*{pre}{#1}%
  \setrmkeys{general,gTypes,gLabels,dots,blobs}%
  \begin{tikzpicture}
    [line width=1pt,
    baseline={([yshift=-0.5ex+\cmdKV@general@yshift]current bounding box.center)},
    scale=\cmdKV@general@scale,
    rotate=\cmdKV@general@rotate,
    font=\cmdKV@general@font]
    \fill[\cmdKV@blobs@blobA] (-0.4,0) ellipse (0.2 and 0.4);
    \fill[\cmdKV@blobs@blobB] (0.4,0) ellipse (0.2 and 0.4);
    \draw[\cmdKV@gTypes@eA] ($ (-0.4,0) + (-135:0.26) $) -- ($ (-0.4,0) + (-135:0.7) $) node[label=(-180+\cmdKV@general@rotate):\cmdKV@gLabels@eLA] {};
    \draw[\cmdKV@gTypes@eB] ($ (-0.4,0) + (135:0.26) $) -- ($ (-0.4,0) + (135:0.7) $) node[label=(180+\cmdKV@general@rotate):\cmdKV@gLabels@eLB] {};
    \draw[\cmdKV@gTypes@eC] ($ (0.4,0) + (45:0.26) $) -- ($ (0.4,0) + (45:0.7) $) node[label=(\cmdKV@general@rotate):\cmdKV@gLabels@eLC] {};
    \draw[\cmdKV@gTypes@eD] ($ (0.4,0) + (-45:0.26) $) -- ($ (0.4,0) + (-45:0.7) $) node[label=(\cmdKV@general@rotate):\cmdKV@gLabels@eLD] {};
    \draw[label distance=1,\cmdKV@gTypes@iA] ($ (-0.4,0) + (60:0.32) $) to[bend left] node[label=(90+\cmdKV@general@rotate):\cmdKV@gLabels@iLA] {}($ (0.4,0) + (120:0.32) $);
    \draw[label distance=1,\cmdKV@gTypes@iB] ($ (0.4,0) + (-120:0.32) $) to[bend left]  node[label=(-90+\cmdKV@general@rotate):\cmdKV@gLabels@iLB] {}($ (-0.4,0) + (-60:0.32) $);
    \draw[dotted] (0,0.25) -- (0,-0.25);
    \node[label=(180+\cmdKV@general@rotate):\cmdKV@gLabels@iLC] at (-0.6,0) {};
    \node[label=(\cmdKV@general@rotate):\cmdKV@gLabels@iLD] at (0.6,0) {};
    \ifKV@dots@lDots
      \draw[dotted] ($(-0.4,0) + (-150:0.7)$) to[bend left] ($(-0.4,0) + (150:0.7)$);%
    \fi
    \ifKV@dots@rDots
      \draw[dotted] ($(0.4,0) + (30:0.7)$) to[bend left] ($(0.4,0) + (-30:0.7)$);%
    \fi
    #2
  \end{tikzpicture}
  \gSetStdTypes{line}
}

\def\cBoxB{\@ifnextchar[\@cBoxB{\@cBoxB[]}}
\def\@cBoxB[#1]#2{%
  \setkeys*{pre}{#1}%
  \setrmkeys{general,gTypes,gLabels,blobs}%
  \begin{tikzpicture}
    [line width=1pt,
    baseline={([yshift=-0.5ex+\cmdKV@general@yshift]current bounding box.center)},
    scale=\cmdKV@general@scale,
    rotate=\cmdKV@general@rotate,
    font=\cmdKV@general@font]
    \fill[\cmdKV@blobs@blobA] (0,-0.3) ellipse (0.4 and 0.2);
    \fill[\cmdKV@blobs@blobB] (0,0.3) ellipse (0.4 and 0.2);
    \draw[\cmdKV@gTypes@eA] ($ (0,-0.3) + (180:0.4) $) -- ($ (0,-0.3) + (-170:0.7) $) node[label=(180+\cmdKV@general@rotate):\cmdKV@gLabels@eLA] {};
    \draw[\cmdKV@gTypes@eB] ($ (0,0.3) + (180:0.4) $) -- ($ (0,0.3) + (170:0.7) $) node[label=(180+\cmdKV@general@rotate):\cmdKV@gLabels@eLB] {};
    \draw[\cmdKV@gTypes@eC] ($ (0,0.3) + (0:0.4) $) -- ($ (0,0.3) + (10:0.7) $) node[label=(\cmdKV@general@rotate):\cmdKV@gLabels@eLC] {};
    \draw[\cmdKV@gTypes@eD] ($ (0,-0.3) + (0:0.4) $) -- ($ (0,-0.3) + (-10:0.7) $) node[label=(\cmdKV@general@rotate):\cmdKV@gLabels@eLD] {};
    \draw[label distance=1,\cmdKV@gTypes@iA] ($ (0,-0.3) + (135:0.26) $) -- node[label=(-180+\cmdKV@general@rotate):\cmdKV@gLabels@iLA] {}($ (0,0.3) + (-135:0.26) $);
    \draw[label distance=1,\cmdKV@gTypes@iB] ($ (0,0.3) + (-45:0.26) $) -- node[label=(\cmdKV@general@rotate):\cmdKV@gLabels@iLB] {}($ (0,-0.3) + (45:0.26) $);
    \node[label=(90+\cmdKV@general@rotate):\cmdKV@gLabels@iLC] at (0,0.5) {};
    \node[label=(-90+\cmdKV@general@rotate):\cmdKV@gLabels@iLD] at (0,-0.5) {};
    #2
  \end{tikzpicture}
  \gSetStdTypes{line}
}

\def\cBoxBoxA{\@ifnextchar[\@cBoxBoxA{\@cBoxBoxA[]}}
\def\@cBoxBoxA[#1]#2{%
  \setkeys*{pre}{#1}%
  \setrmkeys{general,gTypes,gLabels,blobs}%
  \begin{tikzpicture}
    [line width=1pt,
    baseline={([yshift=-0.5ex+\cmdKV@general@yshift]current bounding box.center)},
    scale=\cmdKV@general@scale,
    rotate=\cmdKV@general@rotate,
    font=\cmdKV@general@font]
    \fill[\cmdKV@blobs@blobA] (-0.4,0) ellipse (0.2 and 0.4);
    \fill[\cmdKV@blobs@blobB] (0.4,0) ellipse (0.2 and 0.4);
    \fill[\cmdKV@blobs@blobC] (1.2,0) ellipse (0.2 and 0.4);
    \draw[\cmdKV@gTypes@eA] ($ (-0.4,0) + (-135:0.26) $) -- ($ (-0.4,0) + (-135:0.7) $) node[label=(-180+\cmdKV@general@rotate):\cmdKV@gLabels@eLA] {};
    \draw[\cmdKV@gTypes@eB] ($ (-0.4,0) + (135:0.26) $) -- ($ (-0.4,0) + (135:0.7) $) node[label=(180+\cmdKV@general@rotate):\cmdKV@gLabels@eLB] {};
    \draw[\cmdKV@gTypes@eC] ($ (1.2,0) + (45:0.26) $) -- ($ (1.2,0) + (45:0.7) $) node[label=(\cmdKV@general@rotate):\cmdKV@gLabels@eLC] {};
    \draw[\cmdKV@gTypes@eD] ($ (1.2,0) + (-45:0.26) $) -- ($ (1.2,0) + (-45:0.7) $) node[label=(\cmdKV@general@rotate):\cmdKV@gLabels@eLD] {};
    \draw[label distance=1,\cmdKV@gTypes@iA] ($ (-0.4,0) + (45:0.26) $) -- node[label=(90+\cmdKV@general@rotate):\cmdKV@gLabels@iLA] {}($ (0.4,0) + (135:0.26) $);
    \draw[label distance=1,\cmdKV@gTypes@iB] ($ (0.4,0) + (45:0.26) $) -- node[label=(90+\cmdKV@general@rotate):\cmdKV@gLabels@iLB] {}($ (1.2,0) + (135:0.26) $);
    \draw[label distance=1,\cmdKV@gTypes@iC] ($ (1.2,0) + (-135:0.26) $) -- node[label=(-90+\cmdKV@general@rotate):\cmdKV@gLabels@iLC] {}($ (0.4,0) + (-45:0.26) $);
    \draw[label distance=1,\cmdKV@gTypes@iD] ($ (0.4,0) + (-135:0.26) $) -- node[label=(-90+\cmdKV@general@rotate):\cmdKV@gLabels@iLD] {}($ (-0.4,0) + (-45:0.26) $);
    \node[label=(180+\cmdKV@general@rotate):\cmdKV@gLabels@iLE] at (-0.6,0) {};
    \node[label=(\cmdKV@general@rotate):\cmdKV@gLabels@iLF] at (0.6,0) {};
    \node[label=(\cmdKV@general@rotate):\cmdKV@gLabels@iLG] at (1.4,0) {};
    #2
  \end{tikzpicture}
  \gSetStdTypes{line}
}

\def\cBoxBoxB{\@ifnextchar[\@cBoxBoxB{\@cBoxBoxB[]}}
\def\@cBoxBoxB[#1]#2{%
  \setkeys*{pre}{#1}%
  \setrmkeys{general,gTypes,gLabels}%
  \begin{tikzpicture}
    [line width=1pt,
    baseline={([yshift=-0.5ex+\cmdKV@general@yshift]current bounding box.center)},
    scale=\cmdKV@general@scale,
    rotate=\cmdKV@general@rotate,
    font=\cmdKV@general@font]
    \fill[\cmdKV@blobs@blobA] (0,0.3) ellipse (0.4 and 0.2);
    \fill[\cmdKV@blobs@blobB] (0,-0.3) ellipse (0.4 and 0.2);
    \fill[\cmdKV@blobs@blobC] (1,0) ellipse (0.2 and 0.4);
    \draw[\cmdKV@gTypes@eA] ($ (0,-0.3) + (180:0.4) $) -- ($ (0,-0.3) + (-170:0.7) $) node[label=(180+\cmdKV@general@rotate):\cmdKV@gLabels@eLA] {};
    \draw[\cmdKV@gTypes@eB] ($ (0,0.3) + (180:0.4) $) -- ($ (0,0.3) + (170:0.7) $) node[label=(180+\cmdKV@general@rotate):\cmdKV@gLabels@eLB] {};
    \draw[\cmdKV@gTypes@eC] ($ (1,0) + (45:0.26) $) -- ($ (1,0) + (45:0.7) $) node[label=(\cmdKV@general@rotate):\cmdKV@gLabels@eLC] {};
    \draw[\cmdKV@gTypes@eD] ($ (1,0) + (-45:0.26) $) -- ($ (1,0) + (-45:0.7) $) node[label=(\cmdKV@general@rotate):\cmdKV@gLabels@eLD] {};
    \draw[label distance=1,\cmdKV@gTypes@iA] ($ (0,-0.3) + (135:0.26) $) -- node[label=(-180+\cmdKV@general@rotate):\cmdKV@gLabels@iLA] {}($ (0,0.3) + (-135:0.26) $);
    \draw[label distance=1,\cmdKV@gTypes@iB] ($ (0,0.3) + (0:0.4) $) -- node[label=(90+\cmdKV@general@rotate):\cmdKV@gLabels@iLB] {} ($ (1,0) + (135:0.26) $);
    \draw[label distance=1,\cmdKV@gTypes@iC] ($ (1,0) + (-135:0.26) $) -- node[label=(-90+\cmdKV@general@rotate):\cmdKV@gLabels@iLC] {} ($ (0,-0.3) + (0:0.4) $);
    \draw[label distance=1,\cmdKV@gTypes@iD] ($ (0,0.3) + (-45:0.26) $) -- node[label=(\cmdKV@general@rotate):\cmdKV@gLabels@iLD] {}($ (0,-0.3) + (45:0.26) $);
    \node[label=(90+\cmdKV@general@rotate):\cmdKV@gLabels@iLE] at (0,0.5) {};
    \node[label=(\cmdKV@general@rotate):\cmdKV@gLabels@iLF] at (1.2,0) {};
    \node[label=(-90+\cmdKV@general@rotate):\cmdKV@gLabels@iLG] at (0,-0.5) {};
    #2
  \end{tikzpicture}
  \gSetStdTypes{line}
}

\def\cNPBox{\@ifnextchar[\@cNPBox{\@cNPBox[]}}
\def\@cNPBox[#1]#2{%
  \setkeys*{pre}{#1}%
  \setrmkeys{general,gTypes,gLabels,blob}%
  \begin{tikzpicture}
    [line width=1pt,
    baseline={([yshift=-0.5ex+\cmdKV@general@yshift]current bounding box.center)},
    scale=\cmdKV@general@scale,
    rotate=\cmdKV@general@rotate,
    font=\cmdKV@general@font]
    \fill[\cmdKV@blobs@blobA] (-0.6,0) ellipse (0.2 and 0.4);
    \fill[\cmdKV@blobs@blobB] (0.4,0) ellipse (0.2 and 0.4);
    \draw[\cmdKV@gTypes@eA] ($ (-0.6,0) + (-180:0.2) $) -- ($ (-0.6,0) + (-180:0.6) $) node[label=(-180+\cmdKV@general@rotate):\cmdKV@gLabels@eLA] {};
    \draw[\cmdKV@gTypes@eB] ($ (-0.6,0) + (0:0.2) $) -- (-0.18,0) node[label=(\cmdKV@general@rotate):\cmdKV@gLabels@eLB] {};
    \draw[\cmdKV@gTypes@eC] ($ (0.4,0) + (45:0.26) $) -- ($ (0.4,0) + (45:0.7) $) node[label=(\cmdKV@general@rotate):\cmdKV@gLabels@eLC] {};
    \draw[\cmdKV@gTypes@eD] ($ (0.4,0) + (-45:0.26) $) -- ($ (0.4,0) + (-45:0.7) $) node[label=(\cmdKV@general@rotate):\cmdKV@gLabels@eLD] {};
    \draw[label distance=1,\cmdKV@gTypes@iA] ($ (-0.6,0) + (60:0.3) $) -- node[label=(90+\cmdKV@general@rotate):\cmdKV@gLabels@iLA] {}($ (0.4,0) + (135:0.26) $);
    \draw[label distance=1,\cmdKV@gTypes@iB] ($ (0.4,0) + (-135:0.26) $) -- node[label=(-90+\cmdKV@general@rotate):\cmdKV@gLabels@iLB] {}($ (-0.6,0) + (-60:0.3) $);
    \node[label=(\cmdKV@general@rotate):\cmdKV@gLabels@iLD] at (0.6,0) {};
    #2
  \end{tikzpicture}
  \gSetStdTypes{line}
}

\def\cFourPtRec{\@ifnextchar[\@cFourPtRec{\@cFourPtRec[]}}
\def\@cFourPtRec[#1]#2{%
  \setkeys*{pre}{#1}%
  \setrmkeys{general,gTypes,gLabels,blobs}%
  \begin{tikzpicture}
    [line width=1pt,
    baseline={([yshift=-0.5ex+\cmdKV@general@yshift]current bounding box.center)},
    scale=\cmdKV@general@scale,
    rotate=\cmdKV@general@rotate,
    font=\cmdKV@general@font]
    \fill[\cmdKV@blobs@blobA] (-0.5,0) ellipse (0.2 and 0.3);
    \fill[\cmdKV@blobs@blobB] (0.5,0) ellipse (0.3 and 0.4);
    \filldraw[fill=white,draw=black] (0.55,0.15) circle (0.1);
    \filldraw[fill=white,draw=black] (0.45,-0.1) circle (0.1);
    \draw[\cmdKV@gTypes@eA] ($(-0.5,0) + (135:0.23)$) -- ($(-0.5,0) + (135:0.5)$) node[label=(180+\cmdKV@general@rotate):\cmdKV@gLabels@eLA] {};
    \draw[\cmdKV@gTypes@eB] ($(-0.5,0) + (-135:0.23)$) -- ($(-0.5,0) + (-135:0.5)$) node[label=(180+\cmdKV@general@rotate):\cmdKV@gLabels@eLB] {};
    \draw[\cmdKV@gTypes@eC] ($(0.5,0) + (45:0.35)$) -- ($(0.5,0) + (45:0.7)$) node[label=(\cmdKV@general@rotate):\cmdKV@gLabels@eLC] {};
    \draw[\cmdKV@gTypes@eD] ($(0.5,0) + (-45:0.35)$) -- ($(0.5,0) + (-45:0.7)$) node[label=(\cmdKV@general@rotate):\cmdKV@gLabels@eLD] {};
    \draw[label distance=1,\cmdKV@gTypes@iA] ($(-0.5,0) + (30:0.23)$) -- node[label=(90+\cmdKV@general@rotate):\cmdKV@gLabels@iLA] {} ($(0.5,0) + (157:0.3)$);
    \draw[label distance=1,\cmdKV@gTypes@iB] ($(0.5,0) + (-157:0.3)$) -- node[label=(-90+\cmdKV@general@rotate):\cmdKV@gLabels@iLB] {} ($(-0.5,0) + (-30:0.23)$);
    #2
 \end{tikzpicture}
  \gSetStdTypes{line}
}

\makeatother

\numberwithin{equation}{section}

\newcommand{\nn}{\nonumber}

\newcommand\beq{\begin{equation}}
\newcommand\eeq{\end{equation}}
\newcommand\beal{\begin{aligned}}
\newcommand\eeal{\end{aligned}}
\newcommand\bea{\begin{eqnarray}}
\newcommand\eea{\end{eqnarray}}

\newcommand\dd{{\mathrm d}}
\usepackage{xcolor}

\newcommand{\bell}{{\boldsymbol \ell}}
\newcommand{\bu}{{\boldsymbol u}}
\newcommand{\bb}{{\boldsymbol b}}
\newcommand{\bk}{{\boldsymbol k}}

\newcommand{\bp}{{\boldsymbol p}}

\newcommand{\bq}{{\boldsymbol q}}

\newcommand{\bx}{{\boldsymbol x}}

\newcommand{\cD}{\mathcal{D}}
\newcommand\cE{\mathcal{E}}

\newcommand\cS{\mathcal{S}}

\newcommand\Mp{M_{\rm Pl}}
\newcommand\cL{\mathcal{L}}

\makeatletter
\newcommand{\Biggg}{\bBigg@{3.5}}
\makeatother

\begin{document}
\preprint{DESY\,20-077\\\phantom{~} \hfill SLAC-PUB-17529}
\title{\center Post-Minkowskian Effective Field Theory\\ [0.2cm] for Conservative Binary Dynamics}

\author[a]{\large Gregor K\"alin}
\author[b]{\large and Rafael A. Porto}
\affiliation[a]{SLAC National Accelerator Laboratory, Stanford University, Stanford, CA 94309, USA}
\affiliation[b]{Deutsches Elektronen-Synchrotron DESY, Notkestrasse 85, 22607 Hamburg, Germany}
\emailAdd{greka@slac.stanford.edu}\emailAdd{rafael.porto@desy.de}
\abstract{
  We develop an Effective Field Theory (EFT) formalism to solve for the conservative dynamics of binary systems in gravity via Post-Minkowskian (PM) scattering~data.
  Our framework combines a systematic EFT approach to compute the deflection angle in the PM expansion, together with the `Boundary-to-Bound' (B2B) dictionary introduced~in~\cite{paper1,paper2}.
  Due to the nature of scattering processes, a remarkable reduction of complexity occurs both in the number of Feynman diagrams and type of integrals, compared to a direct EFT computation of the potential in a PM scheme. We provide two illustrative examples. Firstly, we compute all the conservative gravitational observables for bound orbits to 2PM, which follow from only one topology beyond leading order.
  The~results agree with those in \cite{paper1,paper2}, obtained through the `impetus formula' applied to the classical limit of the one loop amplitude in Cheung et al.~\cite{cheung}.
  For~the sake of comparison we reconstruct the conservative Hamiltonian to 2PM order, which is equivalent to the one derived in \cite{cheung} from a matching calculation. Secondly, we compute the scattering angle due to tidal effects from the electric- and magnetic-type Love numbers at leading PM order. Using the B2B dictionary we then obtain the tidal contribution to the periastron advance. We also construct a Hamiltonian including tidal effects at leading PM order. Although relying on (relativistic) Feynman diagrams, the EFT formalism developed here does not involve taking the classical limit of a quantum amplitude, neither integrals with internal massive fields, nor additional matching calculations, nor spurious (`super-classical') infrared singularities.  By~construction, the EFT approach can be automatized to all PM orders.
}

\maketitle
\newpage

\section{Introduction} \label{sec:introduction}

The discovery potential in gravitational wave (GW) science\footnote{\url{https://www.gw-openscience.org}}\,\cite{LIGO} relies in our ability to make precise theoretical predictions~\cite{buosathya,tune,music}.
While the late stages of the binary dynamics require numerical modeling, the majority of GW cycles occur during the inspiral regime where perturbative approximations to Einstein's equations, such as the Post-Newtonian (PN) expansion, remain of vital importance to provide a faithful reconstruction of the signal \cite{buosathya}.
It is in this regime where the effective field theory (EFT) approach introduced in \cite{nrgr}, aka Non-Relativistic General Relativity (NRGR), has proven to be very successful to tackle the binary problem in gravity, see e.g. \cite{walterLH,LHme2,foffa,iragrg,grg13,review} for various reviews. In addition to the systematization of the problem of motion into the computation of a series of Feynman diagrams, and naturally incorporating finite-size effects via worldline terms beyond minimal coupling, one of the main virtues of the EFT formalism is the use of the \emph{method of regions}~\cite{Beneke:1997zp}, which allows us to disentangle the relevant physics involving both (off-shell) {\it potential} and (on-shell) \emph{radiation} modes.
The EFT methodology not only allows us to separate the computation of the relevant ingredients for waveform modeling into the `conservative' and `radiative' sectors, respectively, it also naturally handles the spurious infrared (IR) and ultraviolet (UV) divergences that appear from simplifying the resulting integrals by splitting into regions \cite{lamb,apparent}, as well as the UV divergences due to the use of localized sources \cite{nrgr}.
As a result, joining an effort which involves also more `traditional' methods \cite{blanchet,Schafer:2018kuf,Damour:2014jta,Jaranowski:2015lha,Bernard:2015njp,Bernard:2017bvn,Marchand:2017pir}, the present state-of-the-art in NRGR is at the fourth PN (4PN) order of accuracy in the conservative sector for non-spinning bodies \cite{nrgr,nrgr2pn,nrgr3pn,Foffa:2012rn,tail,nrgrG5,apparent,nrgr4pn1,nrgr4pn2}, which includes also contributions induced from (conservative) radiation-reaction effects \cite{tail,apparent,lamb}.\footnote{Radiation effects can be systematically incorporated in NRGR in terms of source and radiative multipoles, see e.g \cite{dis1, andirad,andirad2,adamchad1,radnrgr}. Moreover, spinning compact objects were introduced in \cite{nrgrs} and also extensively studied in the NRGR literature, see e.g. \cite{prl,Porto:2007px,Porto:2007tt,dis2,nrgrss,nrgrs2,nrgrso,rads1,amps,maiaso,maiass,levi,Levi:2020kvb,Levi:2020uwu}.
We will study both radiation and rotation in future work.} This represents the next-to-next-to-next-to-next-to leading order (N$^4$LO), or `four loops', level of precision. Partial results have also been computed in NRGR at higher orders, both in the static sector at 5PN~\cite{5pn1,5pn2} and for hereditary effects at NLO \cite{tail2,tail3}.\vskip 4pt
 
The method of regions is one of the trademarks of the EFT framework in a PN regime, allowing us to expand integrals in the potential region ($k^0 \ll |\bk|$) in powers of $k_0/|\bk|$.  At the level of the computation of the gravitational potential, this is the reason for the (in)famous IR divergences appearing at 4PN, that ultimately cancel out in NRGR once conservative effects from radiation-reaction are incorporated \cite{tail,lamb,apparent,nrgr4pn2}.\footnote{The IR divergences led to the introduction of ambiguity-parameters in other approaches \cite{Damour:2014jta,Jaranowski:2015lha,Bernard:2015njp,Bernard:2017bvn}, yet to be resolved within the PN regime in the ADM formalism of \cite{Schafer:2018kuf}. See \cite{Marchand:2017pir} for an ambiguity-free derivation following dimensional regularization, as advocated in \cite{tail,apparent}, including a re-derivation of the conservative tail effect computed in \cite{tail}.}
Furthermore, all the velocity (relativistic) corrections, both from vertex and propagator corrections, are truncated depending on the factors of Newton's constant~($G$) in each given contribution, rather than incorporated to all orders from the onset.
The reason being due to the virial theorem, relating $GM/r \sim v^2$ for bound states, which implies that only terms scaling as $G^\ell v^{2(n-\ell)}$ (with $\ell \leq n$) are needed to $n$PN order. While some of the factors of $k_0$, from the expansion of the Green's functions in NRGR, wound up producing acceleration-dependent contributions to the gravitational potential which can be traded for higher powers of $G$ using lower order equations of motion, an infinite series of velocity corrections is truncated. Therefore, although the expansion in the potential region is extremely powerful, reducing complex four- into three-dimensional massless integrals, e.g. \cite{nrgrG5}, the natural expansion of field theory in powers of the coupling, or the Post-Minkowskian (PM) expansion in this context, begs us for a framework in which (special) relativistic effects are included to all orders.\vskip 4pt

An obvious option to implement a PM expansion for bound orbits is to resum all the velocity corrections in the NRGR computation of the potential to a given order in~$G$, while dropping terms depending on the accelerations. This, however, turns out to be unnecessarily cumbersome. (See \cite{5pnfoffa} for an attempt at 1PM.) Moreover, since the potential is a gauge-dependent quantity, many (many) different terms can be present depending on the gauge, with Lorentz invariance only manifest in the final expressions. Another option is to attempt to perform the much more difficult exact relativistic integration, keeping only real contributions to the effective potential. As we shall see, however, a simpler procedure can be adopted, in which rather than bringing the PM expansion to NRGR, we will bring instead the EFT approach to the natural habitat of the PM framework. Inspired by the study of relativistic scattering, in this paper we implement an EFT formalism to compute the deflection angle perturbatively in $G$, but to all orders in the relative velocity. Even though, in principle, it may seem like scattering processes are unrelated to bound orbits, a remarkable connection has been recently uncovered in \cite{paper1,paper2}, in the form of a `Boundary-to-Bound' (B2B) dictionary. The map put forward in \cite{paper1,paper2} relates scattering data directly to dynamical invariants for binary systems, without using a Hamiltonian.
Our goal in this paper, as suggested also in \cite{paper1,paper2}, is therefore to combine the virtues of the EFT formalism and B2B dictionary \cite{paper1,paper2} to construct a systematic EFT approach to the conservative dynamics of binary~systems in the PM expansion, without ever resorting to gauge-dependent objects. \vskip 4pt

The idea of mapping (quantum) scattering information into the (classical) physics of binary systems, dating back to early efforts in the 70's, e.g. \cite{iwasaki}, has been reinvigorated by the recent program to connect amplitudes techniques from high-energy physics \cite{jj,elvang,cliff,reviewdc} to the derivation of the two-body gravitational dynamics in the classical regime, e.g. \cite{ira1,cheung,zvi1,zvi2,donal,donalvines, withchad,Holstein:2008sx,Bjerrum-Bohr:2013bxa,Vaidya:2014kza,Guevara:2017csg,Chung:2018kqs,Guevara:2018wpp,simon,Guevara:2019fsj,bohr,cristof1,Arkani-Hamed:2019ymq,Bjerrum-Bohr:2019kec,Chung:2019duq,Bautista:2019tdr,Bautista:2019evw,KoemansCollado:2019ggb,Johansson:2019dnu,Aoude:2020onz,Cristofoli:2020uzm,Chung:2020rrz,Bern:2020gjj,Bern:2020buy,Parra,Antonelli:2019ytb}.
Furthermore, the connection between scattering data and binary dynamics has also been emphasized in the context of the effective one body (EOB) approach \cite{eob}, e.g.~\cite{damour1,damour2,damour3n,damour3,bini1,bini2,Vines:2018gqi,justin2,binit}.
However, in all these cases the derivation of a gauge-dependent and rather lengthy Hamiltonian, or EOB equivalent, has played a central role. This feature was, after all, one of the main reasons that motivated us in \cite{paper1,paper2} to construct the B2B map directly between the (much simpler) gauge-invariant observables. The B2B dictionary was originally introduced in \cite{paper1}, via the construction of a radial action depending on the analytic continuation of scattering~data. Although the radial action was first discussed in the context of the PN expansion, see e.g.~\cite{9912}, the analysis in \cite{paper1,paper2} unveiled for the first time the astonishingly simple structure in the~PM~framework instead. In~its first incarnation, the B2B radial action was built upon a remarkable connection between the analytic continuation of the scattering amplitude in the classical limit and the momentum of the particles in the center-of-mass frame, which we dubbed the `impetus formula' \cite{paper1}. We~later extended the dictionary in \cite{paper2}, unraveling an unexpected relationship between the deflection angle and periastron advance, once again via analytic continuation, which allowed~us to reconstruct the B2B map  entirely in terms of the scattering angle without invoking the impetus formula. The analysis in \cite{paper1,paper2} has thus uncovered a surprisingly simple hidden structure of the radial action in the PM conservative sector, readily mapping scattering (boundary) data to dynamical invariants for generic (bound) orbits. While the B2B dictionary neatly translates all the gauge-invariant scattering information into observables for binary systems, it has been implemented so far via the classical limit of the amplitude, either through the impetus formula or from the scattering angle, computed in \cite{zvi1,zvi2} to 3PM order (or two loops). Motivated by the prowesses of NRGR, in this paper we provide an alternative framework to collect the necessary scattering data to input in the B2B dictionary, using instead an EFT approach adapted to the computation of the impulse and scattering angle in the PM scheme.\vskip 4pt

The powerful formalism developed in \cite{cheung,zvi1,zvi2} has demonstrated its ability to obtain physical information from scattering amplitudes to high PM orders. Yet, in its present form, it relies on taking the classical limit of a quantum amplitude, performing a matching calculation to an effective theory built with a local potential interaction and, perhaps more importantly, relativistic integrals involving internal massive propagators yielding spurious IR divergences. The singular terms, due to so-called `super-classical' contributions, ultimately cancel out either during the matching calculation in \cite{cheung, zvi1,zvi2} or equivalent after including `Born iterations'~\cite{Holstein:2008sx,cristof1,Cristofoli:2020uzm}.
However, they signal a generic feature of computations involving the classical limit of quantum amplitudes.\footnote{This issue is present also when dealing directly with gauge-invariant quantities, as shown in \cite{donal,donalvines} for the derivation of the scattering angle. In such case the cancelation occurs instead between two independent contributions to the impulse, linear and quadratic in the amplitude.}
Hence, while the method of \emph{generalized unitarity}~\cite{Bern:1994zx,Bern:1994cg} and the \emph{double copy} technique~\cite{Bern:2008qj,Bern:2010ue} allow to bypass the need of a large set of Feynman diagrams \cite{zvi2},\footnote{The gravitational scattering amplitude at two loops has been recently computed using standard Feynman techniques in \cite{Cheung:2020gyp} and shown to agree with the results of \cite{cheung,zvi1,zvi2}.} and the impetus formula \cite{paper1} can sidestep the need of a matching calculation to extract the Hamiltonian, it is still worthwhile to develop an independent systematic framework to compute classical observables for bound orbits from scattering data, without the type of integrals and subtle limits that appear in the amplitude program. \vskip 4pt

In an EFT approach to gravity for extended objects, the bodies are treated as external localized sources endowed with a series of (Wilson) coefficients that parameterize finite-size effects \cite{nrgr}. Therefore, we do not include mass-dependent propagators, and as long as loops of the gravitational field are avoided, we always remain in the classical realm. As a result, the derivation of the scattering~angle is reduced into two- and three-dimensional integrals involving only massless~propagators. Not~only these are easier to compute, we do not encounter spurious super-classical IR singularities, such as e.g. the box diagram at one loop \cite{cheung,cristof1}.\footnote{As we mentioned, other type of spurious divergences may appear when dealing with potential modes \cite{tail,apparent,nrgr4pn2}. However, we expect these to explicitly cancel out in a fully relativistic framework that also incorporates conservative radiation-reaction effects. We will return to this in future~work.}
There are, however, various similarities between the approach developed in \cite{cheung,zvi1,zvi2} and the one introduced here. For instance, one of the key aspects of our formalism consists on isolating the conservative contribution from the potential region.
This is achieved by providing a prescription to perform the energy integrals, which turns out to be similar to the one in \cite{cheung,zvi1,zvi2}, except for the (lack of) `anti-matter' poles and massive internal fields. Moreover, as we shall see, cancelations also occur in our derivation of the scattering angle. For example, the vanishing of a term which we may identify with the crossed-box diagram in \cite{zvi2}.\vskip 4pt

An EFT approach for the PM expansion has many virtues, yet the disadvantage with respect to the amplitudes program of \cite{cheung,zvi1,zvi2} is the persistent reliance on the perturbative machinery of Feynman diagrams, when it comes to purely classical computations. However, because of the notorious simplifications of the scattering process, even at the classical level, a comparatively small number of diagrams are needed, significantly reducing the amount of combinatorial complexity. For instance, only two are required to 2PM, which we compute in detail in this paper, whereas an additional five enter at 3PM order. Therefore, the EFT formalism in the PM scheme provides an alternative systematic machinery to compute the PM conservative dynamics of binary systems, potentially reaching the same level of success as NRGR in the PN expansion --- currently at N$^4$LO (akin of 5PM accuracy) --- using the existent technology in the field \cite{nrgr4pn1,nrgr4pn2}, and elsewhere \cite{Smirnov}. Yet, since the number of diagrams and necessary steps will at some point start to escalate quickly, it is expected that the double copy and other on-shell techniques from the amplitude program may be able to reduce further the level of complexity in the derivation of the scattering angle. (Another possibility is to recast Einstein's gravity as in \cite{Cheung:2017kzx}, to simplify the number of Feynman diagrams.) Unfortunately, at the moment, we have not found a simple way to incorporate these tools into the EFT formalism, although ideas from the classical doubly copy, e.g. \cite{Don1,Luna:2016hge,Goldberger:2016iau, Walter,Li:2018qap,CSH,Plefka,Plefka:2019hmz,Kim:2019jwm,Goldberger:2019xef,Alfonsi:2020lub}, may ultimately provide a hybrid framework to march smoothly towards higher orders.\vskip 4pt

This paper is organized as follows. In \S\ref{sec:b2b} we review the B2B map, emphasizing the construction of the radial action from PM scattering data. In \S\ref{sec:pmeft} we introduce the effective theory for conservative dynamics, the worldline and bulk action, as well as the integration prescription for potential modes.
We then show how to use the effective theory to compute the impulse and deflection angle to all PM orders.
In \S\ref{sec:obs} we demonstrate the power of the effective theory by (re)deriving the impulse and scattering angle to 2PM in a few simple steps, from which we obtain all the dynamical invariants for binary system through the B2B dictionary.
For the sake of comparison, we reconstruct a 2PM Hamiltonian for the two-body dynamics, which we show is equivalent to the one computed in \cite{cheung,zvi1,zvi2}. In \S\ref{sec:finite} we compute the leading PM contribution to the scattering angle from electric- and magnetic-type tidal effects, which are obtained from insertions of higher dimensional operators in the worldline action. Following the B2B map we then obtain the leading correction to the periastron advance due to (spin-independent) finite-size effects at leading PM order. We also construct the associated Hamiltonian including tidal effects. We conclude with a summary and the road ahead in \S\ref{sec:disc}, and a few comment on the relationship between the action \& impulse versus the amplitude \& eikonal in the classical limit, in App.~\ref{eik}.\vskip 4pt

{\it Conventions}: We use the mostly minus signature: $\eta_{\mu\nu} = {\rm diag}(+,-,-,-)$. The Minkowski product between four-vectors is denoted as $k \cdot x = \eta_{\mu\nu} k^\mu x^\nu$, while we use $\bk \cdot \bx = \delta^{ij} \bk^i \bx^j$ for the Euclidean version, with bold letters representing ${\bf 3}$-vectors. We use $\bk_\perp$ for vectors in the plane perpendicular to the direction of the scattering particles.  We use the shorthand notation $\int_k \equiv \int d^4k/(2\pi)^4$, $\int_{\bk} \equiv \int d^3\bk/(2\pi)^3$, and $\int_{\bk_\perp} \equiv \int d^2\bk_\perp/(2\pi)^2$. We also absorb factors of $2\pi$ into the $\delta$ functions: $\hat\delta(x) \equiv 2\pi \delta(x)$. For divergent integrals we use dimensional regularization, such that the number of dimensions is replaced by $D= d-2\epsilon$, with $d$ either $4,3$ or~$2$. We use $\Mp^{-1} \equiv \sqrt{32\pi G}$ for the Planck mass in $\hbar=c=1$ units. 
We~denote  $M=m_1+m_2$ the total mass, $\mu = m_1m_2/M$ the reduced mass, and $\nu \equiv \mu/M$ the symmetric mass ratio. 

\section{Boundary-to-Bound} \label{sec:b2b}
In this section we briefly review the ingredients introduced in \cite{paper1,paper2} to compute gravitational observables for binary systems using scattering data. For the sake of comparison, we also illustrate the reconstruction of the Hamiltonian from the scattering angle. 

\subsection{From Angles to Action $\ldots$}
In this paper we will not use the impetus formula introduced in the B2B map of \cite{paper1} to construct the radial action. Instead, we use the representation discovered~in \cite{paper2} from the relationship between the scattering angle and periastron advance, yielding
\beq
i_r \equiv \frac{{\cal S}_r}{GM\mu} =  \frac{\hat p_\infty}{\sqrt{-\hat p_\infty^2}}
\chi^{(1)}_j - j \left(1 + \frac{2}{\pi} \sum_{n=1}  \frac{\chi^{(2n)}_j}{(1-2n)j^{2n}}\right)  \,.\label{eq:ir}
\eeq
The PM coefficients of the scattering angle are defined through
\beq
\frac{\chi}{2} = \sum_{n=1} \chi^{(n)}_j/j^n\,,\label{eq:pmangle}
\eeq 
with $j = J/(GM\mu)$ the reduced angular momentum. The energy of the two-body system can be written as
\beq
E =M + \mu\cE = M(1+\nu \cE)\,,\label{eq:E}
\eeq
with the momentum at infinity (in the center-of-mass frame) given by
\beq
p_\infty = \mu \frac{\sqrt{\gamma^2-1}}{\Gamma} = \mu\, \hat p_\infty \,,\label{pinf}
\eeq
with 
\beq
\begin{aligned}
\label{Egam}
\gamma &\equiv {p_1\cdot p_2 \over m_1 m_2} = u_1\cdot u_2 = \frac{E^2-m_1^2-m_2^2}{2m_1 m_2}  = 1 + \cE + \frac{\nu \cE^2}{2}\,,\\
\Gamma &\equiv E/M = \sqrt{1+2\nu(\gamma-1)}\,.
\end{aligned}
\eeq
The representation in angular-momentum space may be obtained from the expansion in impact parameter,
\beq
\frac{\chi}{2} = \sum_{n=1} \chi^{(n)}_b \left(\frac{GM}{b}\right)^n\,,\label{eq:pmangleb}
\eeq 
by replacing $j^{-1} = \hat p^{-1}_\infty \frac{GM}{b}$ in \eqref{eq:pmangle} such that $\chi^{(n)}_j = \hat p_\infty^n \chi^{(n)}_b$.

\subsection{$\ldots$ to Binary Observables} 
Once the radial action is reconstructed from scattering data, and after analytic continuation to negative binding energies, $\cE<0$, the gravitational observables are obtained via differentiation. For instance, the periastron advance and periastron-to-periastron period are given by \cite{paper1,paper2}
\beq
\frac{\Delta\Phi(j,\cE)}{2\pi} = -  {\partial  \over \partial j} (i_r+j)  = \frac{1}{\pi} \sum_{n=1}  \frac{2\chi_j^{(2n)}(\cE)}{j^{2n}} =  \frac{1}{2\pi} \left(\chi(j,\cE)+\chi(-j,\cE)\right) \,,\label{eq:peria}
\eeq
and
\beq
\frac{T_p}{2\pi} = GM {\partial  \over \partial \cE} i_r(j,\cE) = GE {\partial  \over \partial \gamma} i_r(j,\gamma)\label{eq:tp}\,,
\eeq
respectively.
From here we can obtain the azimuthal and radial frequencies, defined through
\beq
\Omega_r (j,\cE)\equiv \frac{2\pi}{T_p}\,, \quad
\Omega_p (j,\cE) \equiv \frac{\Delta\Phi}{T_p}\,,
\eeq
\beq
\Omega_{\phi} \equiv \Omega_r +\Omega_p= \frac{2\pi}{T_p} \left(1+\frac{\Delta\Phi}{2\pi}\right)\,.
\eeq
We can also compute the redshift function, $\langle z_a\rangle$, using the first-law \cite{letiec}, yielding \cite{paper2}
\beq
\delta \cS_r(J,\cE,m_a) = - \left(1+\frac{\Delta\Phi}{2\pi}\right)\delta J 
+ \frac{\mu}{\Omega_r}\delta\cE - \sum_a \frac{1}{\Omega_r}\left(\langle z_a\rangle-{\partial E(\cE,m_a) \over \partial m_a}\right)\delta m_a\,.
\eeq

\subsection{Hamiltonian}\label{sec:H}

The construction of the B2B radial action allows us to bypass the need of a Hamiltonian, or the very equations of motion, to derive all the gravitational observables for the binary system in the conservative sector. It is possible, however, to reconstruct a Hamiltonian from which these observables may also be computed. The procedure was described in \cite{paper1}, and relies on Firsov's  solution to the scattering problem \cite{firsov}. It starts by solving for the $f_n$ coefficients in the PM expansion of the (square of the) momentum in the center-of-mass frame,
\beq
\bp^2 = p_\infty^2 \left(1 + \sum_{n=1} f_n \left(\frac{GM}{r}\right)^n\right)\,,\label{pinfpm}
\eeq
as a function of the scattering angle. The connection, as well as the inverse formula, was unfolded in \cite{paper1} to all orders in $G$. It reads
\beq
\label{eq:fi}
f_n = \sum_{\sigma\in\mathcal{P}(n)}g_\sigma^{(n)} \prod_{\ell} \left(\widehat{\chi}_b^{(\sigma_{\ell})}\right)^{\sigma^{\ell}}\,,
\eeq
where
\beq
\widehat{\chi}_b^{(n)} \equiv \frac{2}{\sqrt{\pi}}\frac{\Gamma(\frac{n}{2})}{\Gamma(\frac{n+1}{2})}\chi^{(n)}_b\,,
\eeq
and $\mathcal{P}(n)$ is the set of all integer partitions of $n$. Each partition is described by $n = \sigma_{\ell} \sigma^{\ell}$ (implicit summation) with mutually different $\sigma_{\ell}$'s.
Introducing the notation $\Sigma^\ell \equiv \sum_{\ell} \sigma^{\ell}$, the coefficients are given by
\beq
g_\sigma^{(n)} = \frac{2(2-n)^{\Sigma^{\ell} - 1}}{\prod_{\ell} (2\sigma^{\ell})!!}\,.
\eeq
We can also invert the relation in \eqref{eq:fi}, solving for the scattering angle as a function of the  momentum, yielding
  \beq
\chi_b^{(n)} = \frac{\sqrt{\pi}}{2} \Gamma\left(\frac{n+1}{2}\right)\sum_{\sigma\in\mathcal{P}(n)}\frac{1}{\Gamma\left(1+\frac{n}{2} -\Sigma^\ell\right)}\prod_{\ell} \frac{f_{\sigma_{\ell}}^{\sigma^{\ell}}}{\sigma^{\ell}!}\label{eq:fi2}\,.
\eeq
See \cite{paper1} for more details.\vskip 4pt Once the $f_n$'s are known, we can reconstruct a Hamiltonian as follows. For convenience, we write the expressions in terms of $P_n \equiv p_\infty^2 M^n f_n$, which is more suitable to define the gravitational potential. We begin with the equation for $E$ as a function of $p_\infty^2$,
\beq
  E = \sqrt{p_\infty^2+m_1^2}+\sqrt{p_\infty^2+m_2^2}\,,\label{pinf1}
\eeq 
which we then use to construct a Hamiltonian in `isotropic' gauge, defined as
\beq
E = H (r,\bp^2)= \sum_{i=0}^{\infty}  \frac{c_i(\bp^2)}{i!} \left(\frac{G}{r}\right)^i\,,
\eeq
in the PM expansion. (Notice we use a slightly different normalization than~\cite{cheung,zvi1,zvi2} for the $c_i$ coefficients.) Using \eqref{pinfpm} we have the condition
\beq
\sqrt{\bp^2-\sum_{i=1}^{\infty} P_i(E) \left(\frac{ G}{r}\right)^i+m_1^2}+\sqrt{\bp^2-\sum_{i=1}^\infty P_i (E) \left(\frac{ G}{r}\right)^i+m_2^2}
= \sum_{i=0}^{\infty}  \frac{c_i(\bp^2)}{i!} \left(\frac{G}{r}\right)^i\,, \label{eq:E1}
\eeq 
that can be solved iteratively for the $c_i$'s in terms of recursive relations. At zeroth order we have:
\beq
c_0(\bp^2) = E(\bp^2) = E_1(\bp^2)+E_2(\bp^2) \equiv \sqrt{\bp^2+m_1^2} + \sqrt{\bp^2+m_2^2}\,,\label{c0}
\eeq
and up to 2PM order,
\beq
\begin{aligned}
c_1 (\bp^2) &= -\frac{1}{2E\xi} P_1(E)= -\frac{\nu^2 M^2 }{\xi \Gamma^2} \, \frac{(\gamma^2-1)\chi_b^{(1)}(E)}{\Gamma}\\
c_2(\bp^2)  & = -\frac{1}{E\xi}P_2(E)+\frac{(3\xi-1)P^2_1(E)}{4 E^3 \xi^3}+\frac{P_1(E)P'_1(E)}{2E^2\xi^2}
\\ 
&=  -\frac{\nu^2M^3}{\xi \Gamma^2}\left(\frac{4(\gamma^2-1)}{\pi}  \, \frac{\chi_b^{(2)}(E)}{\Gamma} -\frac{(3\xi-1)\nu^2}{\xi^2\Gamma^3}\left( (\gamma^2-1) \frac{\chi_b^{(1)}(E)}{\Gamma}\right)^2\right. \\
&- \left. \frac{2\nu^2 (\gamma^2-1)}{\Gamma \xi}\frac{\chi_b^{(1)}(E)}{\Gamma} \frac{d}{dE}\left(\frac{E(\gamma^2-1)\chi_b^{(1)}(E)}{\Gamma^3}\right)\right)\label{cnb}
\end{aligned}
\eeq
where $\xi(\bp) \equiv E_1(\bp)E_2(\bp)/E^2(\bp)$. We have written the coefficients directly in terms of the scattering angle as a function of the energy to 2PM. At the end of the day, all the functions must be understood as defined in terms of the momentum through $E(\bp)$ in \eqref{c0}. General expressions to all orders can be found in \cite{paper1}.

\section{Post-Minkowskian Effective Theory} \label{sec:pmeft}

In this section we develop an EFT approach to obtain the impulse and scattering angle in the PM framework. We provide a set of simplified Feynman rules for the Einstein-Hilbert and worldline action and discuss the computation of the conservative contributions from potential modes as well as the required integration procedure. 

\subsection{Point-Particle Sources}\label{sec:action}

Following the EFT approach developed in \cite{nrgr}, we construct a worldline action to describe both constituents of the two-body problem,
\beq
S_{\rm pp} = - \sum_a m_a \int \dd\sigma_a \sqrt{g_{\mu\nu}(x^\alpha_{a}(\sigma)) v_{a}^\mu(\sigma_a) v_{a}^\nu (\sigma_a)} + \cdots\,,\label{1}
\eeq
where $v^\mu = \frac{dx^\mu}{d\sigma}$, and the ellipses account for finite-size effects (as well as counter-terms). For instance, using the proper time, $\tau$, we have
\beq
\cdots = \int \dd\tau_a\, \left(c^{(a)}_R R(x_a) + c^{(a)}_V R_{\mu\nu}(x_a)v_a^\mu v_a^\nu + c^{(a)}_{E^2} E_{\mu\nu}(x_a) E^{\mu\nu}(x_a)+ c^{(a)}_{B^2} B_{\mu\nu}(x_a) B^{\mu\nu}(x_a) \ldots \right)\,,\label{2}
\eeq
and additional operators can be added to include spin effects \cite{review}. The $c_{R,V}$ coefficients do not contribute to physical quantities, since they can be removed by field redefinitions \cite{nrgr}, however they may be needed to properly renormalize the theory removing intermediate UV poles \cite{nrgr4pn2}. The $c_{E^2}$ and $c_{B^2}$ operators, written in terms of the electric and magnetic components of the Weyl tensor, represent the tidal Love number (which vanish for non-rotating black holes \cite{tune}). We will study later on the leading contribution from these tidal operators to the scattering angle in \S\ref{sec:finite}. Other terms can be systematically included in a derivative expansion \cite{review}. Notice that the action is reparameterization invariant.
This allows us to fix the gauge, for instance with the coordinate time $x^0 = \sigma $, which is useful in the PN expansion.\vskip 4pt

In the PM scheme, instead of \eqref{1} it is convenient to work with the Polyakov action,\footnote{The Polyakov action may also be useful to study the high-energy limit of the self-force problem \cite{withchad}, as well as to simplify PN computation \cite{kuntz} using the coordinate time.}
\beq
S_{\rm pp}=-\sum_{a=1,2} \frac{m_a}{2} \int \dd\sigma_a \, e_a \left(\frac{1}{e_a^2} g_{\mu\nu}(x^\alpha_{a}(\sigma)) v_{a}^\mu(\sigma_a) v_{a}^\nu (\sigma_a) +  1\right)\,,
\eeq
such that variations with respect to $e_a$ give $e_a=\sqrt{g_{\mu\nu} v^\mu_a v^\nu_a}$, thus recovering \eqref{1}.
The advantage of this action is that, in contrast to PN computations, the PM approximation is suited for using the proper, rather than coordinate, time to parameterize the trajectories. Therefore we will choose the gauge in which  $e_a=1$, yielding $\sigma_a =  \tau_a$. The worldline action becomes simply \beq
S_{\rm pp} = -\sum_a \frac{m_a}{2} \int \dd\tau_a\,  g_{\mu\nu}(x_{a}(\tau_a)) v_{a}^\mu(\tau_a) v_{a}^\nu (\tau_a)\,.\label{act}
\eeq
with the condition
\beq
\label{v2e1}
e^2_a = g_{\mu\nu}(x_a(\tau_a)) v_a^\mu(\tau_a) v_a^\nu(\tau_a) = 1\,.
\eeq
From \eqref{act} the geodesic equation may be written in compact form as
\beq
\frac{d}{d\tau_a} \big(\, g_{\alpha\mu}(x_a(\tau_a)) v_a^\mu (\tau_a)\big)= \frac{1}{2} {\partial   g_{\mu\nu}\over \partial x^\alpha} (x_a(\tau_a)) v^\mu_a(\tau_a) v_a^\nu (\tau_a)\,, 
\eeq 
although in the EFT formalism we will work directly with the effective Lagrangian instead. Notice that, expanding the metric in the weak field approximation,
\beq
g_{\mu\nu} = \eta_{\mu\nu} + \frac{h_{\mu\nu}}{\Mp}\,,
\eeq
the action in \eqref{act} only generates a one-point function. In turn, all the non-linear effects are therefore encoded in the bulk action. We will work in Einstein's gravity, described by the Einstein-Hilbert action
\beq
S_{\rm EH} = -2\Mp^2 \int \dd^4x \sqrt{-g} \, R[g]\,.
\eeq
It is advantageous to optimize the Feynman rules given by this action to obtain the fewest number of terms for each bulk vertex.
For this purpose we use a generalized gauge-fixing condition, as well as judiciously chosen total derivatives. In principle, one can also use field redefinitions \cite{safi}. However, field redefinitions modify the worldline vertices introducing more diagrams. In this paper we choose to keep the worldline coupling unmodified, as in \eqref{act}. \vskip 4pt

At quadratic order we retain the De-Donder gauge, and the two-point Lagrangian
\begin{equation}
\cL_{hh} = \frac{1}{2}\partial_\alpha h^{\mu\nu}\partial^\alpha h_{\mu\nu}- \frac{1}{4}\partial_\mu h \partial^\mu h\,,
\end{equation}
with $h \equiv h^\alpha_\alpha$. From here we find the propagator
\beq
\langle h_{\mu\nu}(x) h_{\alpha\beta}(y) \rangle = \frac{i}{k^2} P_{\mu\nu\alpha\beta} e^{i k \cdot (x-y)}\,,
\eeq
where (in $D$ dimensions)
\beq
P_{\mu\nu\alpha\beta} = \frac{1}{2} \left(\eta_{\mu\alpha}\eta_{\nu\beta} + \eta_{\nu\alpha}\eta_{\mu\beta} - \frac{2}{D-2} \eta_{\mu\nu}\eta_{\alpha\beta}\right)\,. 
\eeq
For the three-point function, the shortest expression that does not modify the worldline coupling contains six terms,
\begin{equation}
  \begin{aligned}
    \Mp \cL_{hhh} =
    &-\frac{1}{2}h^{\mu\nu}\partial_\mu h^{\rho\sigma}\partial_\nu h_{\rho\sigma}
    +\frac{1}{2}h^{\mu\nu}\partial_\rho h \partial^\rho h_{\mu\nu}
    -\frac{1}{8}h \partial_\rho h\partial^\rho h\\
    &+h^{\mu\nu}\partial_\nu h_{\rho\sigma}\partial^\sigma {h_\mu}^\rho
    -h^{\mu\nu}\partial_\sigma h_{\nu\rho}\partial^\sigma {h_\mu}^\rho
    +\frac{1}{4}h \partial_\sigma h_{\nu\rho}\partial^\sigma h^{\nu\rho}\,.
  \end{aligned}
\end{equation}
There exist other inequivalent three-point Lagrangians of the same size.
The given representation was chosen such that {\it i)} it does not contain any double derivatives acting on the same field and {\it ii)} the four-point Lagrangian has the least amount of terms which is found to be
\begin{equation}
  \begin{aligned}
    \Mp^2 \cL_{hhhh} =
    &-\frac{1}{4} h^{\mu \nu } h^{\rho \sigma } \partial _{\alpha }h_{\rho \sigma } \partial ^{\alpha }h_{\mu \nu }+\frac{1}{2} h^{\mu \nu } h^{\rho\sigma } \partial _{\alpha }h_{\nu \sigma } \partial ^{\alpha }h_{\mu \rho }+h_{\mu }{}^{\rho } h^{\mu \nu } \partial _{\alpha }h_{\rho \sigma }\partial ^{\alpha }h_{\nu }{}^{\sigma }\\
    &-\frac{1}{2} h h^{\nu \rho } \partial _{\alpha }h_{\rho \sigma } \partial ^{\alpha }h_{\nu}{}^{\sigma }-\frac{1}{8} h_{\mu \nu } h^{\mu \nu } \partial _{\alpha }h_{\rho \sigma } \partial ^{\alpha }h^{\rho \sigma }+\frac{1}{16} h^2 \partial _{\alpha }h_{\rho \sigma } \partial ^{\alpha }h^{\rho \sigma }\\
    &-\frac{3}{4} h^{\mu \nu } h^{\rho \sigma } \partial _{\nu
    }h_{\sigma \alpha } \partial _{\rho }h_{\mu }{}^{\alpha }-h_{\mu }{}^{\rho } h^{\mu \nu } \partial ^{\alpha }h_{\nu }{}^{\sigma } \partial _{\rho}h_{\sigma \alpha }+\frac{1}{2} h h^{\nu \rho } \partial ^{\alpha }h_{\nu }{}^{\sigma } \partial _{\rho }h_{\sigma \alpha }\\
    &+\frac{1}{2}h_{\mu }{}^{\rho } h^{\mu \nu } \partial _{\nu }h^{\sigma \alpha } \partial _{\rho }h_{\sigma \alpha }-\frac{1}{4} h^{\mu }{}_{\mu } h^{\nu \rho} \partial _{\nu }h^{\sigma \alpha } \partial _{\rho }h_{\sigma \alpha }-\frac{1}{2} h^{\mu \nu } h^{\rho \sigma } \partial ^{\alpha }h_{\mu \rho} \partial _{\sigma }h_{\nu \alpha }\\
    &+h^{\mu \nu } h^{\rho \sigma } \partial _{\rho }h_{\mu }{}^{\alpha } \partial _{\sigma }h_{\nu \alpha }+\frac{1}{4}h_{\mu }{}^{\rho } h^{\mu \nu } \partial ^{\alpha }h_{\nu }{}^{\sigma } \partial _{\sigma }h_{\rho \alpha }-\frac{1}{2} h_{\mu }{}^{\rho } h^{\mu\nu } \partial _{\sigma }h \partial ^{\sigma }h_{\nu \rho }\\
    &+\frac{1}{4} h h^{\nu \rho } \partial _{\sigma }h \partial ^{\sigma }h_{\nu \rho }+\frac{1}{8} h_{\mu \nu } h^{\mu \nu } \partial _{\alpha }h_{\sigma }{}^{\alpha } \partial ^{\sigma}h^{\rho }{}_{\rho }-\frac{1}{16} h^2 \partial _{\alpha }h_{\sigma }{}^{\alpha } \partial ^{\sigma }h\,.
  \end{aligned}
\end{equation}
Also for the four-point function there exist other versions of the same length.\vskip 4pt

The three-point vertex is all we need to 2PM (while the four-point vertex enters at 3PM), and therefore we find these rules more convenient than adding more topologies from non-linear couplings to the sources.
However, at higher orders further simplifications can be achieved by allowing for field redefinitions.
For instance, keeping the same propagator, the three- and four-point functions can be simplified to 3 and 12 terms respectively, at the cost of worldline non-linearities. (The cubic coupling with only two terms is also possible, resulting in three terms for the quadratic action and a modified propagator.) Other type of simplifications are in principle possible, for instance by recasting Einstein's gravity in terms of only cubic couplings \cite{Cheung:2017kzx}, which can reduce the number of diagrams at higher PM orders.

\subsection{Conservative Effective Action} \label{sec:data}

Given the bulk and worldline action, we can `integrate-out' the graviton field, namely the metric, to construct an effective two-body action.
As in the original EFT approach \cite{nrgr,review}, we compute
\beq
e^{i S_{\rm eff}[x_a] } = \int \cD h_{\mu\nu} \, e^{i S_{\rm EH}[h] + i S_{\rm GF}[h] + i S_{\rm pp}[x_a,h]}\,,\label{eff1}
\eeq
where $S_{\rm GF}$ is the gauge fixing term we alluded before and we have omitted total derivatives we have used to simplify the Feynman rules. Field redefinitions of the metric field are also allowed, although we have not used them in the present paper. Despite the path integral representation, we use the saddle-point approximation keeping only the classical contributions to the effective action.
We thus treat the massive particles as non-propagating external sources and include only connected tree level diagrams, i.e. without graviton loops. In other words, the path integral and associated Feynman rules are a convenient and systematic way to encapsulate solutions to Einstein's equations sourced by localized sources, which are then plugged back into the action. This procedure is sometimes referred in the literature as the construction of the `Fokker-action', e.g. \cite{blanchet}.\vskip 4pt As in NRGR, the perturbative expansion in terms of iterated Green's functions will resemble loop-type integrals in field theory.  Yet, no quantum effects are ever computed, although they can be easily incorporated \cite{nrgr}. The advantage of the EFT approach, in comparison with directly tackling Einstein's equations, is that we do not need to solve for the metric explicitly, which is nonetheless unobservable in the conservative sector. Instead, we compute what is often described as the (connected) \emph{vacuum-to-vacuum amplitude} in the presence of sources  (`$-i\log \langle 0|0\rangle^J$') in the saddle-point approximation, simply steering away from closed quantum loops. In doing so, evaluating the effective action on solutions to the classical equations of motion, the power counting in Planck's constant becomes trivial and $\hbar$ turns into nothing but a conversion factor that drops out of the final answer.\vskip 4pt  

\begin{figure}
  \centering
  \begin{subfigure}[b]{0.25\textwidth}
    \centering
    \begin{tikzpicture}
  \draw[boson] (-0.8,0) to[bend right=75] (1.2,0);
  \draw[fill] (-0.8,0) circle (0.05);
  \draw[fill] (1.2,0) circle (0.05);
\end{tikzpicture}
\caption{}
  \end{subfigure}
  \begin{subfigure}[b]{0.14\textwidth}
    \centering
    \begin{tikzpicture}
  \draw[boson] (-0.8,0) to[bend right=75] (1.2,0);
  \draw[boson] (0.2,0) -- (0.2,-0.57);
  \draw[fill] (-0.8,0) circle (0.05);
  \draw[fill] (0.2,0) circle (0.05);
  \draw[fill] (1.2,0) circle (0.05);\end{tikzpicture}
\caption{}
  \end{subfigure}
   \caption{Feynman diagrams which only yield singular integrals in the potential region. The divergences can be set to zero in dimensional regularization, or absorbed into counter-terms.}
  \label{eftfig2}
\end{figure}
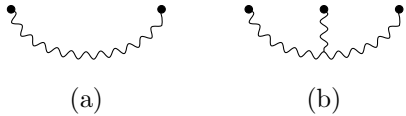

\begin{figure}
  \centering
  \begin{subfigure}[b]{0.08\textwidth}
    \centering
    \begin{tikzpicture}
      [scale=1]
      \draw[boson] (0,0) -- (0,-1.5);
      \filldraw (0,0) circle (0.05);
      \filldraw (0,-1.5) circle (0.05);
    \end{tikzpicture}
    \caption{}
  \end{subfigure}
  \begin{subfigure}[b]{0.1\textwidth}
    \centering
    \begin{tikzpicture}
      [scale=1]
      \draw[boson] (-0.5,0) -- (0,-0.9);
      \draw[boson] (0.5,0) -- (0,-0.9);
      \draw[boson] (0,-0.9) -- (0,-1.5);
      \filldraw (-0.5,0) circle (0.05);
      \filldraw (0.5,0) circle (0.05);
      \filldraw (0,-1.5) circle (0.05);
    \end{tikzpicture}
    \caption{}
  \end{subfigure}
  \begin{subfigure}[b]{0.12\textwidth}
    \centering
    \begin{tikzpicture}
      [scale=1]
      \draw[boson] (-0.7,0) -- (-0.35,-0.5);
      \draw[boson] (0,0) -- (-0.35,-0.5);
      \draw[boson] (0.7,0) -- (0,-0.9);
      \draw[boson] (-0.35,-0.5) -- (0,-0.9);
      \draw[boson] (0,-0.9) -- (0,-1.5);
      \filldraw (-0.7,0) circle (0.05);
      \filldraw (0,0) circle (0.05);
      \filldraw (0.7,0) circle (0.05);
      \filldraw (0,-1.5) circle (0.05);
    \end{tikzpicture}
    \caption{}
  \end{subfigure}  
  \begin{subfigure}[b]{0.12\textwidth}
    \centering
    \begin{tikzpicture}
      [scale=1]
      \draw[boson] (-0.7,0) -- (0,-0.9);
      \draw[boson] (0,0) -- (0,-0.9);
      \draw[boson] (0.7,0) -- (0,-0.9);
      \draw[boson] (0,-0.9) -- (0,-1.5);
      \filldraw (-0.7,0) circle (0.05);
      \filldraw (0,0) circle (0.05);
      \filldraw (0.7,0) circle (0.05);
      \filldraw (0,-1.5) circle (0.05);
    \end{tikzpicture}
    \caption{}
  \end{subfigure}
  \begin{subfigure}[b]{0.12\textwidth}
    \centering
    \begin{tikzpicture}
      [scale=1]
      \draw[boson] (-0.5,0) -- (0,-0.5);
      \draw[boson] (0.5,0) -- (0,-0.5);
      \draw[boson] (0,-1) -- (-0.5,-1.5);
      \draw[boson] (0,-0.5) -- (0,-1);
      \draw[boson] (0,-1) -- (0.5,-1.5);
      \filldraw (-0.5,0) circle (0.05);
      \filldraw (0.5,0) circle (0.05);
    \filldraw (-0.5,-1.5) circle (0.05);
      \filldraw (0.5,-1.5) circle (0.05);
          \end{tikzpicture}
    \caption{}
  \end{subfigure}
 \begin{subfigure}[b]{0.12\textwidth}
    \centering
    \begin{tikzpicture}
      [scale=1]
      \draw[boson] (-0.5,0) -- (0,-0.8);
      \draw[boson] (0.5,0) -- (0,-0.8);
      \draw[boson] (0,-0.8) -- (-0.5,-1.5);
      \draw[boson] (0,-0.8) -- (0.5,-1.5);
       \filldraw (-0.5,0) circle (0.05);
      \filldraw (0.5,0) circle (0.05);
         \filldraw (-0.5,-1.5) circle (0.05);
      \filldraw (0.5,-1.5) circle (0.05);
    \end{tikzpicture}
    \caption{}
  \end{subfigure}
  \begin{subfigure}[b]{0.12\textwidth}
    \centering
    \begin{tikzpicture}
      [scale=1]
      \draw[boson] (-0.5,0) -- (-0.5,-0.75);
      \draw[boson] (-0.5,-0.75) -- (-0.5,-1.5);
      \draw[boson] (0.5,0) -- (0.5,-0.75);
      \draw[boson] (0.5,-0.75) -- (0.5,-1.5);
      \draw[boson] (-0.5,-0.75) -- (0.5,-0.75);
      \filldraw (-0.5,0) circle (0.05);
      \filldraw (0.5,0) circle (0.05);
      \filldraw (-0.5,-1.5) circle (0.05);
      \filldraw (0.5,-1.5) circle (0.05);
    \end{tikzpicture}
    \caption{}
    \label{eftFig1H}
  \end{subfigure}
  \caption{Feynman topologies needed for the computation of the effective action to ${\cal O}(G^3)$. The wavy line represents the propagator (or Green's function), while the black dots are the two worldlines at particle's 1 (bottom) and 2 (top), treated as external sources.}
  \label{eftfig1}
\end{figure}
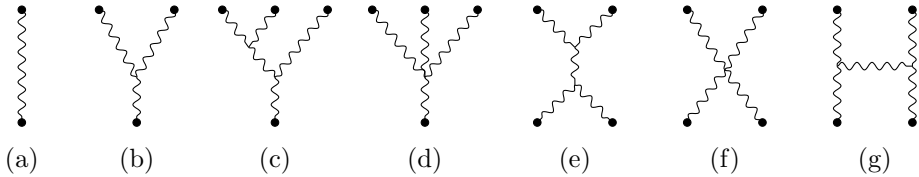

In principle, the integrals that appear in the perturbative expansion of the path integral in \eqref{eff1} include contributions from two regions: \emph{potential} and \emph{radiation} modes \cite{nrgr}.
The latter incorporate the propagating (on-shell) modes that travel to the gravitational wave detector, while the (off-shell) potential modes are responsible for the conservative forces that deflect the particles in a two-body encounter, and also bind the system together in a binary. In this paper we will only consider conservative effects. Since the imaginary part is due to \emph{cuts} induced by radiation modes \cite{nrgr, review}, by concentrating in the conservative sector the effective action will remain real at all stages.
As in any EFT without a cutoff in the momenta, UV divergent integrals appear. However, these are naturally handled by dimensional regularization, and renormalized by counter-terms in a diffeomorphism invariant fashion, see e.g. \cite{nrgr4pn2}. As it is also standard, we will set scaleless integrals to zero in the conservative sector, unless otherwise~noted.
This means we do not keep contributions in which propagators start and end on the same source (although in principle at a different times), as depicted in Fig.~\ref{eftfig2}, since 
these type of diagrams produce scaleless divergent contributions for potential modes.\footnote{As it is well known, hereditary radiative effects can also enter in the conservative sector \cite{tail,apparent}, which requires a careful study of IR/UV divergences and scaleless integrals \cite{nrgr4pn2}. That is partially due to the fact that radiation modes can generate non-zero contributions, e.g. $(x_1^i\dddot x^j_1)_{\rm TF}^2$ from the diagram Fig~
\ref{eftfig2}a, and a tail-type correction from Fig~\ref{eftfig2}b, in the long-wavelength expansion~\cite{andirad}. We will not consider these contributions here and return to this issue in future work.}  Moreover, unlike in the PN expansion, these diagrams cannot be connected to the second particle through worldline non-linearities using the Polyakov action.  In general, only bulk graviton self-interaction are required at higher orders in $G$. Hence, for example, the entire set of diagrams that contribute to the effective action to ${\cal O}(G^3)$ are shown in Fig.~\ref{eftfig1} (plus mirror images). The diagrams (a) and (b) are the only two required to ${\cal O}(G^2)$.
Furthermore, the one-point functions in diagrams (b), (c) and (d) are responsible for the Schwarzschild background to ${\cal O}((Gm_2)^3)$, such that only $(e), (f)$ and $(g)$ carry information beyond the test-particle limit.\vskip 4pt

As it turns out, many contributions to the effective action in the conservative sector are naturally reduced to two- and three-dimensional integrals.
That is the case, as we shall see, due to the appearance of $\delta$ functions in the PM scheme that force the component of the momentum to be orthogonal to the four velocities of the particles at infinity.
(Something similar occurs in the derivation of the classical impulse starting from the scattering amplitude in \cite{donal}.)
However, for another class of diagrams isolating the conservative contribution to the effective action entails a prescription to perform the various $k^0_1, \ldots,  k^0_n$ energy integrals that may appear at $n$PM order. We discuss the procedure below.

\subsection{Potential Region} \label{sec:pot}

The potential modes are natural objects in the PN framework. Contributions from the potential region in PN theory are obtained systematically by expanding the Feynman integrals in powers of $k_0/|\bk|$.
This is exploited in the EFT approach to reduce the complexity of the calculations, resorting to manifest power counting in the velocity expansion \cite{nrgr}. By splitting the four dimensional integrals into regions, the EFT framework has already achieved a high level of accuracy, at 4PN order \cite{tail,nrgr4pn1,nrgr4pn2} (and beyond \cite{5pn1,5pn2}). In this paper, however, we will not perform a small velocity expansion, and use the effective action to compute the scattering angle in relativistic two-body encounters instead. Therefore, our task here is to obtain the contribution from each diagram to the conservative sector in the PM expansion. Namely to incorporate the effects from potential modes at a given order in $G$, but to all orders in velocity. That is to say, we must use the relativistic version of the integrands and propagators, but only retain the contributions to the real part of the effective action which correspond to the conservative poles. In principle, once a contour in the $k^0$ complex plane is chosen, various poles may contribute to a given Feynman diagram. Therefore, in order to properly isolate the relevant (conservative) region of integration, we need a prescription to account for each one of them. We will adapt to our framework the procedure first introduced in \cite{cheung}, and later elaborated further in \cite{zvi1,zvi2,Parra} to extend to higher PM orders. \vskip 4pt

In the EFT example we focus on in this paper, at 2PM, only one energy integral will be needed. As it was discussed in \cite{cheung}, in this case the procedure boils down to evaluating the $k^0$ energy integral by an (oriented) average over poles in the upper/lower half complex plane, that is
\beq
\int \frac{\dd k^0}{2\pi} \left(\cdot\right) = \frac{i}{2}\left[ \sum_{k^0_\star \in H^+} \underset{\, k^0 = k^0_\star}{\rm Res} \left(\cdot\right)\,\,\, - \sum_{k^0_\star \in H^-} \underset{\, k^0 = k^0_\star}{\rm Res} \left(\cdot\right)   \right]\,,\label{poles}
\eeq
while retaining only conservative contributions. This means we only keep the poles in the (potential) region $k^0 \ll |\bk|$. For our purposes here, the prescription in \eqref{poles} will be sufficient. More generally, even though the poles in \cite{cheung,zvi1,zvi2} are different, e.g. no massive lines appear in the EFT computation, a given prescription can in principle be applied to any type of integral. Therefore, by adapting the rules from \cite{zvi2,Parra} to our case we can evaluate all the energy integrals in the EFT approach at any PM order. For instance, by averaging over graviton permutations \cite{sterman}. One of the basic tools is the identity \cite{saotome}
\beq
\delta \left(\sum_i^n \omega_i\right) \sum_{{\rm Perms \, of}\, \omega_i} \frac{1}{\omega_1-i\epsilon}\cdots \frac{1}{\omega_1+\cdots + \omega_{n-1} - i \epsilon}  = (2\pi i)^{n-1} \prod_i^n \delta(\omega_i) \,,\label{perm}
\eeq 
to compute the energy integrals. For example, at 3PM order we may find integrands involving two energies, e.g.
\beq
 \frac{1}{k_1^0 - i\epsilon} \frac{1}{k_1^0 + k_2^0 - i\epsilon} \,.
\eeq
The idea is to re-write it using permutations of $k^0_1,k_2^0,k_3^0$ and imposing the condition $k_3^0+k_2^0+k_1^0=0$, such that \beq
\int \frac{\dd k^0_1}{2\pi}  \frac{\dd k_2^0}{2\pi}   \frac{1}{k_1^0 - i\epsilon} \frac{1}{k_1^0 + k_2^0 - i\epsilon}  = \frac{1}{3!}  (2\pi i )^2 = -\frac{2\pi^2}{3}\,.
\eeq
Let us stress that, although the prescription to deal with energy integrals may turn out to be equivalent to the one in the amplitude program, the origin of the different classical contributions, as well as the associated integrals, will be quite different. For instance, we will not have to deal with super-classical IR divergences, as in \cite{cheung,zvi1,zvi2,donal,cristof1}. (We do have UV poles, which are readily absorbed into counter-terms as we mentioned earlier.) We find, nonetheless, similarities with the calculations in \cite{cheung,zvi1,zvi2,donal}.
Notably, the result for the scattering angle to 2PM can be identified with a (classical) triangle integral, with vanishing contribution from the crossed-box. Yet, unlike what occurs in \cite{cheung,donal,cristof1}, the box diagram does not feature at all in the EFT computation. This is expected, since it would entail a singularity for which we would have no subtraction scheme, e.g. no EFT matching \cite{cheung}, nor `cut diagram' \cite{donal}, nor Born iterations \cite{Holstein:2008sx,cristof1}, as in methods dealing with the scattering amplitude. However, as we shall see in our example, the organization of the various terms entering in the final answer turn out to be remarkably distinct, as it was already seen in \cite{donal}. While this supports the complete independence in methodologies, it also begs for a deeper understanding of the nature of the different classical contributions at higher PM orders.

\subsection{Impulse \& Deflection Angle}

Using the effective action we can then derive the equations of motion from which we can compute the total momentum change, the impulse, as well as the scattering angle.
Without loss of generality, let us study the deflection for particle~1.
In the PM expansion the effective action takes the form,\footnote{At first glance, the resulting effective Lagrangian will appear to be \emph{non-local} in time, involving an integral over~$d\tau_2$. However, as we shall see, the resulting impulse and scattering angle in the conservative sector will be obtained from manifestly local interactions (as long as tail effects are ignored \cite{tail}).} 
\beq
S_{\rm eff} = \sum_n \int \dd\tau_1\, \cL_n [x_1(\tau_1),x_2(\tau_2)]\,,
\eeq
The $\cL_n$'s are the ${\cal O}(G^n)$ contributions to the effective Lagrangian obtained from the sum of all Feynman diagrams. The leading order (kinetic) term is given by, see \eqref{act}, 
\beq
\cL_0 = - \frac{m_1}{2} \eta_{\mu\nu} v_1^\mu(\tau_1) v_1^\nu(\tau_1)\,.
\eeq
Performing the variation of the effective Lagrangian we obtain
\beq
-\eta^{\mu\nu}\frac{\dd}{\dd\tau_1} \left(\frac{\partial \cL_0}{\partial  v^\nu_1}\right) = m_1  \frac{\dd v_1^\mu}{\dd\tau_1}  = -\eta^{\mu\nu}\left(\sum_{n=1}^{\infty} \frac{\partial \cL_n}{\partial x^\nu_1(\tau_1)} - \frac{\dd}{\dd\tau_1} \left(\frac{\partial \cL_n}{\partial  v^\nu_1}\right)\right)\,.\label{eom}
\eeq
From here, and assuming $\cL_{n \geq 1} \to 0$ at infinity, we find\footnote{Notice that, due to our choice of conventions, the \emph{canonical} momentum is given by ${\partial \cL \over \partial v^\mu} = - m  v_\mu$.}
\beq
\Delta p^\mu_1= m_1\Delta v^\mu_1 = - \eta^{\mu\nu} \sum_n \int_{-\infty}^{+\infty} \dd\tau_1 {\partial \cL_n  \over\partial x^\nu_1} \,,\label{spacetime}
\eeq
for the total change of four-momentum; or in 3D space,\beq
\Delta \bp^i_1 = - \Delta \bp_{i1} = \sum_n \int_{-\infty}^{+\infty} \dd\tau_1 {\partial \cL_n  \over\partial \bx^i_1} \,.
\eeq
Similarly for particle 2, which can also be obtained from momentum conservation.\vskip 4pt

The computation of the impulse then follows iteratively, order by order in the PM expansion. At leading order the trajectory is represented by a straight line, and higher PM effects result in a series of corrections, which we parameterize in terms of the proper time as follows \cite{damour1,damour2,donal}
\beq
\label{pmexp}
\begin{aligned}
v^\mu_a(\tau_1) &= u^\mu_a  + \sum_n \delta^{(n)} v^\mu_a(\tau_a) \,,\\
x^\mu_a(\tau_1) &= b^\mu_a + u^\mu_a \tau_a + \sum_n \delta^{(n)} x^\mu_a (\tau_a)\,,
\end{aligned}
\eeq
where $u^\mu_a$ is the incoming velocity at infinity and $b^\mu_a = x^\mu_{a,(0)} - u_a^\mu(x_{a,(0)}\cdot u_a)$, with $x^\mu_{a,(0)}$ the initial position of each particle. For a scattering process we also have \beq g_{\mu\nu}(x_a(\tau_a)) \to \eta_{\mu\nu}\,, \quad \left(\delta^{(n)} x^\mu_a(\tau_a), \delta^{(n)} v^\mu_a(\tau_a) \right) \to 0\quad\quad (\tau_a \to -\infty)\,,\eeq 
such that the condition in \eqref{v2e1} implies $u_a^2=1$, with $u_1\cdot u_2$ then identified with $\gamma$ in \eqref{Egam}. Moreover, notice that $b_a\cdot u_a=0$, and the vector $b^\mu \equiv b^\mu_2-b^\mu_1$ will be thus associated with the impact parameter of the collision in the center-of-mass frame. In~order to comply with standard literature we will incur in some minor abuse of notation and sometimes use $b$ to represent $\sqrt{-b^\mu b_\mu}$ in the final results, as well as for the four-vector appearing in dot products, e.g. $k\cdot b$, during the intermediate computations.\vskip 4pt 

Due to the PM expansion in \eqref{pmexp}, there are various contributions to the impulse in \eqref{spacetime}, involving lower order terms evaluated on the deflected trajectories and expanded to the desired order, i.e.
\beq
\Delta^{(n)} p^\mu_a = \sum_{k \leq n} \Delta^{(n)}_{\cL_{k}} p^\mu_a,\, \label{dp}
\eeq
with
\beq
\Delta^{(n)}_{\cL_k}\, p^\mu_a \equiv  -\eta^{\mu\nu}  \int_{-\infty}^{+\infty}  \dd\tau_a \left({\partial   \over\partial x^\nu_a} \cL_{k} \big[b_a + u_a \tau_a + \sum_{r=0}^{n-k} \delta^{(r)} x_a\big]\right)_{{\cal O}(G^n)}\,.
\eeq
In summary, in order to compute the impulse we first derive the effective Lagrangian to $n$-th order via Feynman diagrams. We then extract the PM corrections to the equations of motion using the standard Euler-Lagrangian procedure.
Hence, we insert the resulting trajectory into the time integral of the derivative of the effective Lagrangian with respect to the position, while keeping contributions up to the desired order. Given that for the impulse at $n$PM the deflection is needed to $(n-1)$PM, we can proceed iteratively in powers of $G$, starting from the undeflected solution at ${\cal O}(G^0)$ in \eqref{pmexp}. Once the impulse is known we can then go to the center-of-mass frame and directly read off the scattering angle using, see e.g. \cite{damour1},
\beq
2\sin\left(\frac{\chi}{2}\right)  = \chi - \frac{1}{24} \chi^3 + {\cal O}(\chi^5) = \frac{|\Delta \bp_{1{\rm cm}} |}{p_\infty}= \frac{\sqrt{-\Delta p_1^2}}{p_\infty}\,,\label{eq:apm}
\eeq
where $p_\infty$ is given in \eqref{pinf}. In the last equation we used that $\Delta p^0_1=0$ to re-write the expression in a covariant fashion using our conventions. 

\section{Conservative Binary Dynamics to 2PM}\label{sec:obs} 

In this section we compute all the dynamical invariants for bound orbits to 2PM order using the B2B map.
The first ingredient is the derivation of the scattering angle using the EFT approach. We illustrate all the steps, starting from the computation of the effective action. We compute the (real part of the) effective Lagrangian following the same procedure as in NRGR, by `integrating out' the metric degrees of freedom, but keeping the propagators fully relativistic. Therefore, unlike NRGR, we will not expand the propagators in powers of $k_0/|\bk|$. We will handle the potential region as we described earlier in \S\ref{sec:pot}. To perform the integrals we will often go to the frame where one of the incoming particles is initially at rest, and align the $x$-axis with the initial velocity of the companion.
We will then use $\bk_\perp$ for vectors in the orthogonal $zy$-plane.

\subsection{Effective Lagrangian}

\subsubsection{Tree}
The leading order contribution to the effective action comes from the `tree' diagram in Fig.~\ref{eftfig1}(a). The computation is straightforward, and we obtain
\bea
 \cL_1 &=& -i \frac{-i}{2} \frac{-i}{2} \int_{-\infty}^{+\infty}  \dd\tau_2 \int_k \frac{i P_{\alpha\beta\mu\nu} }{k^2} v_1^\alpha (\tau_1) v_1^\beta (\tau_1) v_2^\mu (\tau_2)v_2^\nu(\tau_2) e^{i k \cdot \left(x_1(\tau_1)-x_2(\tau_2)\right)}\label{act1}\\
&=&   - \frac{ m_1 m_2}{8\Mp^2} \int_{-\infty}^{+\infty}  \dd\tau_2 \left( 2(v_1(\tau_1)\cdot v_2(\tau_2))^2-v_1^2(\tau_1) v_2^2(\tau_2)\right) \int_k \frac{1}{k^2} e^{i k \cdot \left(x_1(\tau_1)-x_2(\tau_2)\right)}\nn\,.
\eea
We can then read off the contribution to the equations of motion from the tree level action:
\bea
\frac{d v^\mu_1}{\dd\tau_1} &=&\frac{ m_2}{8\Mp^2} \int_{-\infty}^{\infty} \dd\tau_2 \left( 2(v_1(\tau_1)\cdot v_2(\tau_2))^2-v_1^2(\tau_1)v_2^2(\tau_2)\right) \int_k \frac{i k^\mu}{k^2} e^{i k \cdot \left(x_1(\tau_1)-x_2(\tau_2)\right)} \\
&-& \frac{m_2}{4\Mp^2} \int_{-\infty}^{\infty} \dd\tau_2 \, \left(2v_1(\tau_1)\cdot v_2(\tau_2) v_2^\mu(\tau_2)- v_2^2(\tau_2) v_1^\mu(\tau_1)\right) \int_k \frac{i k\cdot v_1(\tau_1)}{k^2} e^{i k \cdot \left(x_1(\tau_1)-x_2(\tau_2)\right)}\nn \\
&-& \frac{ m_2}{4\Mp^2} \int_{-\infty}^{\infty} \dd\tau_2 \, \left((2\frac{\dd v_1(\tau_1)}{\dd\tau_1}\cdot v_2(\tau_2) v_2^\mu(\tau_2)-v_2^2(\tau_2) \frac{\dd v_1^\mu(\tau_1)}{\dd\tau_1}\right) \int_k \frac{1}{k^2} e^{i k \cdot \left(x_1(\tau_1)-x_2(\tau_2)\right)}\nn\,,\label{eqTree}
\eea
which will be useful later on to compute the deflection angle to second order in $G$. Notice the third line includes a term proportional to the acceleration, which can be treated systematically by replacing it with lower order equations of motion. 
\subsubsection{One Loop}

The effective action at ${\cal O}(G^2)$ (or `one loop') has only one contribution, shown in the diagram in Fig.~\ref{eftfig1}(b). Using the cubic vertex described earlier we find
\beq
\begin{aligned}
\cL_2 =& -\frac{m_1m_2^2}{16\Mp^3}  v_1^\alpha(\tau_1) v_1^\beta (\tau_1) \!\int\! \dd\tau_2\!\int\! \dd\tilde\tau_2 \, v_2^\gamma(\tau_2) v_2^\rho(\tau_2) v_2^\sigma (\tilde\tau_2) v_2^\kappa (\tilde \tau_2)  P_{\gamma\rho\tilde\gamma\tilde\rho}(k_1)P_{\sigma\kappa\tilde\sigma\tilde\kappa}(k_2) P_{\alpha\beta\tilde\alpha\tilde\beta}(k_3)  \\
&\times
\int_{k_{1,2,3}}  e^{i k_1\cdot x_1(\tau_1)}e^{i k_2\cdot x_2(\tau_2)}e^{i k_3\cdot x_2(\tilde\tau_2)}\frac{V^{\tilde\gamma\tilde\rho\tilde\sigma\tilde\kappa\tilde\alpha\tilde\beta}_{hhh} (k_1,k_2,k_3) }{k_1^2k_2^2k_3^2}\delta^4(k_1+k_2+k_3) + (1\leftrightarrow 2)\,,\label{cl2}
\end{aligned}
\eeq
where 
\begin{equation}
  \begin{aligned}
    i &V^{abcdef}_{hhh}(p_1,p_2,p_3) = \frac{i}{4 \Mp}\times
      \Big[4p_1\cdot p_2\left(\eta^{af}\eta^{bd}\eta^{ce}+\eta^{ae}\eta^{bd}\eta^{cf}\right)\\
      &+4p_2\cdot p_3\left(\eta^{ae}\eta^{bc}\eta^{df}+\eta^{ac}\eta^{be}\eta^{df}\right)
      +4p_1\cdot p_3\left(\eta^{ad}\eta^{bf}\eta^{ce}+\eta^{ac}\eta^{bf}\eta^{de}\right)\\
      &+(p_1\cdot p_2+p_2\cdot p_3+p_1\cdot p_3)\left(\eta^{ab}\eta^{cd}\eta^{ef}-2\eta^{ae}\eta^{bf}\eta^{cd}-2\eta^{ab}\eta^{ce}\eta^{df}-2\eta^{ac}\eta^{bd}\eta^{ef}\right)\\
      &-4\eta^{ad}\eta^{ce}p_1^f p_2^b-4\eta^{ae}\eta^{bc}p_1^d p_2^f+2\eta^{ac}\eta^{bd}\left(p_1^e p_2^f+p_1^f p_2^e\right)\\
      &-4\eta^{ac}\eta^{de}p_2^f p_3^b-4\eta^{ae}\eta^{cf}p_2^b p_3^d+2\eta^{ce}\eta^{df}\left(p_2^a p_3^b+ p_2^b p_3^a\right)\\
      &-4\eta^{af}\eta^{ce}p_1^d p_3^b-4\eta^{ac}\eta^{be}p_1^f p_3^d+2\eta^{ae}\eta^{bf}\left(p_1^c p_3^d+p_1^d p_3^c\right)
    \Big]
  \end{aligned}
\end{equation}
Notice that, while a few terms are present at first, the orthogonality condition will drastically reduce the number of contributions once the diagram is evaluated on the undeflected solution.  This turns out to be one of the major advantages of working with the scattering angle rather than computing the binding potential directly in a PM expansion. 
\subsubsection{Trajectories} 

To obtain the trajectories we must integrate the equations of motion using the expansion in \eqref{pmexp}. Since we restrict ourselves here to the impulse to 2PM, we only need the first correction to the unperturbed solution. This follows from \eqref{eqTree}, which yields
\beq
\begin{aligned}
\delta^{(1)} v_1^\mu(\tau_1) &= \frac{m_2}{4\Mp^2} \left( \frac{2\gamma^2-1}{2} \eta^{\mu\alpha} - (2\gamma u_2^\mu -u_1^\mu)u_1^\alpha \right) \int_{-\infty}^{\tau_1}\dd\tilde\tau_1\int_q \hat\delta(q\cdot u_2) \frac{i q_\alpha}{q^2}  e^{i (q\cdot u_1 -i\epsilon) \tilde \tau_1}e^{i q\cdot b}\\
&=   \frac{m_2}{4\Mp^2} \left( \frac{2\gamma^2-1}{2} \eta^{\mu\alpha} - (2\gamma u_2^\mu - u_1^\mu) u_1^\alpha \right) \int_q \hat\delta(q\cdot u_2)\frac{iq_\alpha}{q^2}e^{i q\cdot b}\frac{(-i)e^{i (q\cdot u_1-i\epsilon) \tau_1}}{(q\cdot u_1-i\epsilon)}\label{d1v}
\end{aligned}
\eeq
so that, after an additional time integration,
\beq
\delta^{(1)} x_1^\mu(\tau_1) =  - \frac{m_2}{4\Mp^2} \left(\frac{2\gamma^2-1}{2} \eta^{\mu\alpha} - (2\gamma u_2^\mu - u_1^\mu) u_1^\alpha \right) \int_q \frac{i q_\alpha \hat\delta(q\cdot u_2)}{q^2(q\cdot u_1-i\epsilon)^2}e^{i q\cdot b}e^{i (q\cdot u_1-i\epsilon) \tau_1}\,.\label{d1x}
\eeq
We have added a factor of $-i\epsilon$ in all these expressions to ensure the convergence of the time integral at $\tilde\tau \to - \infty$, see e.g. \cite{donal}. Notice the resulting factor of $(q\cdot u -i\epsilon)^{-1}$ resembles the propagator in the heavy-quark effective theory (HQET) \cite{ben}. This will play an important role later on when we compute the impulse in the potential region, and is also directly associated to the large-mass limit of the scattering amplitude in \cite{cheung,cristof1}.\vskip 4pt

The deflection for the second particle can be obtained following the same procedure,
\beq
\begin{aligned}
\delta^{(1)} v_2^\mu (\tau_2)&= \frac{m_1}{4\Mp^2} \left( \frac{2\gamma^2-1}{2} \eta^{\mu\alpha} - (2\gamma u_1^\mu -u_2^\mu)u_2^\alpha \right) \int_{-\infty}^{\tau_2}\!\dd\tilde\tau_2\int_q \!\hat\delta(q\cdot u_1) \frac{-i q_\alpha}{q^2}  e^{-i (q\cdot u_2 + i\epsilon) \tilde \tau_2}e^{i q\cdot b}\\
&= \frac{m_1}{4\Mp^2} \left( \frac{2\gamma^2-1}{2} \eta^{\mu\alpha} - (2\gamma u_1^\mu - u_2^\mu) u_2^\alpha \right)\int_q \hat\delta(q\cdot u_1)\frac{-iq_\alpha}{q^2}e^{i q\cdot b}\frac{(+i)e^{-i (q\cdot u_2-i\epsilon) \tau_2}}{(q\cdot u_2+i\epsilon)}\label{d1v2}
\end{aligned}
\eeq
and
\beq
\delta^{(1)} x_2^\mu (\tau_2)=  \frac{m_1}{4\Mp^2} \left( \frac{(2\gamma^2-1)}{2} \eta^{\mu\alpha} - (2\gamma u_1^\mu - u_2^\mu) u_2^\alpha \right) \int_q \frac{i q_\alpha \hat\delta(q\cdot u_1)}{q^2(q\cdot u_2+i\epsilon)^2}e^{i q\cdot b}e^{-i (q\cdot u_2+i\epsilon) \tau_2}\,.\label{d1x2}
\eeq
As expected, the PM shifts for particle 2 can be simply obtained from relabeling $1 \leftrightarrow 2$ in the corrections to particle 1, together with $q \to -q$. Notice, however, that because of our choice of signature, the factor of $(q\cdot u_1-i\epsilon)$ in \eqref{d1v} and \eqref{d1x} needed for the integral to converge turned into $(q\cdot u_2+i\epsilon)$. As we shall see, this distinction in the position of the poles will become relevant when combining all contributions to the impulse. 

\subsection{Scattering Angle}

We are now in position to compute the scattering angle for the two-body gravitational encounter. For the computation of the impulse we will concentrate on particle 1. A similar computation can be followed for particle 2. 

\subsubsection{Leading Order Impulse}
The leading order impulse follows from \eqref{spacetime} applied to the 1PM effective action in \eqref{act1} and evaluated in the undeflected trajectory in \eqref{pmexp},  
\beq
\Delta^{(1)}_{\cL_1}\, p^\mu_1 =\frac{m_1 m_2}{8\Mp^2}  \left( 2\gamma^2-1\right) \int_k  i k^\mu\frac{\hat\delta(k\cdot u_1) \hat\delta(k\cdot u_2)}{k^2} e^{i k \cdot b}\,.
\eeq
To perform the integral we choose the frame at which particle 1 is initially at rest, $u_1=(1,0,0,0)$. In such coordinates,  the companion has four-velocity $u_2=(\gamma,\beta \gamma,0,0)$, with $\beta \equiv \sqrt{\gamma^2-1}/\gamma$ the incoming relative velocity. Hence, using 
\beq
\int \frac{\dd^D\bk}{(2\pi)^D} \frac{1}{(\bk^2)^n} e^{-i\bk\cdot \bx} = \frac{1}{4^n\pi^{D/2}} \frac{\Gamma[D/2-n]}{\Gamma[n]} (\bx^2)^{(-D/2+n)}\,,\label{FT}
\eeq
for the Fourier transform in $D=2$, we arrive at
\bea
\Delta^{(1)}_{\cL_1}\, p^\mu_1 &=& -\frac{m_1 m_2}{8\Mp^2}  \frac{\left( 2\gamma^2-1\right)}{\gamma\beta}\frac{\partial}{\partial \bb_\perp} \int_{k_\perp} \frac{e^{-i \bk_\perp \cdot \bb_\perp}}{-\bk_\perp^2} = -\frac{m_1 m_2}{8\Mp^2}  \frac{\left( 2\gamma^2-1\right)}{\gamma\beta} \frac{\partial}{\partial \bb_\perp} \left(+\frac{\log|\bb_\perp|}{2\pi}  \right)\nn
\\ &=& -2 m_1m_2 G \frac{\left( 2\gamma^2-1\right)}{\sqrt{\gamma^2-1}}\frac{b^\mu}{|b^2|}  \,.
\eea
In the last line wrote the result in a covariant fashion using the impact parameter four-vector, $b^\mu$, orthogonal to the velocities, which in the rest frame of particle 1 it has components  $b^\mu = (0, \bb_\perp)$, obeying $\bb_\perp\cdot \bu_2=0$. From here, and using \eqref{eq:apm}, we find
\beq
{\chi^{(1)}_b \over \Gamma} =  \frac{(2\gamma^2-1)}{(\gamma^2-1)}\,,\label{chi1b}
\eeq
for the scattering angle at 1PM in impact parameter space, as expected.

\subsubsection{Next-to-leading Order Impulse} 
According to \eqref{dp}, we have two contributions at NLO. We start with the 1PM action in \eqref{act1} using the trajectory expanded to linear order in $G$. The two terms are due to the shifted position and velocity at 1PM,

\beq
\begin{aligned}
  \Delta^{(2)}_{\cL_1} p^\mu_1 =& \frac{m_1 m_2}{4\Mp^2}
  \int_{-\infty}^{\infty}  \dd\tau_1\dd\tau_2 \int_k \frac{i k^\mu}{k^2}
  e^{ik \cdot b + i k\cdot (u_1\tau_1-u_2\tau_2)}
  \bigg[\frac{\left( 2\gamma^2-1\right)}{2} (i k) \cdot \left(\delta^{(1)} x_1(\tau_1)-\delta^{(1)} x_2(\tau_2)\right) \\
&+ \left( 2\gamma u_{2\beta} - u_{1\beta}\right)\cdot \delta^{(1)} v_1^\beta(\tau_1)+\left( 2\gamma u_{1\beta} - u_{2\beta}\right)\cdot \delta^{(1)} v_2^\beta(\tau_1)  \bigg]\,,\label{defl1}
\end{aligned}
\eeq
depending on both the deflection of particle 1 itself and the companion. Using an abuse of language, we refer to the contributions from particle 2 also as `mirror images'. These terms are straightforward to compute by a simple relabeling.
There is, however, one subtle point involving the factor of $q\cdot u_1-i\epsilon$ versus $q\cdot u_2+i\epsilon$ that appears in the deflection for particle~2, and the integration over $q^0$.
For for the sake of illustration, and simplicity, in what follows we will only refer to the effects due to the PM corrections to the motion of particle 1 itself, and deal with mirror images only when we combine all the intermediate results.\vskip 4pt

Inserting the values for the deflected velocity and position given in\eqref{d1v} and \eqref{d1x} into \eqref{defl1}, and massaging the final expression, we have
\bea
\Delta^{(2)}_{\cL_1} p^\mu_1 &=& i\frac{m_1 m^2_2}{128\Mp^4}  \int_{k,\ell}  \left[\left( 2\gamma^2-1\right)^2  \ell^2 -16\gamma^2(k\cdot u_1)^2 \right]  \frac{(\ell^\mu-k^
\mu) \hat\delta(k\cdot u_2)\hat\delta(\ell\cdot u_2)\hat\delta(\ell \cdot u_1)}{k^2(\ell-k)^2(k\cdot u_1 -i\epsilon)^2}e^{i\ell \cdot b}\,,\nn\\
\label{411}
\eea
where we have also discarded contributions which do not lead to a long-range interaction. We next move to the contribution from $\cL_2$, which entails using the unperturbed trajectories in the result quoted in \eqref{cl2}. As we discussed before, we will add the mirror image at the~end. After performing the contractions, and retaining only terms  which lead to a long-range interactions and do not vanish due to the $\delta$ functions, we arrive at the compact expression:
\beq
\Delta^{(2)}_{\cL_2} p^\mu_1 = i\frac{m_1m_2^2}{32\Mp^4}  \int_{k_{1,2}}
\begin{multlined}[t]
  \frac{k_1^\mu \hat\delta(k_1\cdot u_1)\hat\delta(k_2\cdot u_2)\hat\delta(k_1\cdot u_2) }{k_1^2k_2^2(k_1+k_2)^2} e^{i k_1\cdot b}\\
  \times\left(\gamma^2 k_1^2 + (k_2\cdot u_1)^2+(2\gamma^2-1)( k_1\cdot k_2)\right)\,.
  \end{multlined}
\eeq
To compute the integral we start by replacing $k_1\cdot k_2 = \frac{1}{2} \left((k_1+k_2)^2-k_1^2-k_2^2\right) \to -k_1^2/2$, after noticing that is the only term leading to a long-range force. Furthermore, the $k_2$-integral involving $(k_2\cdot u_1)^2$ can be simplified, using \bea
 \int_{k_2} \frac{(k_2\cdot u_1)^2\hat\delta(k_2\cdot u_2)}{k_2^2(k_1+k_2)^2} 
 &=& \frac{3}{8} \frac{\left( k_{1\perp}^\mu k_{1\perp}^\nu -\frac{1}{3}\left(\eta^{\mu\nu}- u_2^\mu u_2^\nu\right)k_{1\perp}^2\right)u_1^\mu u_1^\nu}{\left(1- \frac{ (k_1\cdot u_2)^2}{ k_1^2}\right)^2}  \int_{k_2} \frac{\hat\delta(k_2\cdot u_2)}{k_2^2(k_1+k_2)^2}  \nn \\
  &= &  \frac{k_1^2}{8}\left(\gamma^2-1\right) \int_{k_2} \frac{\hat\delta(k_2\cdot u_2)}{k_2^2(k_1+k_2)^2} + {\cal O}(k_1\cdot u_a)\,, 
\eea
with $k_{1\perp}^\mu = k_1^\mu - u_2^\mu (k_1\cdot u_2)$,
and ignoring $k_1\cdot u_a$ factors due to the overall $\delta$ functions.
Hence, after the innocuous re-labeling of momentum $k_1=\ell$ and $k_2= - k$,
\beq
\Delta^{(2)}_{\cL_2} p^\mu_1 =
i\frac{m_1m_2^2\left(\gamma^2+3\right)}{256\Mp^2}  \int_{k,\ell}\frac{\ell^\mu \hat\delta(\ell\cdot u_1)\hat\delta(\ell \cdot u_2) \hat\delta(k \cdot u_2) }{k^2(\ell-k)^2} e^{i \ell \cdot b} \,. 
\eeq

Before we evaluate the result, which entails adding these two contributions and mirror images, it is instructive to split the two terms above into two other (more suggestive) contributions. Notice that the $(k\cdot u_1)^2$ in the numerator in \eqref{411} cancels out against the denominator. This allows us to complete the squares and replace the vector integral by
\beq
\begin{aligned}
 \int_k \delta(k_2\cdot u_2) \frac{k^\mu}{k^2(\ell-k)^2} &= \frac{\ell^2}{2\ell_\perp^2} \int_k  \delta(k_2\cdot u_2) \frac{\ell^\mu_\perp}{k^2(\ell-k)^2} \to \frac{1}{2} \int_k  \delta(k_2\cdot u_2) \frac{\ell^\mu}{k^2(\ell-k)^2} \,,
\end{aligned}
\eeq
where $\ell^\mu_\perp \equiv \ell^\mu-u_2^\mu\, (u_2\cdot \ell) \to \ell^\mu$, after setting $\ell\cdot u_2=0$ due to the overall $\delta$ functions. From here can then combine terms together and write the impulse at 2PM, modulo mirror images,
\bea
\Delta^{(2)}\,p_1^{\mu }=  \Delta^{(2)}_{\triangle}\,  p_1^{\mu } +\Delta^{(2)}_{u}  p_1^{\mu }\,,\eea
with
\beq
 \Delta^{(2)}_{\triangle}\,  p_1^{ \mu } = -\frac{3 m_1m_2^2(5\gamma^2 -1)}{256 \Mp^4} {\partial \over \partial b_\mu} \int_{k,\ell}     \frac{\hat\delta(k\cdot u_2)\hat\delta(\ell\cdot u_2)\hat\delta(\ell \cdot u_1)}{k^2(\ell-k)^2} e^{i\ell \cdot b}\,,\label{deltatr}
\eeq
and
\beq
\Delta^{(2)}_{u} \, p_1^{ \mu } =  i\frac{m_1 m^2_2}{128\Mp^4}  \int_{k,\ell}  \left( 2\gamma^2-1\right)^2   \frac{(\ell^\mu-k^\mu) \ell^2 \hat\delta(k\cdot u_2)\hat\delta(\ell\cdot u_2)\hat\delta(\ell \cdot u_1)}{k^2(\ell-k)^2(k\cdot u_1 -i\epsilon)^2}e^{i\ell \cdot b}\,.\label{deltau}
\eeq
Notice that the expression in \eqref{deltatr} resembles the derivation of the impulse from the eikonal phase, see App.~\ref{eik}.\vskip 4pt 
To compute the integral we go to the rest frame of particle~2 this time, and using 
\beq
\int \frac{\dd^D\bk}{(2\pi)^D} \frac{1}{((\bk-\bp)^2)^{n_1}(\bk^2)^{n_2}} = \frac{\Gamma[n_1+n_2-D/2]\Gamma[D/2-n_1]\Gamma[D/2-n_2]}{\Gamma[n_1]\Gamma[n_2]\Gamma[D-n_1-n_2]} \frac{(\bp^2)^{D/2-n_1-n_2}}{(4\pi)^{D/2} }\label{bubble}\,,
\eeq
together with the Fourier transform in \eqref{FT}, we obtain
\beq
 \Delta^{(2)}_{\triangle}\,  p_1^\mu  = \frac{3\pi}{4} \frac{\left(5\gamma^2-1\right)}{\sqrt{\gamma^2-1}} \, {\partial \over \partial \bb_\perp} \left(\frac{G^2 m^2_2m_1}{|\bb_\perp|}\right)= -\frac{3\pi m^2_2m_1}{4} \frac{\left(5\gamma^2-1\right)}{\sqrt{\gamma^2-1}} \frac{G^2 b^\mu}{|b^2|^{3/2}}\,,\label{2pmt}
\eeq
where we wrote the result in a covariant fashion in terms of the impact parameter four-vector, $b^\mu$, which in the rest frame of particle 2 has components $b^\mu=(0,\bb_\perp)$, with $\bb_\perp\cdot \bu_1=0$. 
\vskip 4pt
The other term in \eqref{deltau} is a little more involved, also when it comes to adding the contribution from the mirror image, as we show momentarily. It is convenient to first tensor-reduce the integral. Let us  concentrate in the $\int_q$ part,
\beq
\int_q \frac{  (\ell^\mu-q^\mu) \, \hat\delta(q \cdot u_2)}{(\ell -q)^2q^2(q\cdot u_1 -i\epsilon)^2} = A \, \ell^\mu + B \, (\gamma u_2^\mu - \, u_1^\mu)\,,
\eeq
such that
\beq
 \Delta^{(2)}_{u}  \, p_1^{\mu } = \frac{m_1 m^2_2\left( 2\gamma^2-1\right)^2}{128\Mp^4} \left[\int_{\ell}\left( i A(\ell,\gamma) \ell^\mu + i B(\ell,\gamma) \left(\gamma u_2^\mu-u_1^\mu\right)\right)\ell^2 \hat\delta(\ell\cdot u_2) \hat\delta(\ell \cdot u_1)  e^{i \ell \cdot b} \right]  \,, \label{endeta}
\eeq
where we used that, because of all the $\delta$ functions involved, the result from dotting with $u_2$ must vanish. We multiply now by $\ell_\mu$, and use the same trick as before by writing $2\ell\cdot q =\left(-(\ell-q)^2+\ell^2+q^2\right)$ to cancel local terms, so that
\beq
A   =   \frac{1}{\ell^2} \int_q \frac{(\ell^2-(\ell\cdot q)) \hat\delta(q \cdot u_2)}{(\ell -q)^2q^2(q\cdot u_1 -i\epsilon)^2} =  \frac{1}{2} \int_q \frac{ \, \hat\delta(q \cdot u_2)}{(\ell -q)^2q^2(q\cdot u_1 -i\epsilon)^2} \,.\label{aint}
\eeq
For the $B$ coefficient we dot with $u_1$ instead. Using that $\ell\cdot u_a=0$, we find 
\beq
B = -\frac{1}{\gamma^2-1} \int_q \frac{\hat\delta(q \cdot u_2)}{(\ell -q)^2q^2(q\cdot u_1 -i\epsilon)}\,.\label{bint}
\eeq

\vskip 4pt

Notice that both these integrals, $A$ and $B$, resemble the crossed-box and triangle integrals in the large-mass limit of the one loop amplitude in \cite{cheung,zvi1,zvi2}. To compute them we use the prescription we described in \S\ref{sec:pot}. First of all, since we do not pick radiation poles, only the pole at $q\cdot u_1-i\epsilon =0 $ contribute. Furthermore, notice that the $A$ integral converges, and that we have the two poles on the same side of the complex plane. By simply closing the contour in the opposite direction we can thus set $A =0$.
This fact, which is equivalent to the vanishing of the crossed-box contribution at one loop \cite{cheung,zvi1,zvi2}, also explains why there is no $(2\gamma^2-1)^2$ term in the scattering angle at 2PM.\footnote{Notice that in $D>4$ the $A$ integral will give a non-zero contribution to the scattering angle, as it occurs also in the computation using the scattering amplitude \cite{Cristofoli:2020uzm}.}\vskip 4pt For the $B$ integral we find it convenient to move to the rest frame of particle 1 (recall $u_1=(1,0,0,0)$ and $u_2=(\gamma,\gamma\beta,0,0)$). Then, using the prescription in \eqref{poles} and averaging over conservative poles in the upper and lower complex plane, we have
\beq
\begin{aligned}
 iB &= -\frac{i}{\gamma^2-1}\int_{\bq} \frac{\dd q^0}{2\pi} \frac{\hat\delta(q \cdot u_2)}{(\ell -q)^2q^2(q^0-i\epsilon)} = -\frac{i}{(\gamma^2-1)}\frac{i}{2} \int_q  \frac{\hat\delta(q \cdot u_2)\hat\delta(q^0)}{(\ell -q)^2 q^2} \\
 &=    \frac{1}{8\pi \gamma\beta (\gamma^2-1)}  \frac{1}{(-\ell_\perp^2)}\left(\frac{1}{\bar\epsilon}+2\log(-\ell_\perp^2)\right)\,,
\end{aligned}
\eeq
where once again $\ell^\mu_\perp = \ell^\mu- \frac{u^\mu_2 (\ell\cdot u_2)}{u_2^2}$, and we evaluated the integral in $D=4-2\epsilon$ dimensions, 
\beq
\int_{q} \frac{\hat\delta(q\cdot u_2)\hat\delta(q^0)}{(\ell-q)^2 q^2}
 = \frac{1}{4\pi \gamma\beta} \frac{1}{(-\ell_\perp^2)}\left(\frac{1}{\bar\epsilon}+2 \log(-\ell_\perp^2)\right)\,,
\eeq
with some irrelevant factors absorbed into the definition of $\bar\epsilon$. As before, the $\delta$ functions in \eqref{endeta} will ultimately set $\ell^2_\perp \to \ell^2$, such that the contribution from the $1/\bar\epsilon$ pole turns into a contact interaction that can be readily discarded in the classical limit. Therefore, only the logarithmic term survives as a long-range force. We expect this to be a generic feature also at higher PM orders, with intermediate spurious divergences producing analytic contributions, prior to the Fourier transform, which do not survive in the classical limit. (This is reminiscent of the small transfer momentum expansion of the scattering amplitude in \cite{cheung,zvi1,zvi2}.) Finally, using 
 \beq
\int_{\ell}\frac{\ell^2}{-\ell^2} \log (-\ell^2)  e^{-i \ell \cdot b} \delta(\ell\cdot u_1)\delta(\ell\cdot u_2)= +\frac{1}{\pi \gamma \beta} \frac{1}{|\bb_\perp|^2}\,,
\eeq
 we obtain 
\beq \Delta^{(2)}_{u}  \, p_1^\mu = 2\frac{m_1m_2^2\left( 2\gamma^2-1\right)^2}{(\gamma^2-1)^2} \frac{G^2}{|b^2|} \left(\gamma u_2^\mu-u_1^\mu\right)\,,\label{2pmu}
\eeq 
\vskip 4pt

We now come to adding the contributions together and include also the mirror images. For the triangle term in \eqref{deltatr} this is straightforward. First of all, by construction the terms from the total effective Lagrangian, prior to taking the partial derivative from \eqref{spacetime}, must be symmetric under $1 \leftrightarrow 2$. The same property translates to \eqref{deltatr} before applying $\partial_{b_\mu}$, the action of which ultimately changes the relative sign with respect to the mirror image. The one associated with the term in \eqref{2pmu} is a bit more subtle. It is easy to see that the main actor in \eqref{2pmu} is the contribution from the $\eta^{\mu\alpha}$ term in \eqref{d1x}. Notice that keeping $q$ unchanged while relabeling $1\leftrightarrow 2$ results in an overall sign difference with respect to the same term in \eqref{d1x2}, which is then compensated by the relative sign 
 in \eqref{defl1}. Following similar steps we thus arrive at the same junctions, except   
that the deflected trajectory for particle 2 leads to poles at $q^0=-i\epsilon$ for the $A$ and $B$ integrals in \eqref{aint} and \eqref{bint}. Hence, while the $A$ integral vanishes, when it comes to the $B$ integral the pole is now shifted to the lower complex half-plane, resulting in an overall minus sign. Therefore, the mirror image also leads to an anti-symmetrization in $1\leftrightarrow 2$ in \eqref{2pmu}. At the end of the day, combining the terms and adding the 1PM result, we obtain
\beq
\begin{aligned}
\Delta \, p_1^\mu &=  - \frac{G m_1m_2\, b^\mu}{|b^2|} \left(\frac{2\left(2\gamma^2-1\right)}{\sqrt{\gamma^2-1}} + \frac{3\pi}{4} \frac{\left(5\gamma^2-1\right)}{\sqrt{\gamma^2-1}} \frac{G M}{|b^2|^{1/2}}\right)\\
&+ 2\frac{m_1m_2\left( 2\gamma^2-1\right)^2}{(\gamma^2-1)^2} \frac{G^2}{|b^2|} \left((\gamma m_2+m_1) u_2^\mu- (\gamma m_1+m_2) u_1^\mu\right)\,,\label{2pmtot}
\end{aligned}
\eeq
for the total impulse to 2PM order in covariant form. Notice that, because $b \cdot u_a=0$, while the term involving the velocities was the most intricate of the two, it does not enter in the derivation of the 2PM angle. Therefore, only the correction due to \eqref{2pmt} matters, yielding
\beq
\frac{\chi^{(2)}_b}{\Gamma} = \frac{3\pi}{8} \frac{(5\gamma^2-1)}{\gamma^2-1}\,,\label{chi2b}
\eeq
after using \eqref{eq:apm}. The total impulse in \eqref{2pmtot} and resulting scattering angle are in agreement with the result obtained in \cite{west}.

\subsection{Adiabatic Invariants}

Once the scattering angle is known, and translated from impact parameter to angular momentum space, 
\beq
\chi^{(1)}_j = \frac{(2\gamma^2-1)}{\sqrt{\gamma^2-1}}\,, \,\,
\chi^{(2)}_j =  \frac{3\pi}{8} \frac{(5\gamma^2-1)}{\Gamma}\,,
\eeq
we can then use \eqref{eq:ir} to reconstruct the B2B radial action via analytic continuation to $\cE<0$,
\beq
\label{eq:ir2pm}
i^{\rm 2PM}_r(j,\cE) =  \frac{\hat p_\infty}{\sqrt{-\hat p_\infty^2}}
\chi^{(1)}_j(\cE) - j \left(1 - \frac{2}{\pi} \frac{\chi^{(2)}_j({\cE})}{j^{2}} \right) 
= -j + \frac{(2\gamma^2-1)}{\sqrt{1-\gamma^2}} + \frac{3}{4j} \frac{(5\gamma^2-1)}{\Gamma} \,.
\eeq
From here, following the B2B dictionary reviewed in \S\ref{sec:b2b}, we arrive at the same 2PM values for the gravitational observables for bound orbits derived in \cite{paper1,paper2} from the classical limit of the scattering amplitude/angle, namely
\beq
\frac{\Delta\Phi_{\rm 2PM}}{2\pi} = \frac{3}{4j^2}  \frac{(5\gamma^2-1)}{\Gamma}\,,
\eeq
for the periastron-advance, and
\beq
\frac{T^{\rm 2PM}_p}{2\pi GM}=\frac{\gamma(3-2\gamma^2)\Gamma}{(1-\gamma^2)^{3/2}}+\frac{3}{4j}\frac{15\gamma^2\nu-10\gamma(2\nu-1)+\nu}{\Gamma^2}\,, 
\eeq
for the periastron-to-periastron period. From here we obtain for the azimuthal frequency,
\beq
\begin{aligned}
GM \Omega^{\rm 2PM}_\phi &= \left(\frac{T^{\rm 2PM}_p}{2\pi GM}\right)^{-1}\left(1+ \frac{\Delta\Phi_{\rm 2PM}}{2\pi}\right) \\
&=  -\frac{(1-\gamma^2)^{3/2}\Gamma(4j^2\Gamma+15\gamma^2-3)}{j\left(4j\gamma(2\gamma^2-3)\Gamma^3-3(1-\gamma^2)^{3/2}(\nu+5\gamma(2+(3\gamma-4)\nu))\right)}\,.
\label{omegaphi2}
\end{aligned}
\eeq
It is also useful to notice that keeping only up to the 1PM coefficient of the scattering angle in the radial action leads to no periastron-advance, as expected, which implies \cite{paper2}
\beq
x_{\rm 1PM} \equiv (GM\Omega^{\rm 1PM}_\phi)^{2/3}= \left(\frac{T^{\rm 1PM}_p}{2\pi GM}\right)^{-{2/3}}= \frac{(1-\gamma^2)}{((3\gamma-2\gamma^3)\Gamma)^{2/3}}\,.
\eeq

We can also compute the redshift function, (re)obtaining at 2PM \cite{paper2}
\begin{equation}
  \begin{aligned}
  \langle z_1\rangle- \langle z_1^{(0)}\rangle =&
  \frac{
    \Gamma
  }{
    M(1+\Delta)\nu\left(4j\gamma(3-2\gamma^2)\Gamma^3+3(1-\gamma^2)^{3/2}(\nu+5\gamma(2+(3\gamma-4)\nu))\right)
  }\\
  &\times
    \begin{multlined}[t]
      \bigg[8j^2(1-\gamma^2)^{3/2}\Gamma\nu+4j\gamma(2\gamma^2-3)\Gamma(1+\Delta+2(\gamma-1)\nu)\\
        +3(1-\gamma^2)^{3/2}(5\gamma^2(\Delta-1)\nu-10\gamma(1+\Delta-2\nu)-(3+\Delta)\nu)
      \bigg]\,,
    \end{multlined}
  \end{aligned}
\end{equation}
with
\begin{equation}
  \langle z_1^{(0)} \rangle = 1+\frac{(1+\Gamma)(2\nu+\Delta-1)}{2\nu}\,,
\end{equation}
and $\Delta \equiv \sqrt{1-4\nu}$. The expression for $z_2$ results after replacing $\Delta \to -\Delta$.\vskip 4pt

These PM corrections include a series of PN terms (inside the $\gamma$'s) at each order in~$G$.  As we discussed in \cite{paper1,paper2}, the 2PM result for the periastron advance (and period) are \emph{one loop exact}, in the sense that all the PN contributions at $1/j^2$ (and $1/j$) are encapsulated in the above expressions, to all orders in velocities.
Moreover, as we emphasized in \cite{paper1}, the PN expansion of $x_{\rm 1PM}$ for circular orbits recovers the correct coefficient for all of the ${\cal O}(\cE^n \nu^n)$ contributions at $n$PN order, for any value of $n$.

\subsubsection{Circular Orbits}

To obtain the orbital frequency for circular orbits, we can proceed as explained in \cite{paper1} by identifying the orbital elements, and then imposing the vanishing of the eccentricity.
Or,~equivalently, the simplicity of the radial action to 2PM allows us to also impose $i_r=0$, and then solve for~$j$ as a function of $\gamma$ (see e.g. footnote 19 in \cite{paper1}). Notice, however, in that case it is useful to use the 2PM\text{-}resummed version of the radial action \cite{paper2}, 
\beq
i^{\rm 2PM\text{-}res}_r = \left(\frac{B}{\sqrt{-A}} - \sqrt{-C}\right)\,,
\eeq
with
\beq
\begin{aligned}
-A &= -\hat p_\infty^2 = \frac{1-\gamma^2}{\Gamma^2}\,, \\
 B &= \hat p_\infty^2 \chi^{(1)}_b = \frac{2\gamma^2-1}{\Gamma} \,, \\
  -C&= j^2 - \frac{4}{\pi} \hat p_\infty^2  \chi^{(2)}_b = j^2\left(1 - \frac{3(5\gamma^2-1)}{2\Gamma j^2}\right)\,,
\end{aligned}
\eeq
yielding
\beq
i^{\rm 2PM\text{-}res}_r =  \frac{(2\gamma^2-1)}{\sqrt{1-\gamma^2}} - j \sqrt{1 - \frac{3(5\gamma^2-1)}{2\Gamma j^2}}\,.
\eeq
Imposing the condition $i_r=0$ for a circular orbit, we then recover the value
\beq
j_{\rm 2PM}^2 = \frac{(2\gamma^2-1)^2}{(1-\gamma^2)} + \frac{3(5\gamma^2-1)}{2\Gamma}  \,,\label{j2pm}
\eeq
for the angular momentum to 2PM~\cite{paper1}. We can then enter this expression in \eqref{omegaphi2}, resulting in the orbital frequency as a function of $\gamma$, and ultimately the binding energy from \eqref{Egam}. A~simpler route is to use the the first law of black hole dynamics \cite{letiec}, and compute instead~\cite{paper1}
\beq
\begin{aligned}
GM \Omega^{\rm 2PM}_{\rm circ} &=  \left(\Gamma\frac{\dd j_{\rm 2PM}(\gamma)}{\dd\gamma}\right)^{-1} = \frac{2 j_{\rm 2PM}}{\Gamma}\left(  \frac{\dd j_{\rm 2PM}^2(\gamma)}{\dd\gamma}\right)^{-1} \\
&=  \frac{2}{\Gamma} \sqrt{\frac{(2\gamma^2-1)^2}{(1-\gamma^2)} + \frac{3(5\gamma^2-1)}{2\Gamma}}\left( \frac{3(-5\gamma^2 \nu + 10\gamma \Gamma^2 +\nu)}{2\Gamma^3}-\frac{2\gamma(4\gamma^4-8\gamma^2+3)}{(\gamma^2-1)^2}\right)^{-1} .
\end{aligned}
\eeq
Notice that, keeping only the 1PM term, the variation in \eqref{eq:tp} that gives us the orbital period, $\partial_\gamma i_r$, is the same we obtain after solving for $j$ on orbits with $i_r=0$.

\subsection{Hamiltonian}
Due to the B2B map, we did not require a Hamiltonian in order to obtain all the gravitational observables in the conservative sector. However, as described in \S\ref{sec:b2b}, we can easily reconstruct~it to derive also the equations of motion for bound orbits. Using the values for $\chi_b^{(1)}(E)$ and $\chi^{(2)}_b(E)$ in \eqref{chi1b} and \eqref{chi2b}, and the expression in \eqref{cnb}, we find 
\begin{equation}
  \begin{aligned}
    \frac{c_1(\bp^2)}{1!}  = &\frac{\nu^2M^2}{\Gamma^2\xi}(1-2\gamma^2)\,,\\
    \frac{c_2(\bp^2)}{2!}  = &\frac{\nu^2M^3}{\Gamma^2\xi}\left[
      \frac{3}{4}(1-5\gamma^2)+\frac{\nu^2(2\gamma^2-1)^2(3\xi-1)}{2\Gamma^3\xi^2}\right. \\
      &\hspace{1cm}\left.
      +\frac{(2\gamma^2-1)(4\gamma\nu+(1-8\gamma+6\gamma^2)\nu^2)}{\Gamma^3\xi}
      \right]\,,
  \end{aligned}
\end{equation}
for the PM coefficients of the Hamiltonian, which we wrote using the convention of \cite{cheung,zvi1,zvi2}. It is straightforward to show that the Hamiltonian is equivalent to the one in \cite{cheung,zvi1,zvi2}, e.g. Eq.~(10.10) of \cite{zvi2}, written using a different combination for the two  terms involving products of 1PM coefficients.
Notice, nonetheless, that only the first terms in $c_1$ and $c_2$ actually matter, as illustrated by the radial action. The other two in $c_2$ are simply there to cancel the unwanted contributions which appear when deriving the $f_1, f_2$ in \eqref{pinfpm} as a function of the energy. This feature is even more striking at higher PM orders, with the $\chi_j^{(2n)}$'s (and associated $f_n$'s) carrying all the relevant information to compute dynamical invariants.

\section{Leading Tidal Effects}\label{sec:finite}

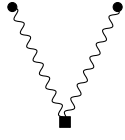
\begin{figure}
  \centering
    \begin{tikzpicture}
      [scale=1]
      \draw[boson] (-0.7,0) -- (0,-1.5);
      \draw[boson] (0.7,0) -- (0,-1.5);
      \filldraw (-0.7,0) circle (0.06);
      \filldraw (0.7,0) circle (0.06);
      \filldraw (-0.07,-1.6) rectangle ++(4pt,4pt);
\end{tikzpicture}
\caption{Feynman diagram with an insertion of a tidal operator represented by the square.}\label{figsize}
  \end{figure}

In this section we provide an example of how finite-size effects are incorporated in our EFT approach for the PM regime. For concreteness, we study the correction due to the electric- and magnetic-type tidal operators in the effective action, see \eqref{1} and \eqref{2}, 
\beq
\sum_a c^{(a)}_{E^2} \int \dd\tau_a\, E_{\mu\nu}(x_a(\tau_a) ) E^{\mu\nu} (x_a(\tau_a)) +  c^{(a)}_{B^2} \int \dd\tau_a\, B_{\mu\nu}(x_a(\tau_a) ) B^{\mu\nu} (x_a(\tau_a)) \,,
 \eeq
 with  $E_{\mu\nu} = R_{\mu\alpha\nu\beta} v^\alpha v^\beta\,$ and $B_{\mu\nu} = R^\star_{\mu\alpha\nu\beta} v^\alpha v^\beta$. In the weak field limit we have
 \beq
 R_{abcd} = \frac{1}{2} \left( \partial_b\partial_c h_{ad}+ \partial_a\partial_d h_{bc} - \partial_a\partial_c h_{bd} - \partial_b\partial_d h_{ac} \right)\,,
 \eeq
and therefore the contribution to the effective action is straightforward, and consists of computing the diagram shown in Fig~\ref{figsize},  including the tidal operators and two mass insertions. 
Since we are interested in the leading PM correction to the impulse, only the undeflected worldlines in \eqref{pmexp} are needed. The result for the effective action reads
 \beq
 \begin{aligned}
\int \dd\tau \cL^{E^2}_{\rm LO} =  \frac{c^{(1)}_{E^2}m_2^2}{64 \Mp^4} \int_{p,\ell}& \,\frac{\hat\delta(p\cdot u_2)}{p^2(p-\ell)^2} \Big((1-2\gamma^2)^2(p\cdot \ell)^2+2(1-4\gamma^2)(p\cdot \ell)(p\cdot u_1)^2+2 (p\cdot u_1)^4  \Big)  \\ \times
 &\hat\delta(\ell\cdot u_1)\hat\delta(\ell\cdot u_2) e^{i\ell\cdot b} + 1\leftrightarrow 2\,,
 \end{aligned}
 \eeq
 for the electric-type, whereas the the magnetic term we find
 \beq
 \begin{aligned}
\int \dd\tau \cL^{B^2}_{\rm LO} =  \frac{c^{(1)}_{B^2}m_2^2}{32 \Mp^4} \int_{p,\ell}& \,\frac{\hat\delta(p\cdot u_2)}{p^2(p-\ell)^2} \Big(2\gamma^2(\gamma^2-1)(p\cdot \ell)^2+(1-4\gamma^2)(p\cdot \ell)(p\cdot u_1)^2+ (p\cdot u_1)^4  \Big)  \\ \times
 &\hat\delta(\ell\cdot u_1)\hat\delta(\ell\cdot u_2) e^{i\ell\cdot b} + 1\leftrightarrow 2\,.
 \end{aligned}
 \eeq 
 As before, we compute the integral in the rest frame of particle 2, which reduces the integration in $p$ to $D=3$, and then use well-known results for the moments of the integral in \eqref{bubble}, see e.g. \cite{Smirnov}. After performing the final (Fourier transform) integral in $\ell$, we find 
\beq
\int \dd\tau \cL^{E^2}_{\rm LO} =  \frac{9\pi }{64} \frac{G^2}{|b^2|^{5/2}}\left(\frac{35\gamma^4-30\gamma^2+11}{\sqrt{\gamma^2-1}}\right) \left(c^{(1)}_{E^2}m_2^2 + c^{(2)}_{E^2}m_1^2\right)\,,
\eeq  
\beq
\int \dd\tau \cL^{B^2}_{\rm LO} =  \frac{9 \pi }{64} \frac{G^2}{|b^2|^{5/2}}\left(\frac{35\gamma^4-30\gamma^2-5}{\sqrt{\gamma^2-1}}\right) \left(c^{(1)}_{B^2}m_2^2 + c^{(2)}_{B^2}m_1^2\right)\,,
\eeq 
for the electric- and magnetic-type contributions to the effective action, respectively. The impulse follows by taking a derivative with respect to the impact parameter. Using eq.~\eqref{eq:apm} we thus get, in terms of the reduced angular momentum ($j = GM\mu J$),
 \beq
 \chi^{E^2}_{\rm LO} (j) =  \frac{45 \pi \lambda_{E^2}}{64}   \frac{(\gamma^2-1)^2\left(35\gamma^4-30\gamma^2+11\right)}{\Gamma^5}\frac{1}{j^6}\,,\label{chie2}
 \eeq
 \beq
 \chi^{B^2}_{\rm LO} (j) =  \frac{45 \pi \lambda_{B^2}}{64}   \frac{(\gamma^2-1)^2\left(35\gamma^4-30\gamma^2-5\right)}{\Gamma^5}\frac{1}{j^6}\,,\label{chie2n}
 \eeq
from tidal effects at leading PM order, with 
 \beq
\lambda_{E^2} \equiv\frac{1}{ G^4 M^5 } \left(C^{(1)}_{E^2}\frac{m_2}{m_1}+C^{(2)}_{E^2}\frac{m_1}{m_2}\right)\,,
 \eeq
 and similarly for $\lambda_{B^2}$. \vskip 4pt The expressions in \eqref{chie2} and \eqref{chie2n} agree with the result in \cite{binit}, see e.g. Eq. (6.2), after we translate between the different conventions.  Notice, as emphasized in \cite{binit}, in the `high energy limit' $\gamma \gg 1$  the scattering angle receives the same relative contribution from both the electric- and magnetic-type tidal effects. From \eqref{eq:peria} and \eqref{chie2} we obtain   
 \beq
 \frac{\Delta\Phi_{\rm LO}^{\rm Tidal}}{2\pi}  = \frac{45 (1-\gamma^2)^2}{64\Gamma^5 j^6} \Big( \lambda_{E^2} \left(35\gamma^4-30\gamma^2+11\right)+ \lambda_{B^2} \left(35\gamma^4-30\gamma^2-5\right)\Big)\,,
 \eeq
 for the periastron advance after analytic continuation in the binding energy to $\gamma < 1$. This agrees with the Newtonian limit  \cite{binit} ($\beta = \sqrt{1-\gamma^2}/\gamma \ll 1$),
  \beq
   \frac{\Delta\Phi^{\rm Tidal}_{\rm LO}}{2\pi} = \frac{ 45 \beta^4}{4j^6}\left(\lambda_{E^2} +{\cal O}(\beta^2)\right)\,.
   \eeq
 
 Once again, we can also reconstruct a Hamiltonian that includes tidal effect at leading PM order, but to all orders in the velocity.
 The derivation is straightforward using the recursion relation from \cite{paper1}, see \S\ref{sec:H}, and we find
 \beq
\begin{aligned}
  \frac{c_6}{6!} &= -\frac{P_6(E)}{2 E \xi} +\cdots 
  = -\frac{16 M^{12}\nu^6\chi_j^{(6)}(E)}{15\pi E p_\infty^4 \nu^6 \xi} +\cdots 
  = -\frac{16M^7\Gamma^3 \nu^2\chi_j^{(6)}(E)}{15 \pi \xi(1-\gamma^2)^2}+\cdots \\
  &= -\frac{3M^7\nu^2}{8\Gamma^2\xi}\left( \lambda_{E^2} \left(35\gamma^4-30\gamma^2+11\right)+ \lambda_{B^2} \left(35\gamma^4-30\gamma^2-5\right)\right)+\cdots\,,
\end{aligned}
\eeq
such that
 \beq
 \frac{H_{\rm LO}^{\rm Tidal}(\bp^2)}{\mu} =  -\frac{15\nu}{8\Gamma^2\xi}\left( \lambda_{E^2} \left(7\gamma^4-6\gamma^2+\frac{11}{5}\right)+ \lambda_{B^2} \left(7\gamma^4-6\gamma^2-1\right)\right) \left(\frac{GM}{r}\right)^6\,,
 \eeq 
at leading order in the PM expansion. Higher order corrections can be easily computed by including non-linear gravitational couplings as well as iterations from the modified worldlines.  
 \section{Discussion \& Outlook}\label{sec:disc}

Building upon NRGR \cite{nrgr} and the B2B map \cite{paper1,paper2}, we developed a systematic framework to compute dynamical invariants for binary system in a PM expansion, to~all orders in velocity. The two main ingredients in our formalism are: {\it i)} An EFT approach to compute the gravitational scattering angle in perturbation theory, and {\it ii)} The B2B radial action, constructed via analytic continuation from the scattering angle. We~illustrated the procedure with two paradigmatic examples. First, we used the EFT formalism to compute the impulse and scattering angle in the conservative sector to ${\cal O}(G^2)$, and subsequently (re-)derived all the associated gravitational observables for bound orbits. Secondly, we computed the leading PM contribution due to tidal effects to the scattering angle and periastron advance. The~results for the conservative dynamics to 2PM agree with the ones obtained through the B2B dictionary using the impetus formula \cite{paper1,paper2}, applied to the classical limit of the one loop amplitude in \cite{cheung}. For the sake of completeness, we reconstructed the Hamiltonian for the two-body dynamics from scattering data to NLO, which agrees with the one obtained in~\cite{cheung} via a matching calculation. We also computed  the leading PM contribution to the Hamiltonian due to tidal effects, which is equivalent to the EOB approach discussed in~\cite{binit}.\vskip 4pt 

One of the advantages of our formalism, in comparison with computing the (more intricate and gauge-dependent) gravitational potential in a PM expansion, is the dependence of the B2B radial action on the gauge-invariant, and much simpler, scattering angle instead. This feature notoriously simplifies the type of diagrams and integrals involved in our case; all the while incorporating the relativistic information that is lacking in a~PN~scheme (when both are kept up to the same order in~$G$). Furthermore, although still relying on Feynman tools, the EFT approach circumvents somewhat the intermediate steps required, thus far, in the program to obtain classical gravitational physics out of quantum amplitudes \cite{ira1,cheung,zvi1,zvi2,donal,donalvines, withchad,Holstein:2008sx,Bjerrum-Bohr:2013bxa,Vaidya:2014kza,Guevara:2017csg,Chung:2018kqs,Guevara:2018wpp,simon,Guevara:2019fsj,bohr,cristof1,Arkani-Hamed:2019ymq,Bjerrum-Bohr:2019kec,Chung:2019duq,Bautista:2019tdr,Bautista:2019evw,KoemansCollado:2019ggb,Johansson:2019dnu,Aoude:2020onz,Cristofoli:2020uzm,Chung:2020rrz,Bern:2020gjj,Bern:2020buy,Parra,Antonelli:2019ytb}. In particular, even though the impetus formula \cite{paper1} removes the need of the additional matching calculation performed in \cite{ira1,cheung,zvi1,zvi2}, or the intricate Born iterations of \cite{Holstein:2008sx,cristof1}, the mass-dependent loop integrals, and spurious IR super-classical divergences, still remain in the amplitude program. On the other hand, only massless integrals (without super-classical singularities) appear in the classical EFT framework. These differences are also manifest when computing gauge-invariant quantities, such as the scattering angle, as we demonstrated in the explicit example in this paper at 2PM order. Only two diagrams are needed for the NLO impulse, shown in Figs.~\ref{eftfig1}ab, and just the (Fourier transform of the) following integral, 
\beq
\int_\bq\label{5.1} \frac{1}{\bq^2(\bq-\bell)^2} = \frac{1}{8|\bell|}\,,
\eeq
was required when restricted to the scattering angle. Needless to say, the (super-classical) IR divergent box diagram never shows up, as expected. This is due to the lack of particle/anti-particle propagators yielding poles on opposite sides of the real~axis. Moreover, the analogous to the crossed-box term, which appears here as a correction from the deflected trajectory in the tree level action, readily vanishes in the potential region. In practice, this means that the exact prescription to perform the energy integral(s) that we used in \eqref{bint} was not needed to compute observables for bound orbits to 2PM. (That is the case because the extra term depending on the velocities in \eqref{2pmtot} is orthogonal to the impact parameter.) \vskip 4pt In contrast, even after removing/ignoring the box diagram by whichever method, the result using the scattering amplitude still requires the mass-dependent one loop triangle integral, with a contour prescription \cite{cheung,zvi1,zvi2,bohr} together with the large-mass and small momentum-transfer limits. At the end of the day, as illustrated in~\cite{cheung,donal}, the relevance of the integral in \eqref{5.1} emerges in the classical limit. (This is expected, since the massive field then collapses to a non-propagating source as in HQET \cite{ben}.) Yet, in our approach, the resulting scattering angle and dynamical invariants to 2PM were obtained directly from~\eqref{5.1}. Moreover, in a classical EFT framework the large-mass limit is implicitly taken, and the small momentum-transfer expansion is equivalent to retaining only non-analytic terms prior to the Fourier transform, which yield long-range interactions. Notice this implied the removal of intermediate divergences that produce only contact terms, as in \eqref{bint}. The final answer for the scattering angle, however, emerged as a combination between a 1PM correction to the trajectory, due to the tree level effective action in Fig.~\ref{eftfig1}a, plus the leading one loop contribution from~Fig.~\ref{eftfig1}b evaluated on the undeflected solution. This combination, also noted in \cite{donal}, illustrates how the amplitude and classical derivations differ when it comes to the organization of all the relevant contributions. The reordering of terms suggests a dual understanding for the origin of the different pieces entering in the classical limit, which may ultimately help us also to incorporate powerful on-shell methods in our approach. \vskip 4pt 

The EFT procedure in the PM expansion described here can be straightforwardly automatized to all orders.
Since (for non-spinning bodies) worldline non-linearities are only produced by finite-size effects, the number of Feynman diagrams is relatively small when restricted to the monopole term, which notoriously simplifies the derivations. Moreover, for a large fraction of the diagrams evaluated on a straight worldline, which produce factors of $\delta(k\cdot u_a)$ after integration in the proper time, the calculation in the potential region resembles the same type of three-dimensional massless integrals that we find in NRGR with static propagators. The associated master integrals are known to four loops \cite{nrgrG5}, akin of 5PM order. However, for another set of diagrams the associated $\delta$ functions yield new integrals which do not appear in the PN expansion. By replacing these $\delta$'s by factors of $(q\cdot u -i\epsilon)^{-1}$, these turn out to be equivalent to the ones resulting from the \emph{iterations} of the deflected worldlines in the computation of the impulse. As we emphasized, the type of integrals from this procedure resemble instead those of HQET, as it is the case for the NLO impulse in \eqref{bint}. Therefore, we expect the various powerful results in the literature, e.g. \cite{Smirnov}, can be then translated to our framework when pursuing higher PM orders. (Similar integrals appear in the {\it soft} expansion discussed in \cite{Parra}.) As we mentioned, notice that the term proportional to \eqref{bint} does not enter in the scattering angle at 2PM. This is a remarkable feature of the B2B map which further simplifies its implementation at higher orders, by allowing us to restrict to corrections to the impulse in the $b^\mu$-direction, when calculating contributions at the highest PM order.\vskip 4pt

The derivation of the 3PM impulse using the EFT formalism is ongoing. Notice, however, that the natural power counting  in $1/j$ of the B2B radial~action in \eqref{eq:ir} requires the 4PM value as well. As we studied in \cite{paper1,paper2}, one can include all the contributions to $\chi^{(4)}_j$ from lower order terms in \eqref{eq:fi2}, i.e. using the impetus formula together with the results from \cite{zvi1,zvi2} to read off the value of the $f_{n \leq 3}$'s, thus yielding a 2PN truncation to the radial~action. Yet,~as we emphasized \cite{paper1,paper2}, from the PM standpoint this truncation does not account for all the velocity contributions at ${\cal O}(1/j^3)$, contrary to what happens at ${\cal O}(1/j)$ with $f_2$ alone.\footnote{Regarding this point, in a recent paper \cite{bini2} the authors claim that the radial action in \cite{paper1,paper2} truncated to 2PM, re-derived here in \eqref{eq:ir2pm}, is a ``PN-incomplete 2PM truncation."  It is straightforward to show that the expression given in \cite{bini2} to 2PM, $I^{\rm loc}_r = -j + I_0^S + I_1^S(\gamma)/(hj)$ with $h=\Gamma$ and~$I_0^S, I_1^S$ in their Eqs. (9.3) and (9.5), coincides with \eqref{eq:ir2pm}. The derivation in \cite{bini2} thus reproduces the~2PM result in~\cite{paper1,paper2}. We think the authors of \cite{bini2} perhaps meant the truncation with only the $f_{n\leq 3}$ contributions to~$\chi^{(4)}_j$. As~we emphasized in \cite{paper1,paper2}, indeed this misses the (so far unknown) $f_4$ term. However, it is a consistent truncation as we showed explicitly by deriving all the observables to 3PM/2PN in \cite{paper2}, including PN-exact terms controlled by \eqref{eq:ir2pm}.} Therefore, the scattering angle at 4PM remains a key ingredient to complete the next order in the B2B map, that is yet to be calculated. The missing piece may be~obtained from the classical limit of the three loop amplitude and/or the 4PM impulse; using the tools from \cite{cheung,zvi1,zvi2,donal,bohr,cristof1}, the EFT framework developed here, or some hybrid approach between the two (plausibly in combination with the classical double copy  \cite{Don1,Luna:2016hge,Goldberger:2016iau, Walter,Li:2018qap,CSH,Plefka,Plefka:2019hmz,Kim:2019jwm,Goldberger:2019xef,Alfonsi:2020lub}, on-shell methods~\cite{Bern:1994zx,Bern:1994cg}, soft limits \cite{Parra}, and simplifications of the Einstein-Hilbert action \cite{Cheung:2017kzx}). In either case, the B2B dictionary~\cite{paper1,paper2} provides a direct (analytic) connection between scattering data and bound states, neatly demonstrating the need to push the computations to (even) higher  orders, to fully incorporate the power of the PM expansion in the dynamics of binary systems. 

\begin{acknowledgments}

  We thank the Munich Institute for Astro- and Particle Physics (MIAPP), supported by the DFG cluster of excellence ``Origin and Structure of the Universe'', and the participants of the program ``Precision Gravity: From the LHC to LISA" for several fruitful discussions. In~particular, Zvi Bern, Poul Damgaard, Guillaume Faye, Alfredo Guevara, Donal O'Connell, Radu Roiban, Chia-Hsien Shen, Mikhail Solon, Jan Steinhoff, and Justin Vines. We~also~thank Clifford Cheung, Zhengwen Liu and Julio Parra-Martinez for useful discussions. R.A.P. acknowledges financial support from the ERC Consolidator Grant ``Precision Gravity: From the LHC to LISA"  provided by the European Research Council (ERC) under the European Union's H2020 research and innovation programme (grant agreement No. 817791), as well as from the Deutsche Forschungsgemeinschaft (DFG, German Research Foundation) under Germany's Excellence Strategy (EXC 2121) `Quantum Universe' (390833306). G.K. is supported by the Knut and Alice Wallenberg Foundation under grant KAW 2018.0441, and is supported in part by the US Department of Energy under contract DE-AC02-76SF00515.

\end{acknowledgments}
\newpage
\appendix

\section{Action \& Impulse vs. Amplitude \& Eikonal}\label{eik}

In our derivation of the total impulse in \eqref{2pmtot} we found two contributions. One of them along the $b^\mu$ direction, reproduced here for the reader's convenience (with mirror image),
\beq
 \frac{3(1-5\gamma^2)}{256 \Mp^4} {\partial \over \partial b_\mu} \left[ m_1m_2^2\int_{k,\ell}     \frac{\hat\delta(k\cdot u_2)\hat\delta(\ell\cdot u_2)\hat\delta(\ell \cdot u_1)}{k^2(\ell-k)^2} e^{i\ell \cdot b} + 1 \leftrightarrow 2\right]\,,\label{deltatrapp}
\eeq
 and another one proportional to the velocities. By inspection, the reader will notice the resemblance of the above equation with the eikonal approximation for the scattering amplitude, see e.g. \cite{Bern:2020buy}. Indeed, in the center-of-mass we have
 \beq
 \begin{aligned}
  \Delta^{(2)}\,  \bp_\perp &= \frac{1}{256 \Mp^4}\frac{3(5\gamma^2-1)}{\sqrt{\gamma^2-1}} {\partial \over \partial \bb_\perp} \left[ m_1m_2^2\int_{\bk,\bell_\perp}     \frac{1}{\bk^2(\bell_\perp-\bk)^2} e^{-i\bell_\perp \cdot \bb_\perp} + 1 \leftrightarrow 2\right] = {\partial \over \partial \bb_\perp} \theta^{(2)}_{\rm eik} \,,\label{deltatrapp2}
\end{aligned}
 \eeq
with
\beq
\theta^{(2)}_{\rm eik} \equiv   \frac{\mu M^2}{256 \Mp^4}\frac{3(5\gamma^2-1)}{\sqrt{\gamma^2-1}} \int_{\bk,\bell_\perp}     \frac{1}{\bk^2(\bell_\perp-\bk)^2} e^{-i\bell_\perp \cdot \bb_\perp} =   \frac{3\pi(5\gamma^2-1)}{4\sqrt{\gamma^2-1}}\frac{\mu (GM)^2}{|\bb_\perp|}\,.
 \eeq
 It is then straightforward to extract the (IR-finite part of the) classical limit of the amplitude. Using the relationship,
 \beq
 \theta^{(2)}_{\rm eik}(\bb_\perp) = \frac{1}{4\mu M \sqrt{\gamma^2-1}} \int_{\bq_\perp} {\cal M}^{(2)}_{\rm cl}(\bq_\perp) e^{-i\bq_\perp\cdot\bb_\perp}\,, 
 \eeq
 we find
 \beq
  {\cal M}^{(2)}_{\rm cl}(\bq) =   (5\gamma^2-1)\frac{6\pi^2G^2\mu^2M^3}{|\bq|}\,, \eeq
 which agrees with the classical limit in \cite{cheung,zvi1,zvi2} (see e.g. first line of Eq. 5.43 in \cite{Bern:2020buy}).\vskip 4pt In general, for contributions in the conservative sector, the impulse in the $b^\mu$ direction in the center-of-mass frame can always be written as    
 \beq
  \Delta p^\mu_\perp =   \int_\ell M(\ell) \delta(\ell\cdot u_1)\delta(\ell\cdot u_2) (i \ell^\mu) e^{i\ell\cdot b}\,.
  \eeq 
The numerator, $M(\ell)$, results from `loop-type' two-point functions with external (transfer) momentum $\ell$, as in \eqref{deltatrapp2}. This expression suggest the definition
  \beq
  S_{\rm eik}(|\bb_\perp|,\gamma,m_1,m_2) \equiv   \int_{\bell_\perp} M(\bell_\perp) e^{-i\bell_\perp\cdot \bb_\perp}\,,
  \eeq
  such that
  \beq
   \Delta \bp_\perp = {\partial \over \partial \bb_\perp} S_{\rm eik}(|\bb_\perp|,\gamma,m_1,m_2)\,.\label{seik}
   \eeq
   The reader will immediately inquire whether the $S_{\rm eik}$, which coincides with the eikonal phase to 2PM, $S_{\rm eik}^{(2)} =\theta_{\rm eik}^{(2)}$, is related to the classical action evaluated on the trajectories. This would not be entirely surprising, after all we can think of the scattering matrix in the classical limit as the exponential of the action on the classical solution, which would be captured by the eikonal approximation.\vskip 4pt As we know already, the Fourier transform of the amplitude encodes the value of the square of the momenta, through the impetus formula \cite{paper1}. At the same time, from the radial action we have
\beq
\frac{\chi +\pi}{2} = -\frac{1}{p_\infty} {\partial \over \partial b}\int^{\infty}_{r_{\rm min}} p_r(r,b,E) \dd r \,,
\eeq   
which combined with \eqref{seik} and \eqref{eq:apm} implies 
\beq
S_{\rm eik} = 2\left(b\frac{\pi}{2} + \int_{r_{\rm min}}^{\infty} p_r(r,b,E) \dd r\right)\,,
\eeq
to one loop order. It is straightforward to show that $S_{\rm eik}$ then coincides with (twice) the phase shift in the WKB/classical approximation, which in turn gives us the eikonal phase as we just discovered to 2PM order. At higher orders, however, the relationship is a bit more subtle, among other things because of the mismatch between the impact parameter entering in \eqref{seik} and the one in the stationary phase approximation in the eikonal approach, e.g. \cite{Parra}.\vskip 4pt  

We can, on the other hand, attempt to connect directly the impulse rather than the scattering angle. For instance, we know that on solution to the equations of motion, the derivative of the action w.r.t. the end point gives us the momenta at that point in time 
  \beq
  \bp_a(\tau_a) = {\partial S[x_a] \over \partial \bx_a(\tau_a)} \,.
  \eeq
Since at $\tau_a \to - \infty$ we have $\bp_a\cdot \bb_\perp = 0$, the impulse along $b^\mu$ is determined by the limit 
  \beq
  \Delta \bp_{a\perp}  = \lim_{\tau_a \to \infty} {\partial S[x_a] \over \partial \bx_{a \perp} (\tau_a )}\label{dpt} \,.
  \eeq
  Of course, on the undeflected trajectory, $\bx_a = \bb_a + \bu_a \tau_a$, we find
  \beq
  \Delta^{(1)} \bp_{a\perp}  = {\partial S_1[x_a] \over \partial \bb_a} \,.
  \eeq
  This tells us that, indeed, the function in \eqref{seik} is the action at 1PM evaluated on the undeflected trajectory: $S^{(1)}_{\rm eik} = S_1$.  On the other hand, the solution is shifted at 2PM, $\bb_a \to \bb_a + \delta^{(1)} \bx_a$, with $\delta^{(1)}\bx_a\cdot \bb_a \neq 0$, which implies that the derivative in \eqref{seik} and \eqref{dpt} are not the same, so that $S \neq S_{\rm eik}$ beyond leading order. This can be seen explicitly in e.g. \eqref{defl1}, where one notices that we can pull out the derivative w.r.t. the impact parameter, but only at the cost of adding a subtraction involving ${\partial \over \partial {\bb}} \delta^{(1)} \bx_a$ (and also for the velocities).
  Hence, while from Lorentz invariance and the condition $b\cdot u_a=0$ we can always construct the function $S_{\rm eik}$ in the conservative sector, the connection to the action evaluated on the solution is less direct. We will explore the properties of $S_{\rm eik}$ in more detail elsewhere.

\bibliographystyle{JHEP}
\bibliography{refpart3}

\end{document}